\documentclass[aps,prl,twocolumn,showpacs,superscriptaddress]{revtex4}  
\usepackage{graphicx}  
\usepackage{dcolumn}   
\usepackage{bm}        
\usepackage{amssymb}   
\usepackage{amsmath}

\begin{document}

\newcommand{\dzero}     {D\O}
\newcommand{\ppbar}     {\mbox{$p\bar{p}$}}
\newcommand{\ttbar}     {\mbox{$t\bar{t}$}}
\newcommand{\bbbar}     {\mbox{$b\bar{b}$}}
\newcommand{\ccbar}     {\mbox{$c\bar{c}$}}
\newcommand{\pythia}    {{\sc{pythia}}\xspace}
\newcommand{\alpgen}    {{\sc{alpgen}}\xspace}
\newcommand{\geant}     {\sc{geant}}
\newcommand{\met}       {\mbox{$\not\!\!E_T$}}
\newcommand{\rar}       {\rightarrow}
\newcommand{\eps}       {\epsilon}
\newcommand{\bs}        {{\it b $\rightarrow$ s$\gamma$}}
\newcommand{\tb}        {{\it t $\rightarrow$ b}}
\newcommand{\coss}      {\mbox{$\cos\theta^*$}}
\newcommand{\cossp}     {\mbox{\rm{cos}$\theta^{\star}_p$}}
\newcommand{\ptlep}     {$P_T^{lepton}$}
\newcommand{\ljets} {$\ell+$jets}

\hspace{5.2in} \mbox{Fermilab-Pub-10/484-E}

\title{Measurement of the W boson helicity in top quark decays using  5.4 fb$^{\boldsymbol{-1}}$ of $\boldsymbol{p\bar{p}}$ collision data}
\affiliation{Universidad de Buenos Aires, Buenos Aires, Argentina}
\affiliation{LAFEX, Centro Brasileiro de Pesquisas F{\'\i}sicas, Rio de Janeiro, Brazil}
\affiliation{Universidade do Estado do Rio de Janeiro, Rio de Janeiro, Brazil}
\affiliation{Universidade Federal do ABC, Santo Andr\'e, Brazil}
\affiliation{Instituto de F\'{\i}sica Te\'orica, Universidade Estadual Paulista, S\~ao Paulo, Brazil}
\affiliation{Simon Fraser University, Vancouver, British Columbia, and York University, Toronto, Ontario, Canada}
\affiliation{University of Science and Technology of China, Hefei, People's Republic of China}
\affiliation{Universidad de los Andes, Bogot\'{a}, Colombia}
\affiliation{Charles University, Faculty of Mathematics and Physics, Center for Particle Physics, Prague, Czech Republic}
\affiliation{Czech Technical University in Prague, Prague, Czech Republic}
\affiliation{Center for Particle Physics, Institute of Physics, Academy of Sciences of the Czech Republic, Prague, Czech Republic}
\affiliation{Universidad San Francisco de Quito, Quito, Ecuador}
\affiliation{LPC, Universit\'e Blaise Pascal, CNRS/IN2P3, Clermont, France}
\affiliation{LPSC, Universit\'e Joseph Fourier Grenoble 1, CNRS/IN2P3, Institut National Polytechnique de Grenoble, Grenoble, France}
\affiliation{CPPM, Aix-Marseille Universit\'e, CNRS/IN2P3, Marseille, France}
\affiliation{LAL, Universit\'e Paris-Sud, CNRS/IN2P3, Orsay, France}
\affiliation{LPNHE, Universit\'es Paris VI and VII, CNRS/IN2P3, Paris, France}
\affiliation{CEA, Irfu, SPP, Saclay, France}
\affiliation{IPHC, Universit\'e de Strasbourg, CNRS/IN2P3, Strasbourg, France}
\affiliation{IPNL, Universit\'e Lyon 1, CNRS/IN2P3, Villeurbanne, France and Universit\'e de Lyon, Lyon, France}
\affiliation{III. Physikalisches Institut A, RWTH Aachen University, Aachen, Germany}
\affiliation{Physikalisches Institut, Universit{\"a}t Freiburg, Freiburg, Germany}
\affiliation{II. Physikalisches Institut, Georg-August-Universit{\"a}t G\"ottingen, G\"ottingen, Germany}
\affiliation{Institut f{\"u}r Physik, Universit{\"a}t Mainz, Mainz, Germany}
\affiliation{Ludwig-Maximilians-Universit{\"a}t M{\"u}nchen, M{\"u}nchen, Germany}
\affiliation{Fachbereich Physik, Bergische  Universit{\"a}t Wuppertal, Wuppertal, Germany}
\affiliation{Panjab University, Chandigarh, India}
\affiliation{Delhi University, Delhi, India}
\affiliation{Tata Institute of Fundamental Research, Mumbai, India}
\affiliation{University College Dublin, Dublin, Ireland}
\affiliation{Korea Detector Laboratory, Korea University, Seoul, Korea}
\affiliation{CINVESTAV, Mexico City, Mexico}
\affiliation{FOM-Institute NIKHEF and University of Amsterdam/NIKHEF, Amsterdam, The Netherlands}
\affiliation{Radboud University Nijmegen/NIKHEF, Nijmegen, The Netherlands}
\affiliation{Joint Institute for Nuclear Research, Dubna, Russia}
\affiliation{Institute for Theoretical and Experimental Physics, Moscow, Russia}
\affiliation{Moscow State University, Moscow, Russia}
\affiliation{Institute for High Energy Physics, Protvino, Russia}
\affiliation{Petersburg Nuclear Physics Institute, St. Petersburg, Russia}
\affiliation{Stockholm University, Stockholm and Uppsala University, Uppsala, Sweden }
\affiliation{Lancaster University, Lancaster LA1 4YB, United Kingdom}
\affiliation{Imperial College London, London SW7 2AZ, United Kingdom}
\affiliation{The University of Manchester, Manchester M13 9PL, United Kingdom}
\affiliation{University of Arizona, Tucson, Arizona 85721, USA}
\affiliation{University of California Riverside, Riverside, California 92521, USA}
\affiliation{Florida State University, Tallahassee, Florida 32306, USA}
\affiliation{Fermi National Accelerator Laboratory, Batavia, Illinois 60510, USA}
\affiliation{University of Illinois at Chicago, Chicago, Illinois 60607, USA}
\affiliation{Northern Illinois University, DeKalb, Illinois 60115, USA}
\affiliation{Northwestern University, Evanston, Illinois 60208, USA}
\affiliation{Indiana University, Bloomington, Indiana 47405, USA}
\affiliation{Purdue University Calumet, Hammond, Indiana 46323, USA}
\affiliation{University of Notre Dame, Notre Dame, Indiana 46556, USA}
\affiliation{Iowa State University, Ames, Iowa 50011, USA}
\affiliation{University of Kansas, Lawrence, Kansas 66045, USA}
\affiliation{Kansas State University, Manhattan, Kansas 66506, USA}
\affiliation{Louisiana Tech University, Ruston, Louisiana 71272, USA}
\affiliation{Boston University, Boston, Massachusetts 02215, USA}
\affiliation{Northeastern University, Boston, Massachusetts 02115, USA}
\affiliation{University of Michigan, Ann Arbor, Michigan 48109, USA}
\affiliation{Michigan State University, East Lansing, Michigan 48824, USA}
\affiliation{University of Mississippi, University, Mississippi 38677, USA}
\affiliation{University of Nebraska, Lincoln, Nebraska 68588, USA}
\affiliation{Rutgers University, Piscataway, New Jersey 08855, USA}
\affiliation{Princeton University, Princeton, New Jersey 08544, USA}
\affiliation{State University of New York, Buffalo, New York 14260, USA}
\affiliation{Columbia University, New York, New York 10027, USA}
\affiliation{University of Rochester, Rochester, New York 14627, USA}
\affiliation{State University of New York, Stony Brook, New York 11794, USA}
\affiliation{Brookhaven National Laboratory, Upton, New York 11973, USA}
\affiliation{Langston University, Langston, Oklahoma 73050, USA}
\affiliation{University of Oklahoma, Norman, Oklahoma 73019, USA}
\affiliation{Oklahoma State University, Stillwater, Oklahoma 74078, USA}
\affiliation{Brown University, Providence, Rhode Island 02912, USA}
\affiliation{University of Texas, Arlington, Texas 76019, USA}
\affiliation{Southern Methodist University, Dallas, Texas 75275, USA}
\affiliation{Rice University, Houston, Texas 77005, USA}
\affiliation{University of Virginia, Charlottesville, Virginia 22901, USA}
\affiliation{University of Washington, Seattle, Washington 98195, USA}
\author{V.M.~Abazov} \affiliation{Joint Institute for Nuclear Research, Dubna, Russia}
\author{B.~Abbott} \affiliation{University of Oklahoma, Norman, Oklahoma 73019, USA}
\author{B.S.~Acharya} \affiliation{Tata Institute of Fundamental Research, Mumbai, India}
\author{M.~Adams} \affiliation{University of Illinois at Chicago, Chicago, Illinois 60607, USA}
\author{T.~Adams} \affiliation{Florida State University, Tallahassee, Florida 32306, USA}
\author{G.D.~Alexeev} \affiliation{Joint Institute for Nuclear Research, Dubna, Russia}
\author{G.~Alkhazov} \affiliation{Petersburg Nuclear Physics Institute, St. Petersburg, Russia}
\author{A.~Alton$^{a}$} \affiliation{University of Michigan, Ann Arbor, Michigan 48109, USA}
\author{G.~Alverson} \affiliation{Northeastern University, Boston, Massachusetts 02115, USA}
\author{G.A.~Alves} \affiliation{LAFEX, Centro Brasileiro de Pesquisas F{\'\i}sicas, Rio de Janeiro, Brazil}
% ...
%
%
\collaboration{The D0 Collaboration\footnote{with visitors from
$^{a}$Augustana College, Sioux Falls, SD, USA,
$^{b}$The University of Liverpool, Liverpool, UK,
$^{c}$SLAC, Menlo Park, CA, USA,
$^{d}$ICREA/IFAE, Barcelona, Spain,
$^{e}$Centro de Investigacion en Computacion - IPN, Mexico City, Mexico,
$^{f}$ECFM, Universidad Autonoma de Sinaloa, Culiac\'an, Mexico,
and 
$^{g}$Universit{\"a}t Bern, Bern, Switzerland.%
}} \noaffiliation
\vskip 0.25cm
\date{November 28, 2010}

\begin{abstract}
We present a measurement of the helicity of the W boson produced in top quark decays using \ttbar ~decays in the $\ell+$jets and dilepton final states selected from a sample of 5.4 fb$^{-1}$ of collisions recorded using the D0 detector at the Fermilab Tevatron \ppbar\ collider.  We measure the fractions of longitudinal and right-handed $W$ bosons to be $f_0 = 0.669 \pm 0.102 \mbox{ } [ \pm 0.078 \hbox{ (stat.)} \pm  0.065 \hbox{ (syst.)}]$ and $f_+ = 0.023 \pm 0.053\mbox{ }[\pm 0.041 \hbox{ (stat.)} \pm  0.034  \hbox{ (syst.)}]$, respectively.  This result is consistent at the 98\% level with the standard model.  A measurement with $f_0$ fixed to the value from the standard model yields $f_+ = 0.010 \pm 0.037 \mbox{ }[\pm 0.022 \hbox{ (stat.)} \pm 0.030 \hbox{ (syst.). }]$
\end{abstract}

\pacs{14.65.Ha, 14.70.Fm, 12.15.Ji, 12.38.Qk, 13.38.Be, 13.88.+e}
\maketitle 


\section{\label{sec:intro}Introduction}
The top quark is the heaviest known fundamental particle and was discovered in 1995~\cite{cdftopobs,d0topobs} at the Tevatron proton-antiproton collider at Fermilab. The dominant top quark production mode at the Tevatron is
$p\bar{p} \rightarrow t\bar{t}X$.  Since the time of discovery, over 100 times more integrated luminosity has been collected, providing a large number of \ttbar\ events with which to study the properties of the top quark. In the standard model (SM), the branching ratio for the top quark to decay to a $W$ boson and a $b$ quark is $> 99.8$\%. The on-shell $W$ boson from the top quark decay has three possible helicity states, and we define the fraction of $W$ bosons produced in these states as $f_0$ (longitudinal), $f_-$ (left-handed), and $f_+$ (right-handed). In the SM, the top quark decays via the $V-A$ charged weak current interaction, which strongly suppresses right-handed $W$ bosons and predicts $f_0$ and $f_-$ at leading order in terms of the top quark mass ($m_t$), $W$ boson mass ($M_W$), and $b$ quark mass ($m_b$) to be~\cite{fval}
\begin{eqnarray}
f_0 =  \frac{(1-y^2)^2-x^2(1+y^2)}{(1-y^2)^2+x^2(1-2x^2+y^2)} \\
f_- = \frac{x^2(1-x^2+y^2+\sqrt{\lambda})}{(1-y^2)^2+x^2(1-2x^2+y^2)} \\
f_+=\frac{x^2(1-x^2+y^2-\sqrt{\lambda})}{(1-y^2)^2+x^2(1-2x^2+y^2)}
\end{eqnarray}
where $x=M_W/m_t$,  $y=m_b/m_t$, and $\lambda = 1+x^4+y^4-2x^2y^2-2x^2-2y^2$. With the present measurements of $m_t = 173.3 \pm 1.1$ GeV/$c^2$~\cite{topmass} and  $M_W=80.399 \pm 0.023$ GeV$/c^2$~\cite{wmass}, and taking $m_b$ to be 5 GeV/$c^2$, the SM expected values are $f_0$=0.698, $f_-$=0.301, and $f_+=4.1 \times10^{-4}$. The absolute uncertainties on the SM expectations, which arise from uncertainties on the particle masses as well as contributions from higher-order effects, are $\approx (0.01 - 0.02)$ for $f_0$ and $f_-$, and ${\cal O}(10^{-3})$ for $f_+$~\cite{fval}. 

In this paper, we present a measurement of the $W$ boson helicity fractions $f_0$ and $f_+$ and constrain the fraction $f_-$ through the unitarity requirement of $f_- + f_+ + f_0 = 1$. Any significant deviation from the SM expectation would be an indication of new physics, arising from either a deviation  from the expected $V-A$ coupling of the $tWb$ vertex or the presence of non-SM events in the data sample. The most recently published results are summarized in Table~\ref{tab:prevMeas}.

\begin{table}
\caption{\label{tab:prevMeas} Summary of the most recent $W$ boson helicity measurements from the D0~ \cite{prevd0result} and CDF~\cite{prevcdfresult} collaborations.  The first uncertainty is statistical and the second systematic.   
}
\begin{tabular}{cl}
\hline
\hline 
D0, 1 fb$^{-1}$ \cite{prevd0result} & $f_0 = 0.425 \pm 0.166 \pm 0.102,$        \\ 
                              &  $f_+ = 0.119 \pm 0.090 \pm 0.053$      \\
                              &  $f_+$ fixed: $f_0 = 0.619 \pm 0.090 \pm 0.052$      \\
                              &  $f_0$ fixed: $f_+ = -0.002 \pm 0.047 \pm 0.047$       \\ \hline 
CDF,  2.7 fb$^{-1}$ \cite{prevcdfresult}                      &  $f_0 = 0.88 \pm 0.11 \pm 0.06,$             \\
                              &    $f_+ = -0.15 \pm 0.07 \pm 0.06$            \\
                              &   $f_+$ fixed: $f_0 = 0.70 \pm 0.07 \pm 0.04$   \\  
                              &    $f_0$ fixed: $f_+ = -0.01 \pm 0.02 \pm 0.05$  \\ \hline \hline
\end{tabular}
\end{table}


The extraction of the $W$ boson helicities is based on
the measurement of the angle  $\theta^{\star}$ between the opposite of the
direction of the top quark and the direction of the down-type fermion
(charged lepton or $d$, $s$ quark) decay product of the $W$ boson in the $W$
boson rest frame.  The dependence of the distribution of \coss on the $W$ boson helicity fractions is given by 
\begin{eqnarray}
\omega(c) \propto 2(1-c^2)f_0 + (1-c)^2 f_- + (1+c)^2 f_+
\label{eq:expcost}
\end{eqnarray}
with $c=\coss$. After selection of a \ttbar\ enriched sample the four-momenta of the \ttbar\ decay products in each event are reconstructed as described below, permitting the calculation of \coss. Once the \coss\ distribution is measured, the values of $f_0$ and $f_+$ are extracted with a binned Poisson likelihood fit to the data. The measurement presented here is based on \ppbar ~collisions at a center-of-mass energy $\sqrt s$ = 1.96 TeV corresponding to an integrated luminosity  of 5.4~fb$^{-1}$, five times more than the amount used in the result in Ref.~\cite{prevd0result}. 

\section{\label{sec:detector} Detector}

The D0 Run II detector~\cite{d0nim} is a multipurpose detector which consists of three primary systems: a central tracking system, calorimeters, and a muon spectrometer. We use a standard right-handed coordinate system. The nominal collision point is the center of the detector with coordinate (0,0,0). The direction of the proton beam is the positive $+z$ axis. The $+x$ axis is horizontal, pointing away from the center of the Tevatron ring. The $+y$ axis points vertically upwards.  The polar angle, $\theta$, is defined such that $\theta = 0$ is the $+z$ direction. Usually, the polar angle is replaced by the pseudorapidity $\eta = - \ln \tan\left(\frac{\theta}{2}\right)$. The azimuthal angle, $\phi$, is defined such that $\phi =0$ points along the $+x$ axis, away from the center of the Tevatron ring. 

The silicon microstrip tracker (SMT) is the innermost part of the tracking system and has a six-barrel longitudinal structure, where each barrel consists of a set of four layers arranged axially around the beam pipe.  A fifth layer of SMT sensors was installed near the beam pipe in 2006~\cite{smtl0}.  The data set recorded before this addition is referred to as the ``Run IIa'' sample, and the subsequent data set is referred to as the ``Run IIb'' sample.   Radial disks are interspersed between the barrel segments.   The SMT provides a spatial resolution of approximately 10 $\mu$m in $r-\phi$ and 100 $\mu$m in $r-z$ (where $r$ is the radial distance in the $x$-$y$ plane) and covers $|\eta| < 3$. The central fiber tracker (CFT) surrounds the SMT and consists of eight concentric carbon fiber barrels holding doublet layers of scintillating fibers (one axial and one small-angle stereo layer),  with the outermost barrel covering $|\eta| < 1.7$. The solenoid surrounds the CFT and provides a 2 T uniform axial magnetic field.\\
The liquid-argon/uranium calorimeter system is housed in three cryostats, with the central calorimeter (CC) covering $|\eta|<1.1$ and two end calorimeters (EC) covering $1.5 < |\eta| < 4.2$.  The calorimeter is made up of unit cells consisting of an absorber plate and a signal board; liquid argon, the active material of the calorimeter, fills the gap. The inner part of the calorimeter is the electromagnetic (EM) section and the outer part is the hadronic section.\\
 The muon system is the outermost part of the D0 detector and covers $|\eta| < 2$. It is primarily made of two types of detectors, drift tubes and scintillators, and consists of three layers (A,B and C). Between layer A and layer B, there is magnetized steel with a 1.8 T toroidal field.

\section{\label{sec:samples} Data and Simulation samples}

At the Tevatron, with proton and anti-proton bunches colliding at intervals of 396 ns, the collision rate is about 2.5 MHz. Out of these $2.5\times10^{6}$ beam crossings per second at D0, only those that produce events which are identified by a three-level trigger system as having properties matching the characteristics of physics events of interest are retained, at a rate of $\sim$100~Hz~\cite{d0nim,l1cal2b}. This analysis is performed using events collected with the triggers applicable for $\ell+$jets and dilepton final states between April 2002 and June 2009, corresponding to a total integrated luminosity of 5.4 fb$^{-1}$.  Analysis of the Run IIa sample, which totals about 1 fb$^{-1}$, was presented in Ref.~\cite{prevd0result}.  Here we describe the analysis of the Run IIb data sample and then combine our result with the result from Ref.~\cite{prevd0result} when reporting our measurement from the full data sample.

The Monte Carlo (MC)  simulated samples used for modeling the data are generated with {\sc alpgen}~\cite{ref:alpgen} interfaced to {\sc pythia}~\cite{ref:pythia} for parton shower simulation, passed through a detailed detector simulation based on {\sc geant}~\cite{geant}, overlaid with data collected from a random subsample of beam crossings to model the effects of noise and multiple interactions, and reconstructed using the same algorithms that are used for data. For the signal (\ttbar) ~sample, we must model the distribution of \coss\
corresponding to any set of values for the $W$ boson helicity fractions,  a 
task that is complicated by the fact that {\sc alpgen}  can only produce linear combinations of $V-A$ and $V+A$ $tWb$ couplings. Hence, for this analysis, we use samples that are either purely $V-A$  or purely   $V+A$, and use a reweighting procedure (described below) to form models of arbitrary helicity states. {\sc alpgen} is also used for generating all $V+$jets processes where $V$ represents the vector bosons. {\sc pythia} is used for generating diboson ($WW$, $WZ$, and $ZZ$) backgrounds in the dilepton channels. Background from multijet production is modeled using data.

\section{\label{sec:eventselection1}Event Selection}

We expect {\it a priori} that our measurement will be limited by statistics, so our analysis strategy aims to maximize the acceptance for \ttbar\ events.  
The selection is done in two steps. In the first step, a loose initial selection using data quality, trigger, object identification, and kinematic
criteria is applied to define a sample with the characteristics of \ttbar\ events. Subsequently, a multivariate likelihood discriminant is defined to separate the \ttbar ~signal from the background in the data.  We use events in the $\ell+$jets and dilepton \ttbar\ decay channels, which are defined below.

 

In the $\ell+$jets decay $\ttbar ~\rightarrow~W^{+}~W^{-}\bbbar ~\rightarrow~\ell\nu~qq^{'} \bbbar$, events contain one charged lepton (where lepton here refers to an electron or a muon), at least four jets with two of them being $b$ quark jets, and significant missing transverse energy \met (defined as the opposite of the vector sum of the transverse energies in each calorimeter cell, corrected for the energy carried by identified muons and energy added or subtracted due to the jet energy calibration described below) . The event selection requires at least four jets with transverse momentum $p_T > 20$ GeV/$c$ and $|\eta| < 2.5$ with the leading jet $p_T > 40$ GeV/$c$. At least one lepton is required with $p_T > 20$ GeV/$c$ and $|\eta| < $ 1.1 (2.0) for electrons (muons).  Requirements are also made on the value of \met\ and the angle between the \met\ vector and the lepton (to reduce the contribution of events in which mismeasurement of the lepton energy gives rise the spurious \met): in the $e+$jets channel the requirement is  $\met > 20$ GeV and $\Delta\phi(e,\met) > 0.7\pi - 0.045\cdot\met\hbox{/GeV}$, and in the $\mu+$jets channel the requirement is  $\met > 25$ GeV and $\Delta\phi(\mu,\met) > 2.1 - 0.035\cdot\met\hbox{/GeV}$.   In addition, for the $\mu+$jets channel, the invariant mass of the selected muon and any other muon in the event is required to be outside of the $Z$ boson mass window ($< 70$ GeV/$c^2$ or $> 100$ GeV/$c^2$).

For the dilepton decay channel,  $\ttbar ~\rightarrow W^{+} W^{-} \bbbar ~\rightarrow \bar{\ell}\nu\ell^\prime\bar{\nu^\prime} \bbbar$, the signature is two leptons of opposite charge, two $b$ quark jets, and significant \met.  The event selection requires at least two jets with  $p_T > 20$ GeV/$c$ and $|\eta| < 2.5$ and two leptons (electron or muon) with $p_T > 20$ GeV/$c$. The muons are required to have $|\eta| < 2.0$, and the electrons are required to have $|\eta| < 1.1$ or $1.5 < |\eta| < 2.5$.

Jets are defined using a mid-point cone algorithm~\cite{jetalg} with radius 0.5.  Their energies are first calibrated to be equal, on average, to the sums of the energies of the particles within the jet cone.  This calibration accounts for the energy response of the calorimeters, the energy that crosses the cone boundary due to the transverse shower size, and the additional energy from event pileup and multiple $p\bar{p}$ interactions in a single beam crossing.  The energy added to or subtracted from each jet in due to the above calibration is propagated to the calculation of \met. Subsequently, an additional correction to for the average energy radiated by gluons  outside of the jet cone is applied to the jet energy.
Electrons are identified by their energy deposition and shower shape in the calorimeter combined with information from the tracking system.  Muons are identified using information from the muon detector and the tracking system. We require the (two) highest-$p_T$ lepton(s) to be isolated from other tracks and calorimeter energy deposits in the $\ell+$jets (dilepton) channel.  For all channels, we require a well-reconstructed $p\bar{p}$ vertex (PV) with the distance in $z$ between this vertex  and the point of closest approach of the lepton track being less than 1 cm. 

 The main sources of background after the initial selection in the $\ell+$jets channel are $W+$jets and multijet production; in the dilepton channels they are $Z$ boson and diboson production as well as multijet and $W$+jets production. Events with fewer leptons than required (multijet events, or $W+$jets events in the dilepton channel) can enter the sample when  jets are either misidentified as leptons or contain a lepton from semileptonic quark decay that passes the electron likelihood or muon isolation criterion. In all cases they are modeled using data with relaxed lepton identification or isolation criteria. The multijet contribution to the $\ell+$jets final states in the initially-selected sample is estimated from data following the method described in Ref.~\cite{matrix}. This method relies on the selection of two data samples, one (the tight sample) with the standard lepton criteria, and the other (the loose sample) with relaxed isolation or identification criteria. The numbers of events in each sample are:

\begin{align}
N_{\rm loose} &= \phantom{\varepsilon_{sig}} N^{
\ttbar + W}+\phantom{\varepsilon_{\rm MJ}} N^{\rm MJ}  \label{eq:matrix1} \\
N_{\rm tight}   &= \varepsilon_{\ell}   N^{
\ttbar + W}+\varepsilon_{\rm MJ} N^{\rm MJ} \label{eq:matrix2}
\end{align}
Here the coefficient $\varepsilon_{\ell}$ is the efficiency for isolated leptons in \ttbar\ or $Wjjjj$ events to satisfy the standard lepton requirements, while $\varepsilon_{\rm MJ}$ is the efficiency for a jet in multijet events to satisfy those requirements.  We measure   $\varepsilon_{\ell}$ in $Z\rightarrow\ell\ell$ control samples and  $\varepsilon_{\rm MJ}$ in multijet control samples.  Inserting the measured values, we solve Eqs. ~\ref{eq:matrix1} and~\ref{eq:matrix2} to obtain the number of
multijet events ($N^{\rm MJ}$) and
      the number of events with isolated leptons ($N^{\ttbar + W}$).   In the dilepton channels we model the background due to jets being
misidentified as isolated leptons using data events where both leptons have the same charge.  This background originates from multijets events with two jets misidentified as leptons and from
$W+$jets events with one jet misidentified as a lepton.\\

To separate the \ttbar\ signal from these sources of background, we define a multivariate likelihood and retain only events above a certain threshold in the value of that likelihood.  The set of variables used in the likelihood and the threshold value are optimized separately for each \ttbar\ decay channel.  The first step in the optimization procedure is to identify a set of candidate variables that may be used in the likelihood.  The set we consider is:

\begin{itemize}

\item{{\bf Aplanarity $\boldsymbol{{\cal A}}$}, defined as 3/2 of the smallest eigenvalue of
the normalized momentum tensor for the jets (in the $\ell$+jets channels) or jets and leptons (in the dilepton channels).  The aplanarity ${\cal A}$ is a measure
of the deviation from flatness of the event, and \ttbar~ events tend to
have larger values than background.}

\item{ {\bf Sphericity $\boldsymbol{{\cal S}}$}, defined as 3/2 of the sum of the two smallest eigenvalues
of the normalized momentum tensor for the jets (in the $\ell$+jets channels) or jets and leptons (in the dilepton channels).  This variable is a measure of the isotropy of the energy flow in the event, and \ttbar~ events tend to have larger values than background.}

\item{$\boldsymbol{H_T}$, introduced in Refs.~\cite{cdftopevidence} and~\cite{d0runItopsearch}, is defined as the scalar sum of the jets' $p_T$ values.
Jets arising from gluon radiation often have lower $p_T$ than jets in \ttbar~ events, so 
background events tend to have smaller values of $H_T$ than signal.}

\item{{\bf Centrality $\boldsymbol{{\cal C}}$}, defined as $\frac{H_T}{H_E}$ where $H_E$ is the sum of all
jet energies.  The centrality ${\cal C}$ is similar to $H_T$ but normalized in a way to
minimize dependence on the top quark mass.}

\item{$\boldsymbol{{K_{T\text{\bf min}}^\prime}}$, defined as $\Delta R_{jj{\rm min}}\cdot\frac{E_{T{\rm min}}}{E_T^W}$, where $\Delta R_{jj{\rm min}}$ is the distance in $\eta-\phi$ space
between the closest pair of jets,  $E_{T\rm min}$ is the lowest jet $E_T$ value 
in the pair, and $E_T^W$ is the transverse energy of the leptonically-decaying $W$ boson (in the dilepton
channels $E_T^W$  is the magnitude of the vector sum of the \met\ and
leading lepton $p_T$). Only the four
leading-$E_T$ jets are considered in computing this variable.  Jets arising
from gluon radiation (as is the case for most of the  background) tend to have
lower values of $K_{T{\rm min}}^\prime$.}

\item{$\boldsymbol{{m_{jj{\text{min}}}}}$, defined as the smallest dijet mass of pairs of selected jets.  This
variable is sensitive to gluon radiation and tends to be smaller for  background than signal.}

\item{$\boldsymbol{h}$, defined as the scalar sum of all the selected jet and lepton energies.  Jets arising from gluon radiation often have lower energy than jets in \ttbar~ events, and 
leptons arising from the decay of heavy flavor jets often have lower energy than leptons from $W$ boson decay, so 
background events tend to have smaller values of $h$ than signal.}

\item{{\bf  $\boldsymbol{ \chi^2_k}$}, defined as the $\chi^2$ for a kinematic fit of $\ell+$jets
final states to the \ttbar\ hypothesis.  Signal events tend to have
smaller $\chi^2$ values than background.  This variable is not used for dilepton events, for which 
a kinematic fit is underconstrained.}

\item{$\boldsymbol{\Delta\phi(\hbox{\bf lepton}, \met)}$, defined as the angle between the leading lepton and
the \met. $W+$jets events with \met\ arising from mismeasured
lepton $p_{T}$ tend to have $\Delta\phi(\hbox{lepton}, \met) \approx 0$ or $\pi$.}

\item{{\bf $\boldsymbol{b}$ jet content of the event}.  Due to the long lifetime of the $b$ quark, tracks within jets arising from $b$ quarks have different properties (such as larger impact parameters with respect to the PV and the presence of secondary decay vertices) than tracks within light-quark or gluon jets.  The  consistency of a given jet with the hypothesis that the jet was produced by a $b$ quark is quantified with a neural network (NN) that considers several properties of the tracks contained within the jet cone~\cite{bidNIM}.  In the $\ell+$jets channels, we take the average of the NN values NN$_b$ of the two most $b$-like jets to form a variable called NN$_{b{\rm avg}}$, and in the dilepton channels we take the NN$_b$ values of the two most $b$-like jets as separate variables NN$_{b1}$  (the largest NN$_{b}$ value) and NN$_{b2}$ (the second-largest NN$_b$ value). For top quark events, these variables tend to be close to one, while for events containing only light jets they tend to be close to zero.}

\item{$\boldsymbol {\met}$ {\bf or}  $\boldsymbol{\chi^2_Z}$}. For the $e\mu$ and $ee$ channels only, \met\ is considered as a variable in the likelihood discriminant.   In the $\mu\mu$ channel, where
 spurious \met\  can arise from mismeasurement of the muon momentum, we instead use $\chi^2_Z$, the $\chi^2$ of a kinematic fit to the $Z\rightarrow\mu\mu$ hypothesis.  

\item{\bf Dilepton mass }$\boldsymbol{m_{\ell\ell}}.$  Also for the dilepton channels only, the invariant mass of the lepton pairs is considered as a variable in the classical likelihood.  The motivation is to discriminate against $Z$ boson production.
\end{itemize}

We consider all combinations of the above variables to select the optimal set to use for each \ttbar\ decay channel.  For a given combination of variables, the likelihood ratio $L_t$ is defined as


\begin{eqnarray}
L_t = \frac{\exp\left\{\sum_{i=1}^{N_{\rm var}} [\ln(\frac{S}{B})_i^{\text{fit}}]\right\}}
{\exp\left\{\sum_{i=1}^{N_{\rm var}} [\ln(\frac{S}{B})_i^{\text{fit}}]\right\}+ 1},
\label{eq:claslhood}
\end{eqnarray}
where $N_{\rm var}$ is the number of input variables used in the likelihood, and $(\frac{S}{B})_i^{\text{fit}}$ is the ratio of the parameterized signal and background probability density functions. We consider all possible subsets of the above variables to be used in $L_t$ and scan across all potential selection criteria on $L_t$.  For each $L_t$ definition and prospective selection criterion, we compute the following  figure of merit (FOM):
\begin{eqnarray}
  {\rm FOM} = \frac{N_S}{\sqrt{N_S + N_B + \sigma^2_{B}}},
\label{eq:FOM}
\end{eqnarray}
where $N_S$ and $N_B$ are the numbers of signal and background events expected to satisfy the 
$L_t$ selection.\\   

The term $\sigma_{B}$ reflects the uncertainty in the background selection efficiency arising from any mis-modeling of the input variables in the MC.  To assess $\sigma_{B}$, we compare each variable in data and MC in background-dominated samples.  The background-dominated samples are created by forming a multivariate likelihood ratio (Eq.~\ref{eq:claslhood}) that does not use the variable
under study, nor any variable that is strongly correlated with it, where the criterion is a correlation coefficient between $-$0.10 and 0.10.   We select events that have low 
values of this likelihood,  and are therefore unlikely to be \ttbar\ events, such that 95\% of MC \ttbar\ events are rejected.  Because the \ttbar\ contribution to the selected data sample is negligible, we can directly compare the background model to data.  The impact of any mis-modeling on the  likelihood distribution is assessed  by taking the ratio of the observed to the expected distributions as a function of each variable and fitting this to a polynomial. The result is that for each variable $i$ we build a function $k_i$ that encodes the data/MC discrepancies in that variable. For each simulated background event, we reweight each likelihood according to the data/MC differences. For example, for a likelihood that uses $n$ of the possible variables, the likelihood is given a weight
\begin{eqnarray}
w=\prod_{i=1}^{n} k_i(v_i).
\end{eqnarray}
The quantity  $\sigma_{B}$ is the difference in the predicted background yield when the unweighted and weighted $L_t$ distributions are used for background.  This uncertainty is propagated through the analysis as one component of the total uncertainty in the background yield.

\begin{table}[hhh]
\caption{\label{tab:optimization} The set of variables chosen for use in $L_t$ for the $e$+jets and $\mu+$ jets channels.  The numbers of background and $t\bar{t}$ events in the initially-selected data, as determined from a fit to the $L_t$ distribution, are also presented.}
\begin{tabular}{lr@{$\,\pm \,$}lr@{$\,\pm \,$}l}
\hline
\hline 
                                  &  \multicolumn{2}{c}{$e+$jets}  & \multicolumn{2}{c}{$\mu+$jets}\\ \hline
Events passing initial selection                 & \multicolumn{2}{c}{1442} & \multicolumn{2}{c}{1250}\\
\hline
Variables in best $L_t$        & \multicolumn{2}{c}{${\cal C}$} & \multicolumn{2}{c}{${\cal C}$}\\
                     & \multicolumn{2}{c}{${H_T}$} & \multicolumn{2}{c}{${H_T}$}  \\
                   & \multicolumn{2}{c}{${K_{T\text{min}}^\prime}$} & \multicolumn{2}{c}{${K_{T\text{min}}^\prime}$} \\
                  & \multicolumn{2}{c}{NN$_{b{\rm avg}}$} & \multicolumn{2}{c}{NN$_{b{\rm avg}}$}\\
                      & \multicolumn{2}{c}{$\mathbf \chi^2_k$} & \multicolumn{2}{c}{$h$}  \\
                   & \multicolumn{2}{c}{${m_{jj{\text{min}}}}$} & \multicolumn{2}{c}{}\\
                  & \multicolumn{2}{c}{Aplanarity} &  \multicolumn{2}{c}{} \\
\hline
$N$ (\ttbar)         & 592.6 & 31.8 & 612.7 & 31.0\\
$N$ ($W+$jets)   & 690.2 & 21.8 & 579.8 & 18.6\\
$N$ (multijet)  & 180.3 & 9.9 & 6.5 &  4.9\\
\hline
\hline
\end{tabular}
\end{table}

\begin{table*}[htdp]
\caption{\label{tab:ll_ltfits}The set of variables chosen for use in $L_t$ for the dilepton channels.  The number of background and $t\bar{t}$ events in the initially-selected data, as determined from a fit to the $L_t$ distribution, are also presented.}
\begin{center}
\begin{tabular}{lr@{$\,\pm \,$}llr@{$\,\pm \,$}llr@{$\,\pm \,$}l}
\hline
\hline
& \multicolumn{2}{c}{$e\mu$} &  &  \multicolumn{2}{c}{$ee$} & &\multicolumn{2}{c}{$\mu\mu$} \\ \hline
Events passing initial selection               & \multicolumn{2}{c}{323}  & &  \multicolumn{2}{c}{3275} &  & \multicolumn{2}{c}{5740}   \\ \hline
 Variables in optimized $L_t$ & \multicolumn{2}{c}{${\cal A}$,${\cal S}$,$h$,$m_{jj\text{min}}$} & $\mbox{ }$ &\multicolumn{2}{c}{${\cal A}$,${\cal S}$,$m_{jj\text{min}}$} & $\mbox{ }$ & \multicolumn{2}{c}{${\cal A}$,${\cal S}$,$m_{jj\text{min}}$,$K_{T\text{min}}^\prime$}\\
                  & \multicolumn{2}{c}{$K_{T\text{min}}^\prime$,\met,NN$_{b1}$,$m_{\ell\ell}$} & &\multicolumn{2}{c}{\met,NN$_{b1}$,$m_{\ell\ell}$} & & \multicolumn{2}{c}{$\chi^2_Z$,NN$_{b1}$}\\ 
\hline
$N$ (\ttbar)     & 178.7 & 15.6 &  & 74.9 & 10.7 &   & 86.0 & 13.8  \\
$N$ (background) &  144.3 & 14.5 & &  3200 & 57 &  & 5654 & 76 \\ \hline
\hline
\end{tabular}
\end{center}
\end{table*}

\begin{figure*}[tbp]
  \includegraphics[scale=0.4]{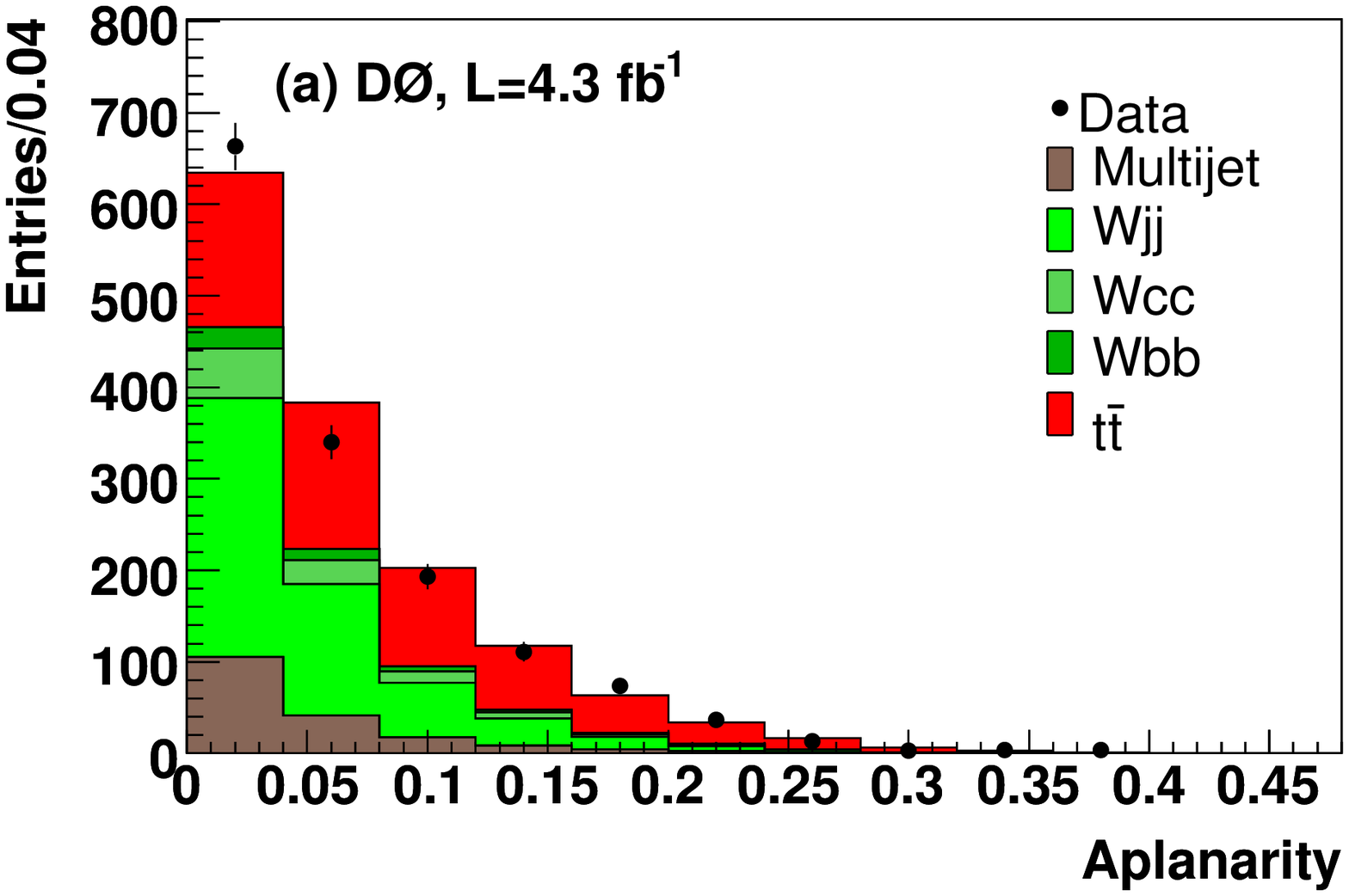}
  \includegraphics[scale=0.4]{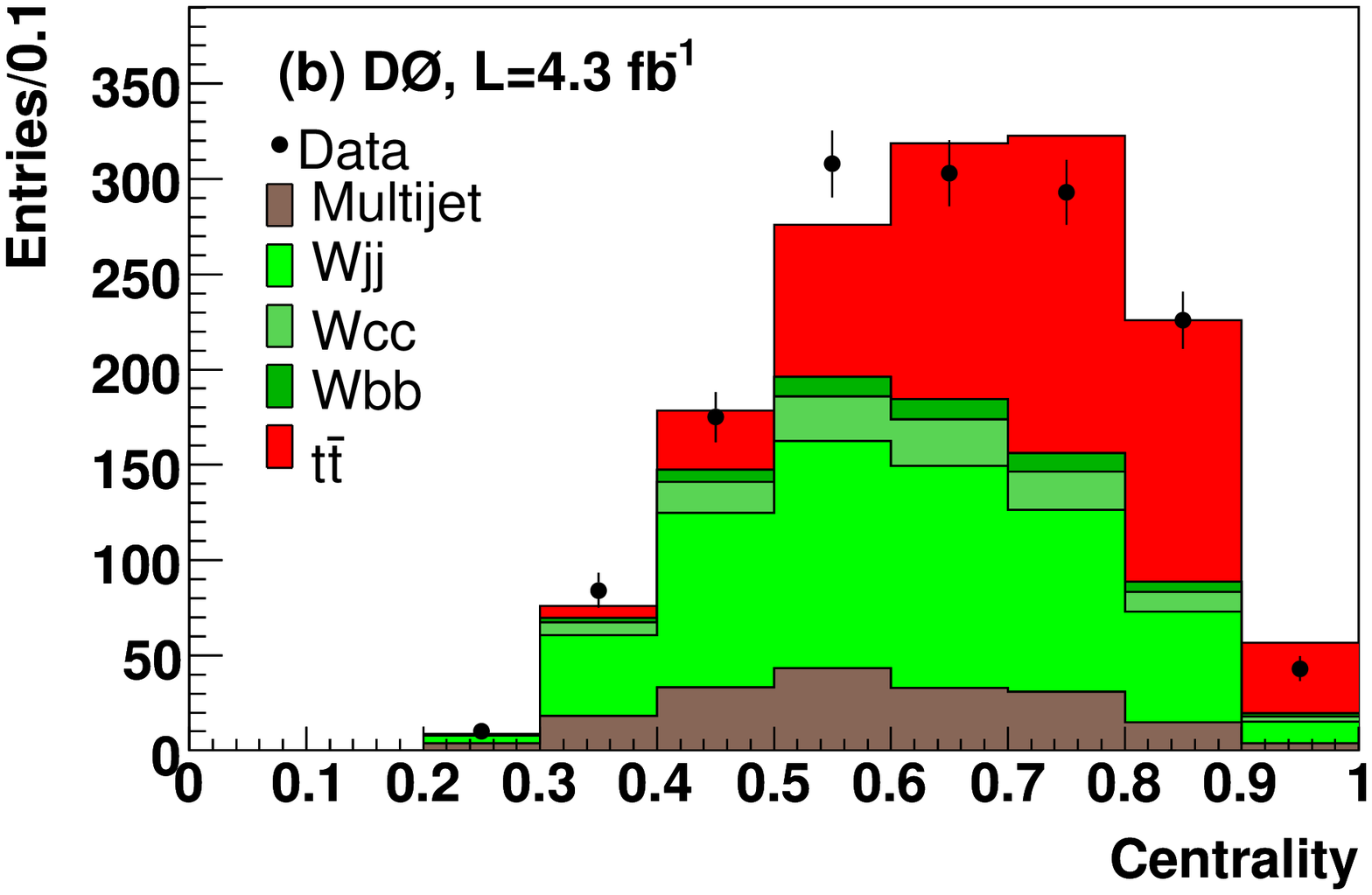}
  \includegraphics[scale=0.4]{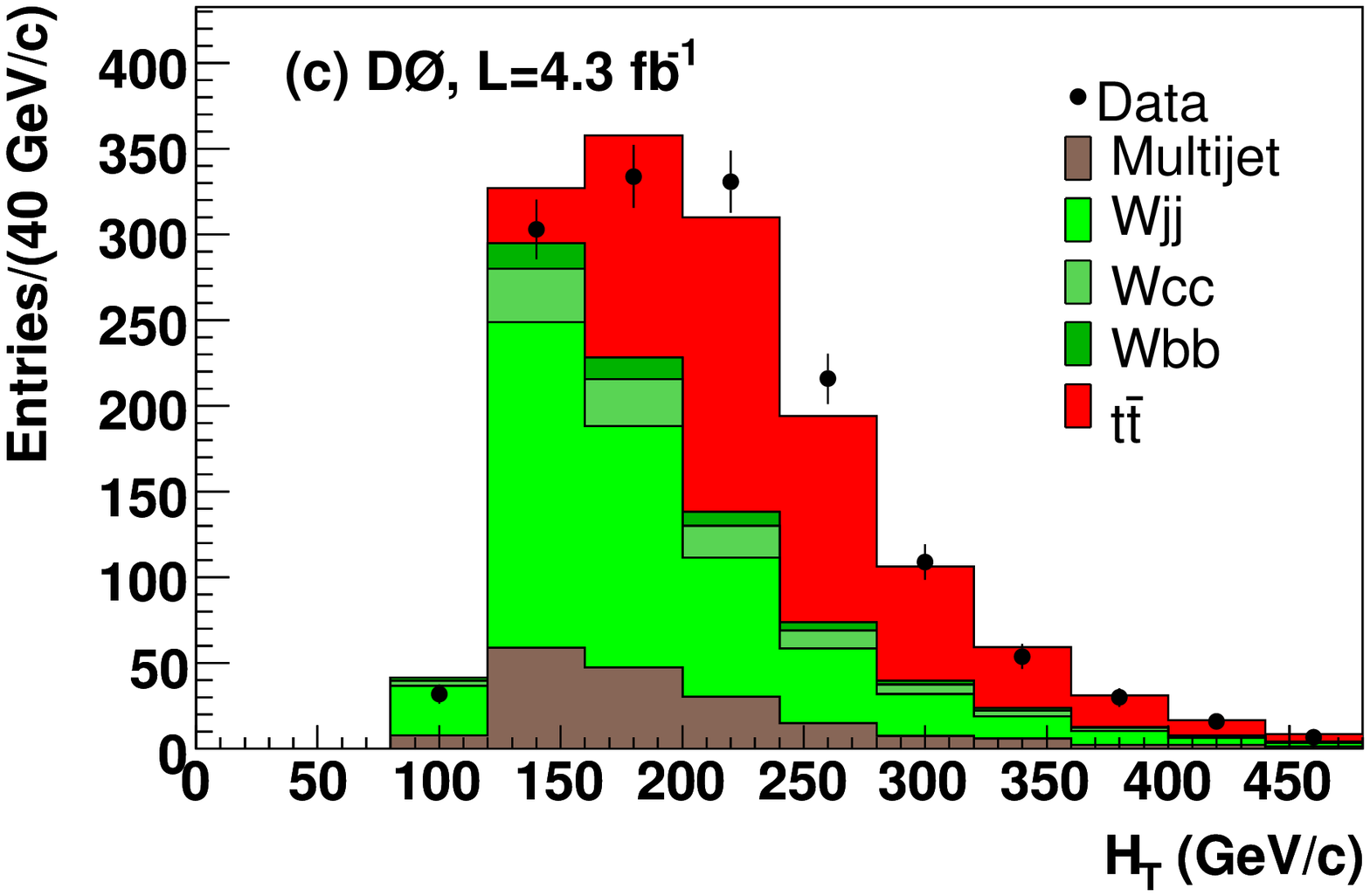}
  \includegraphics[scale=0.4]{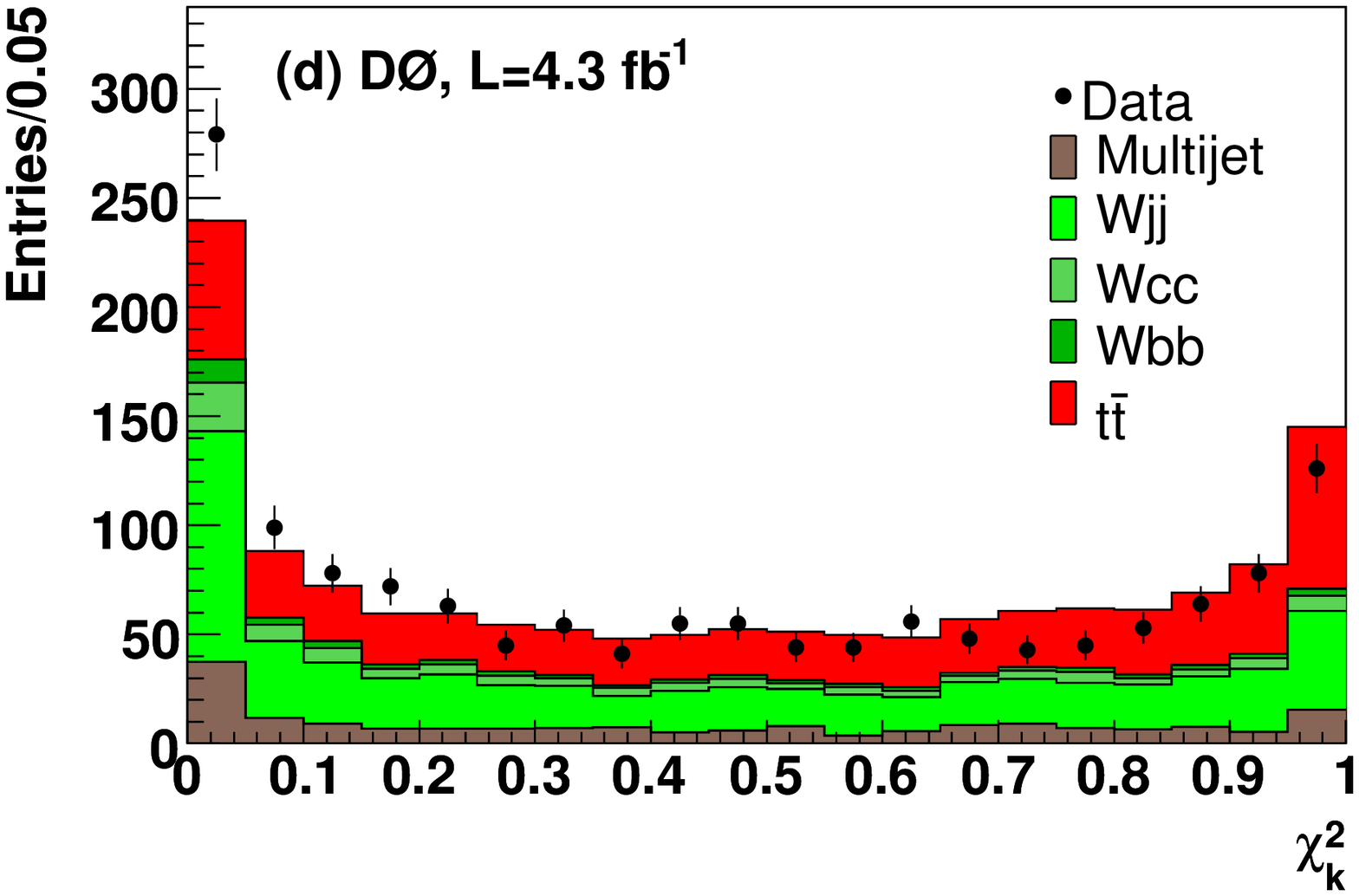}
  \includegraphics[scale=0.4]{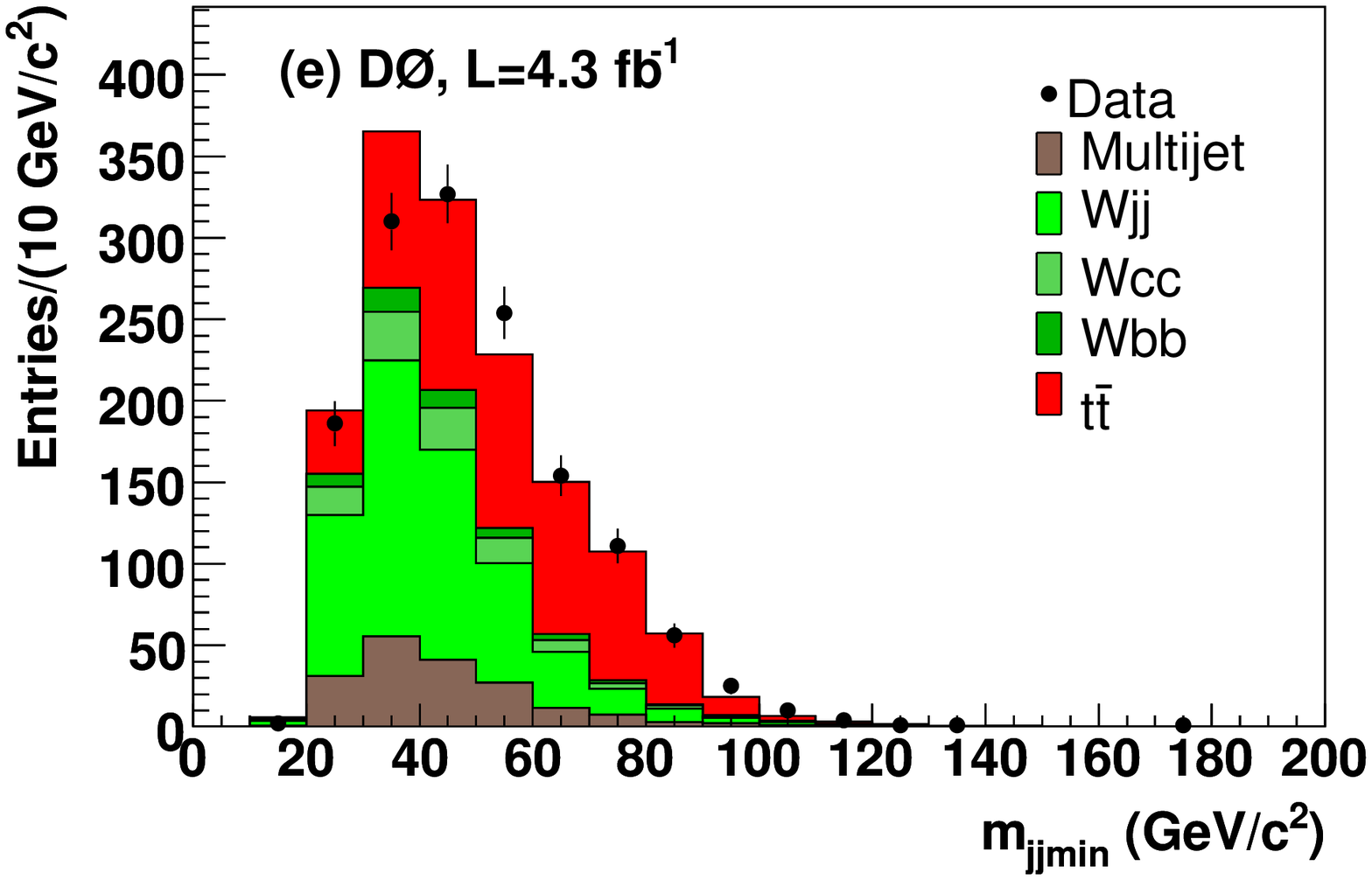}
  \includegraphics[scale=0.4]{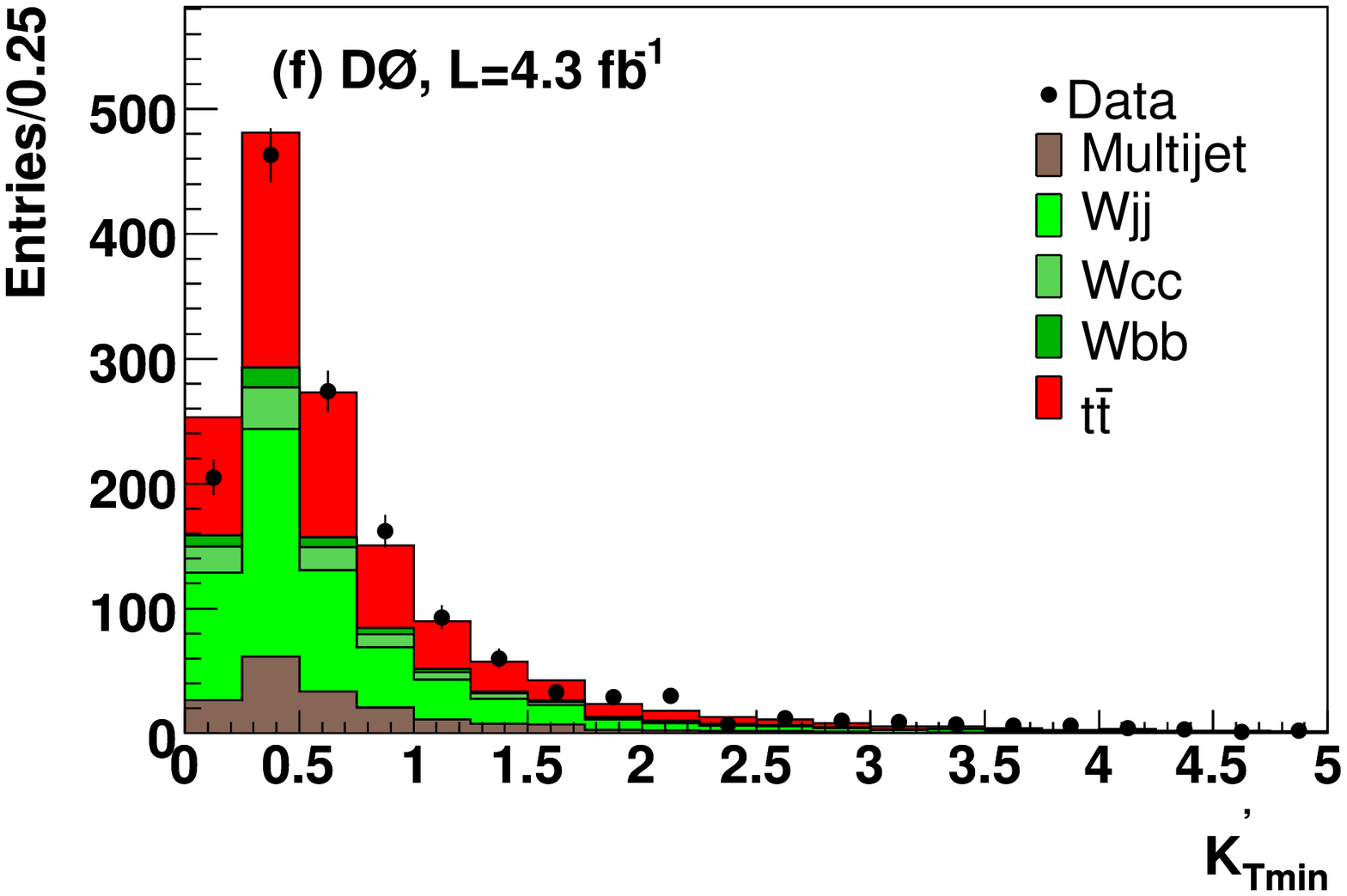}
  \includegraphics[scale=0.4]{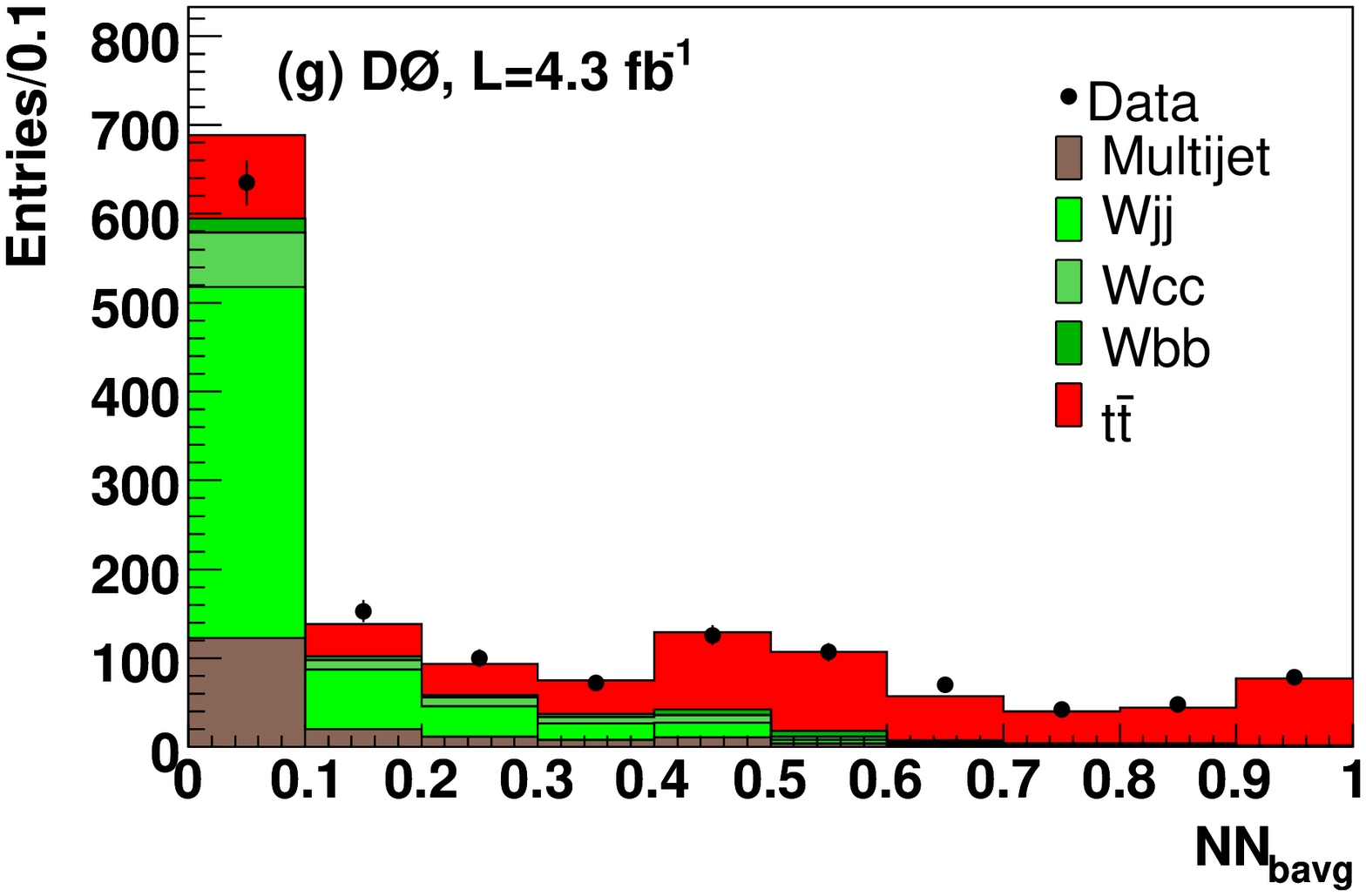}
  \caption{\label{fig:input_ejets}  (Color online) Comparison of data and MC of the variables for preselected events, chosen for the best likelihood discriminant $L_{t}$ in the $e+$jets channel: (a) ${\cal A}$, (b) ${\cal C}$, (c) $H_T$, (d) $\chi_k^2$, (e) $m_{jj\text{min}}$, (f) ${K_{T\text{min}}^\prime}$, and (g) NN$_{b\rm{avg}}$.  The uncertainties on the data points are statistical only. }
\end{figure*}

\begin{figure*}[tbp]
  \includegraphics[scale=0.4]{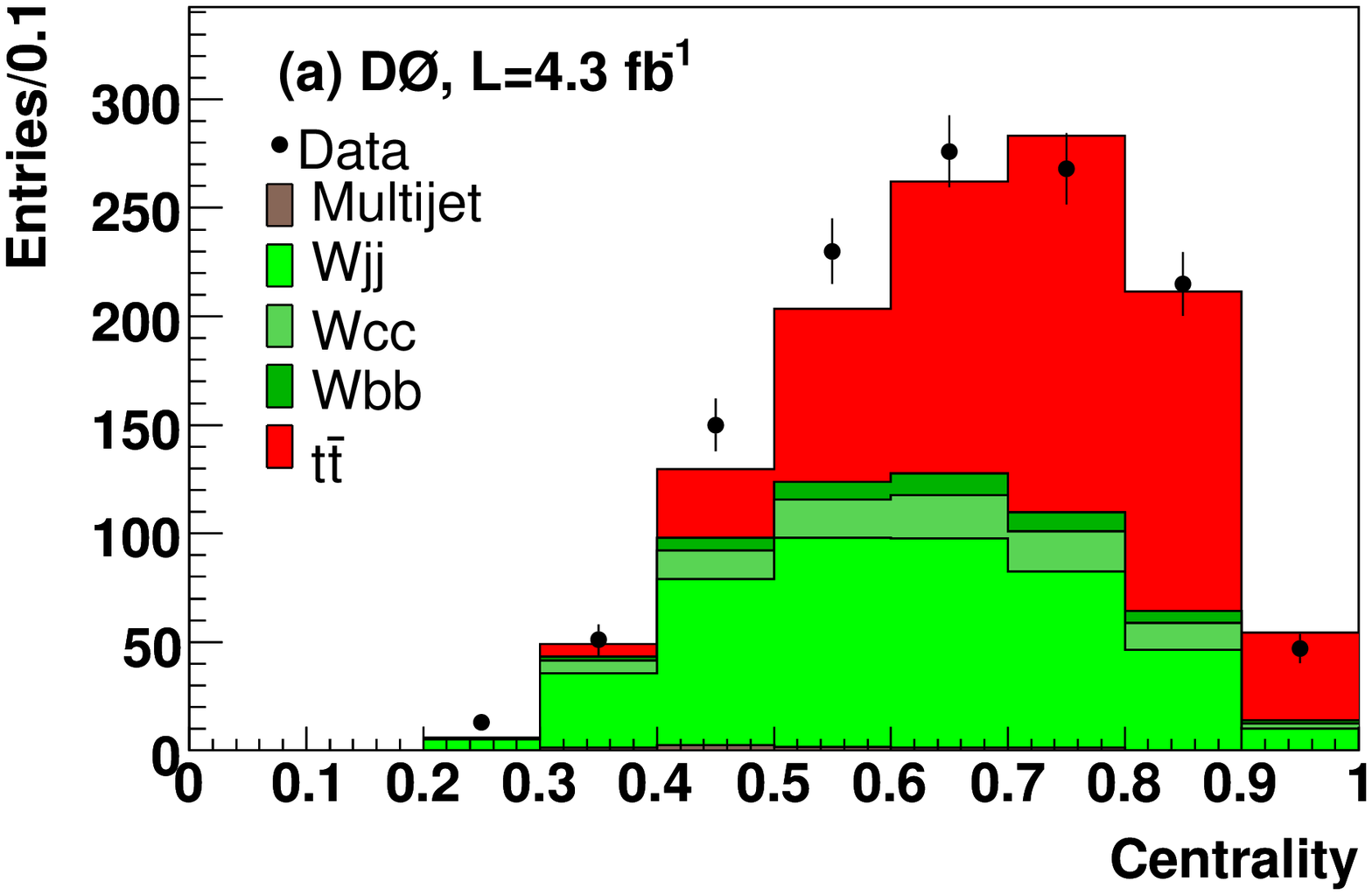}
  \includegraphics[scale=0.4]{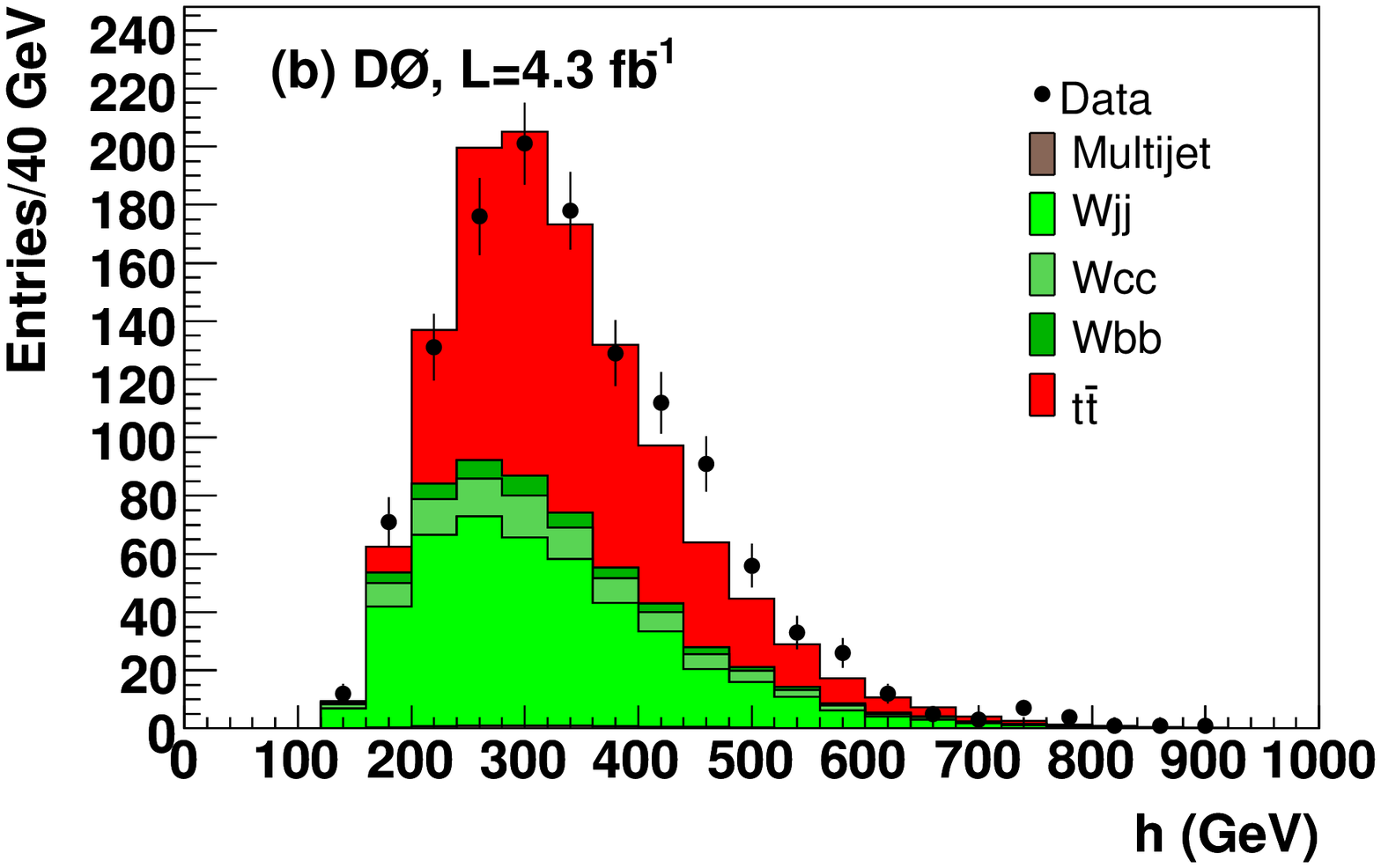}
  \includegraphics[scale=0.4]{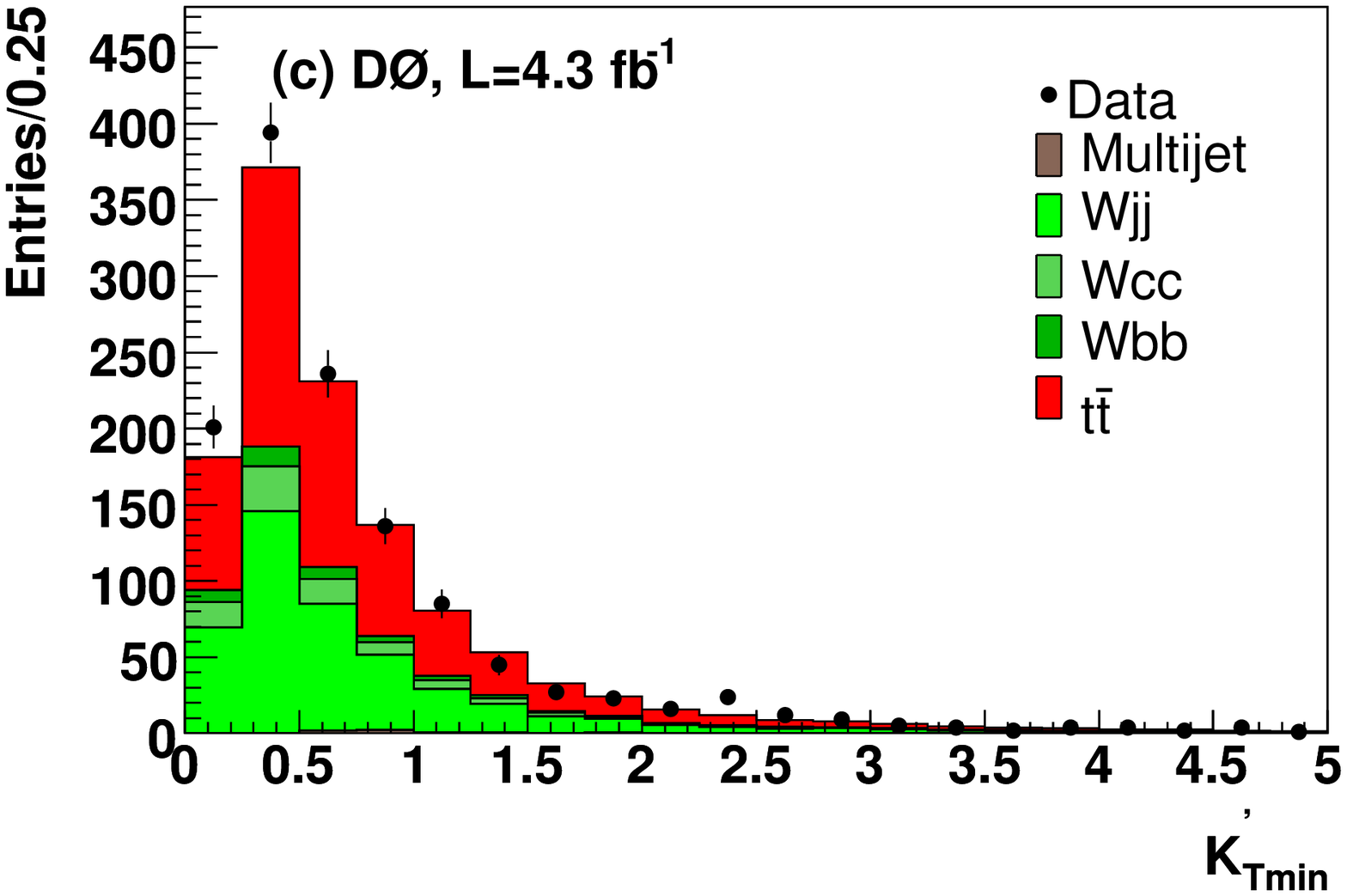}
  \includegraphics[scale=0.4]{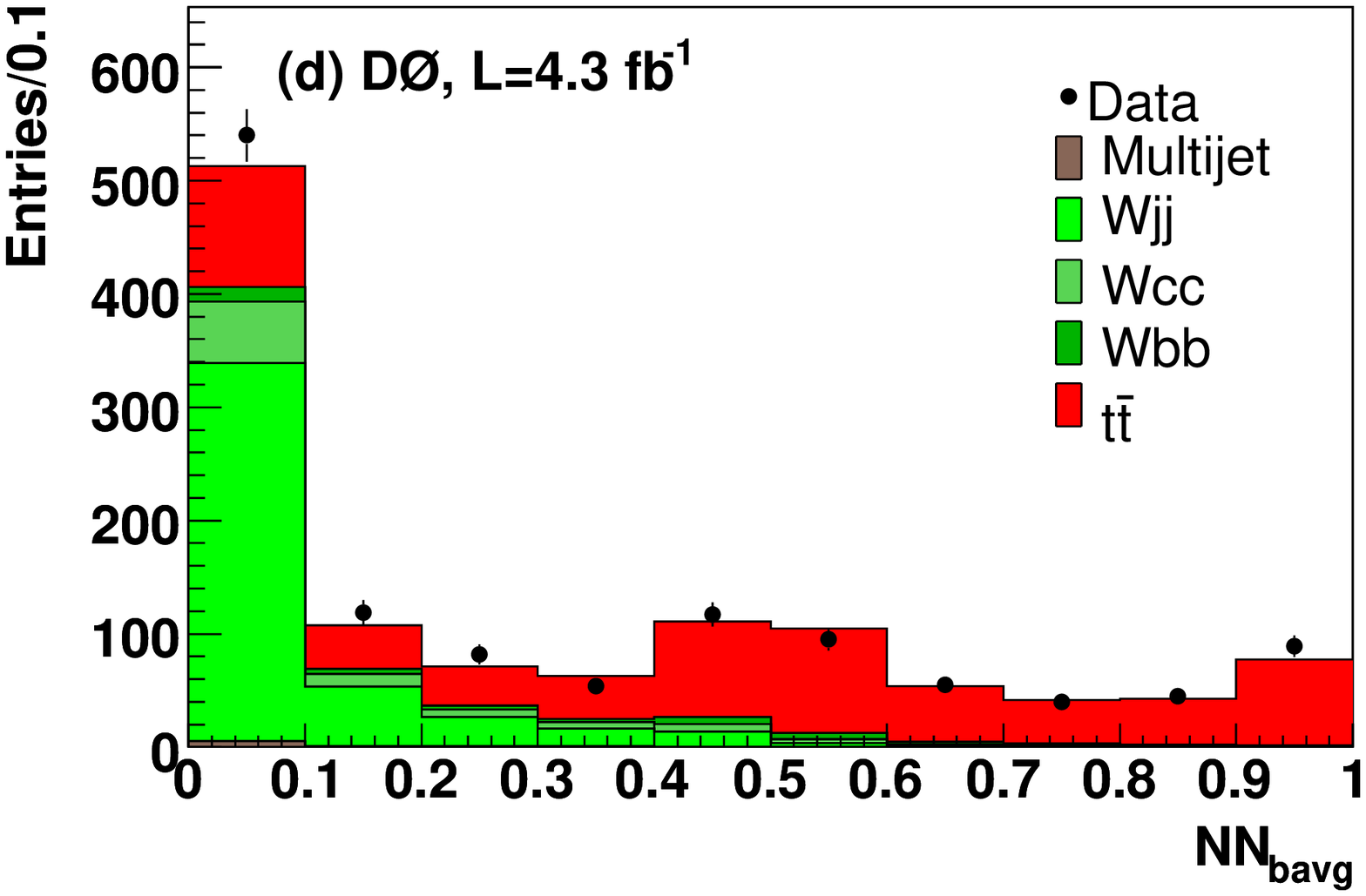}
 \includegraphics[scale=0.4]{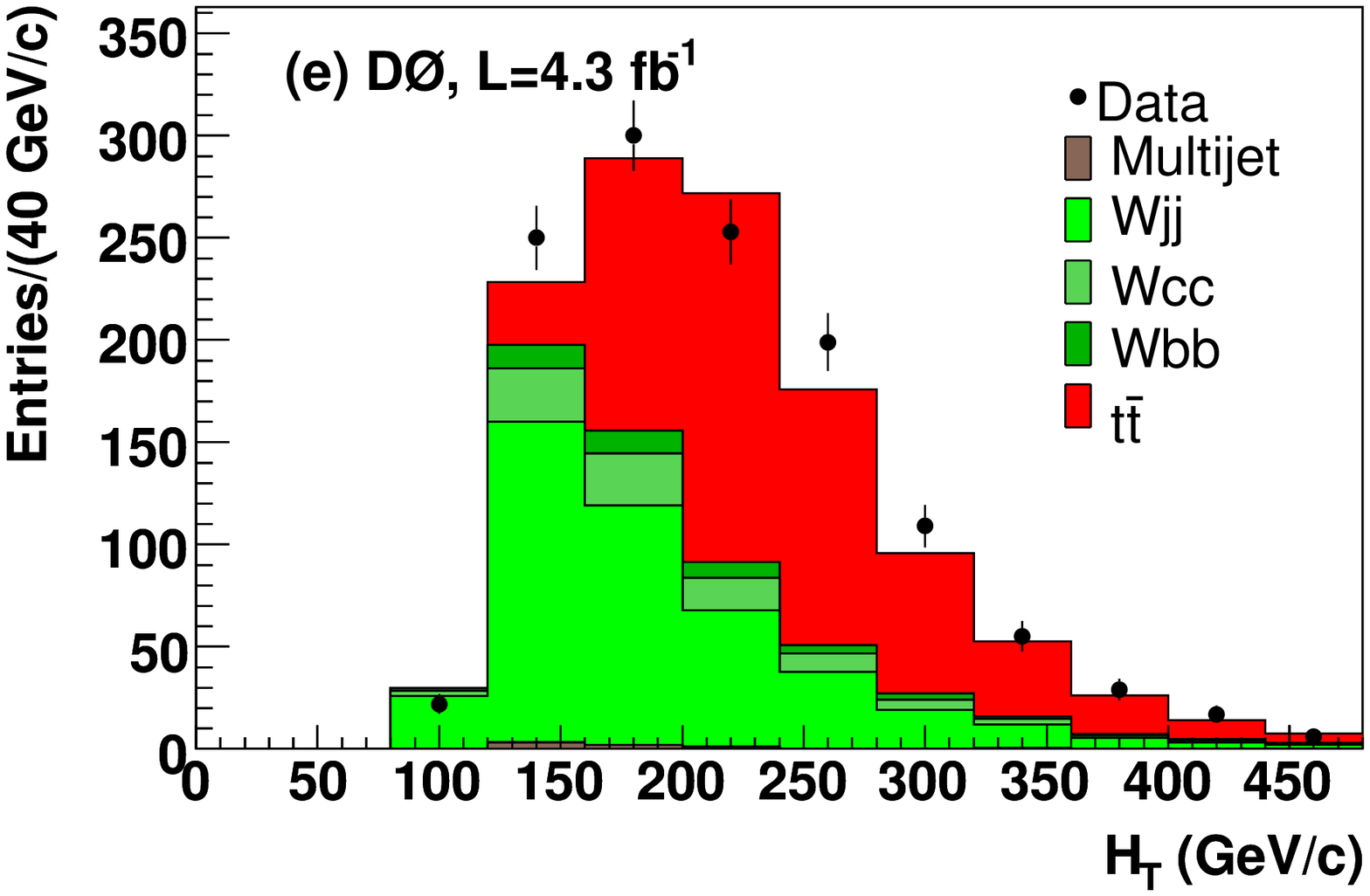} 
\caption{\label{fig:input_mujets}  (Color online) Comparison of data and MC of the variables for preselected events, chosen for the best likelihood discriminant $L_{t}$ in the $\mu+$jets channel: (a) ${\cal C}$,(b) $h$,  (c)  ${K_{T\text{min}}^\prime}$, (d) NN$_{b{\rm avg}}$ and (e) $H_T$. The uncertainties on the data points are statistical only.}
\end{figure*}

\begin{figure*}
\begin{center}
\includegraphics[scale=0.4]{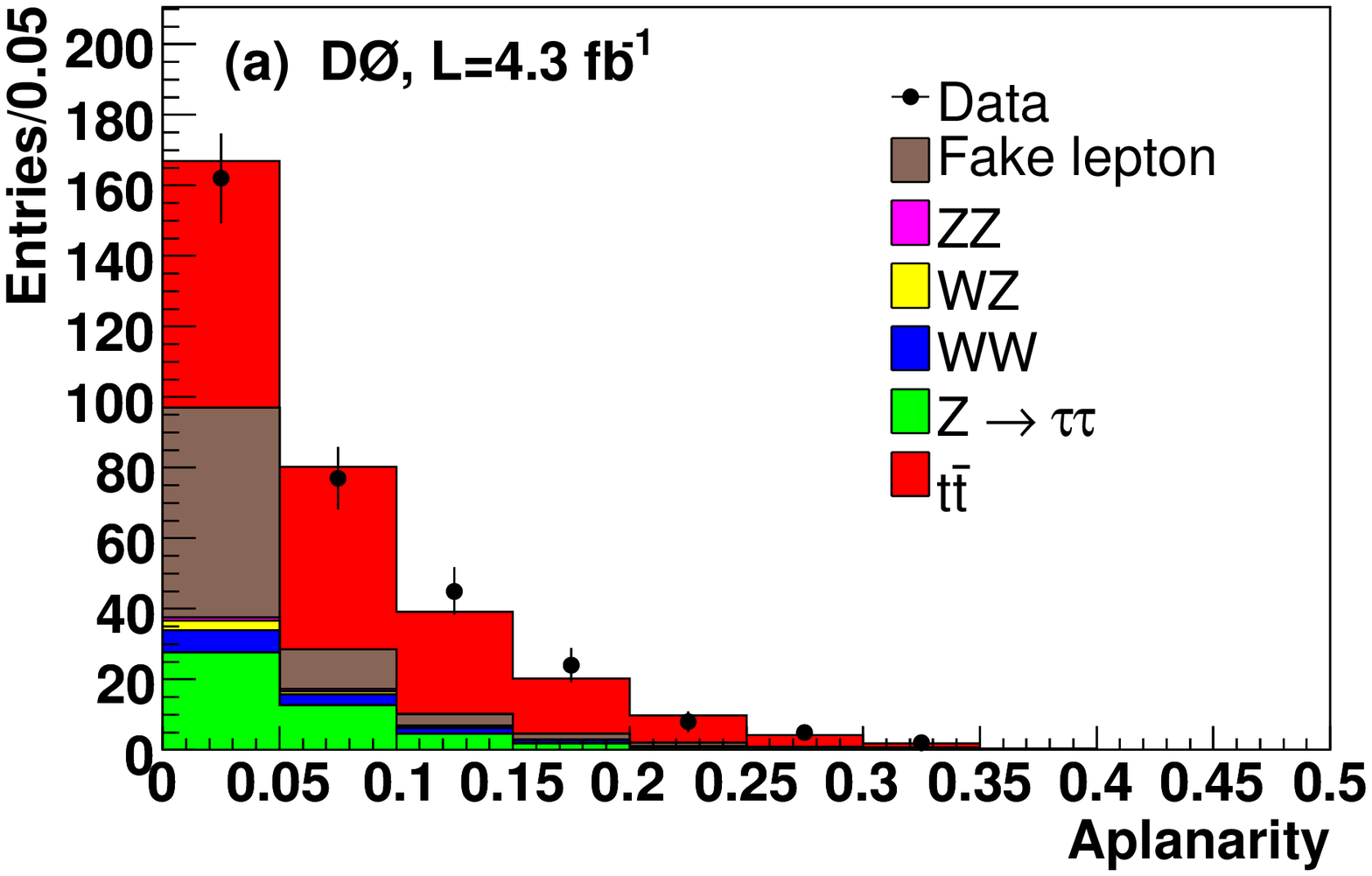}
\includegraphics[scale=0.4]{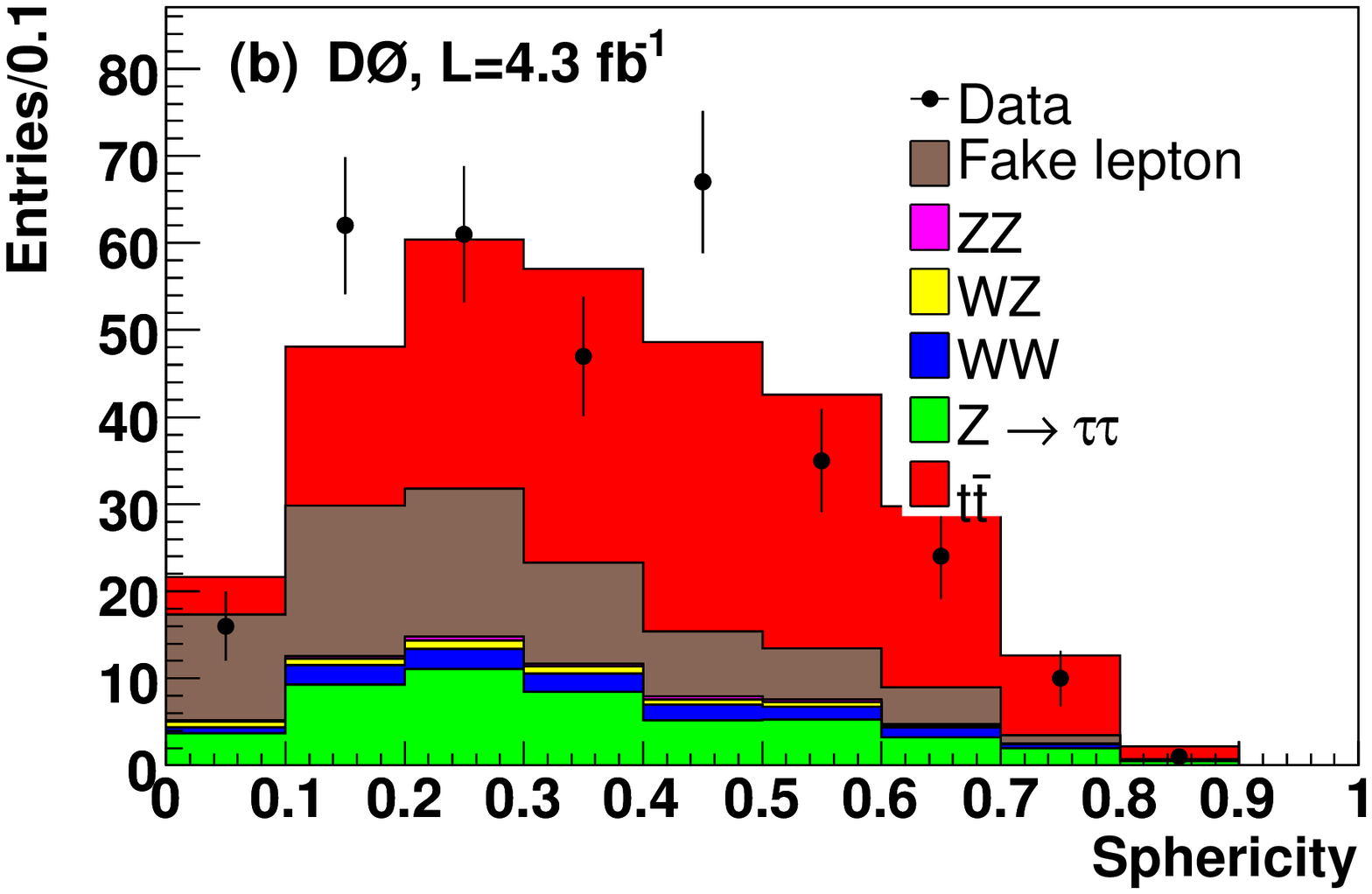}
\includegraphics[scale=0.4]{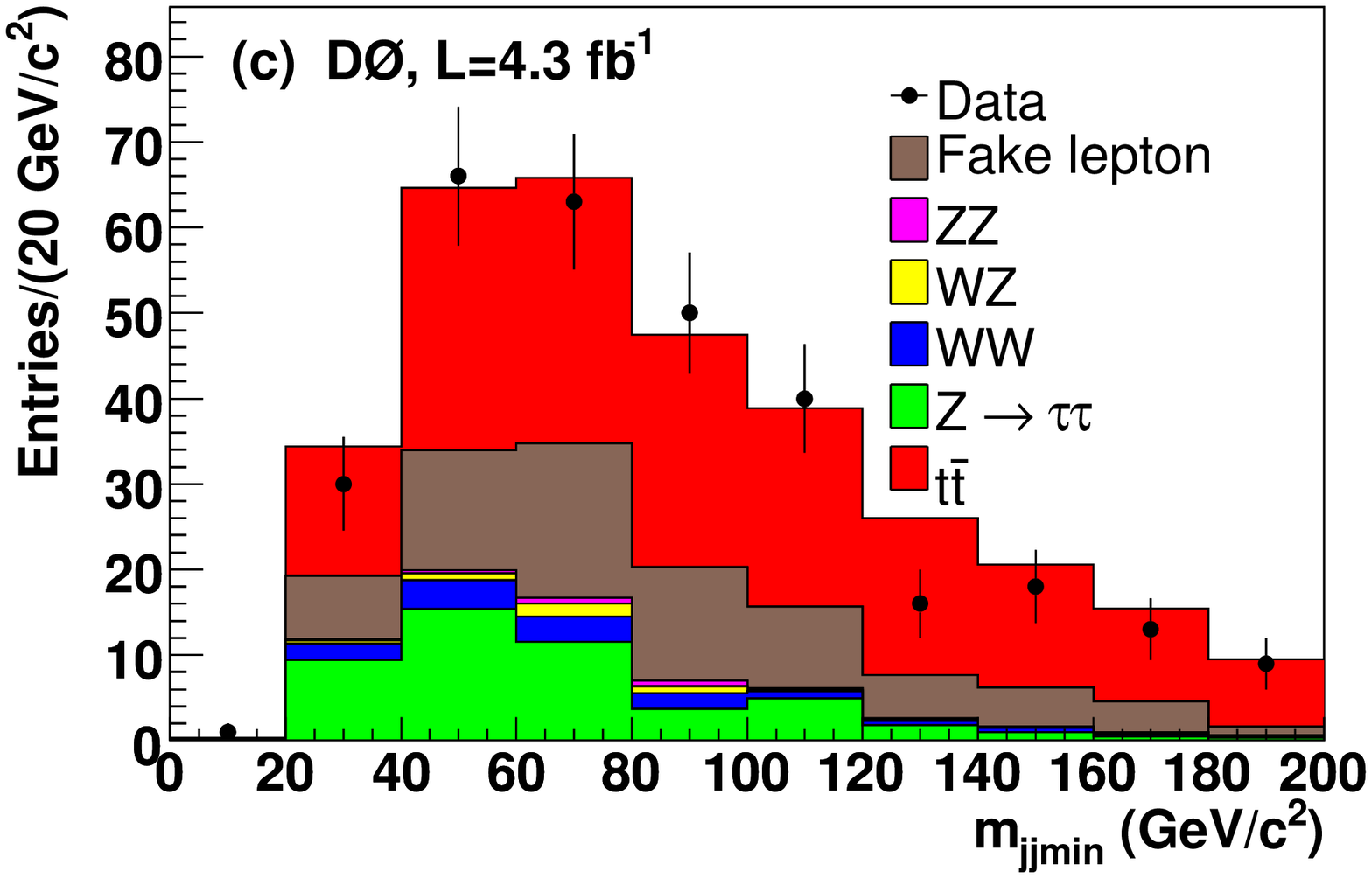}
\includegraphics[scale=0.4]{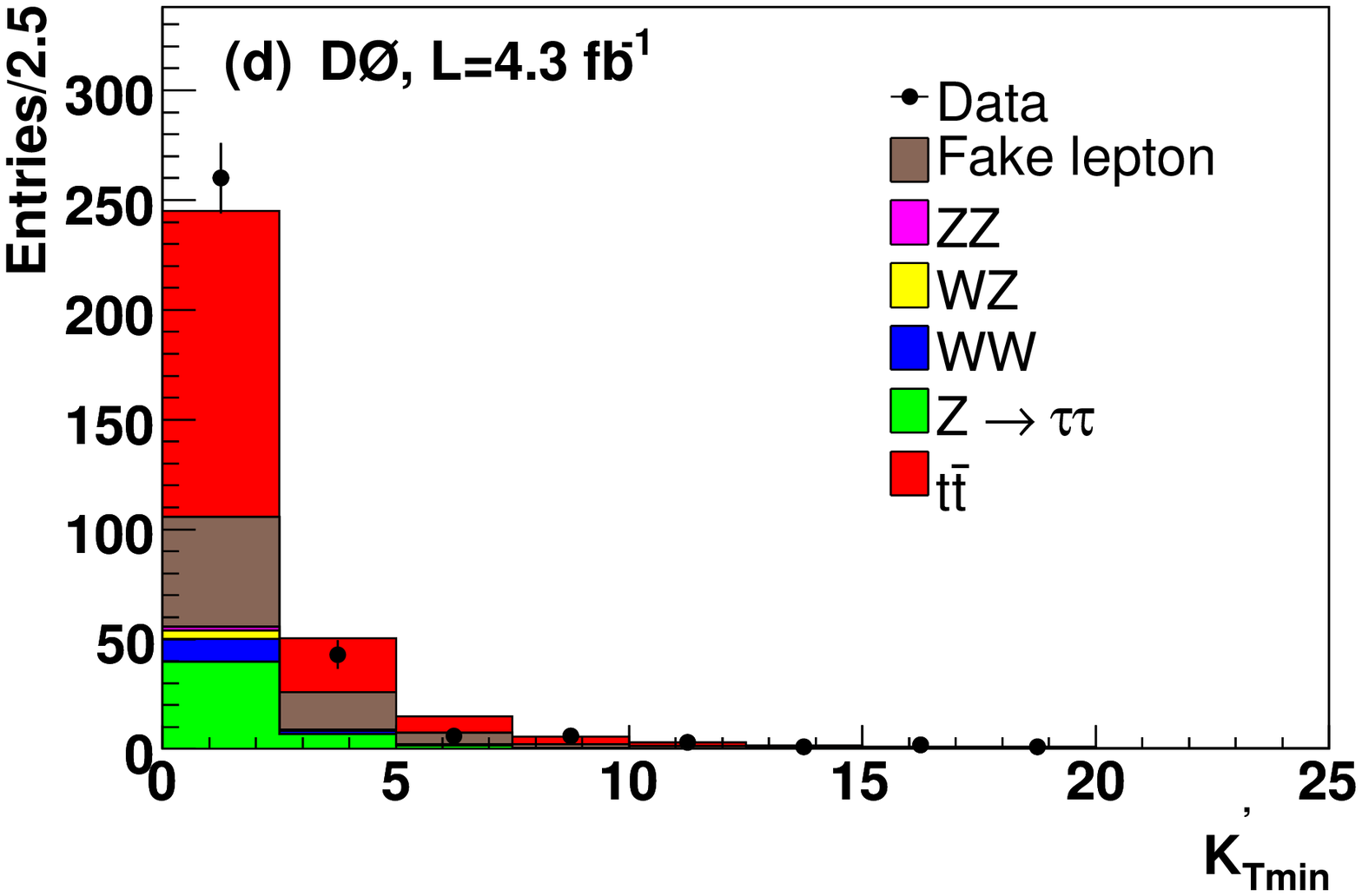}
\includegraphics[scale=0.4]{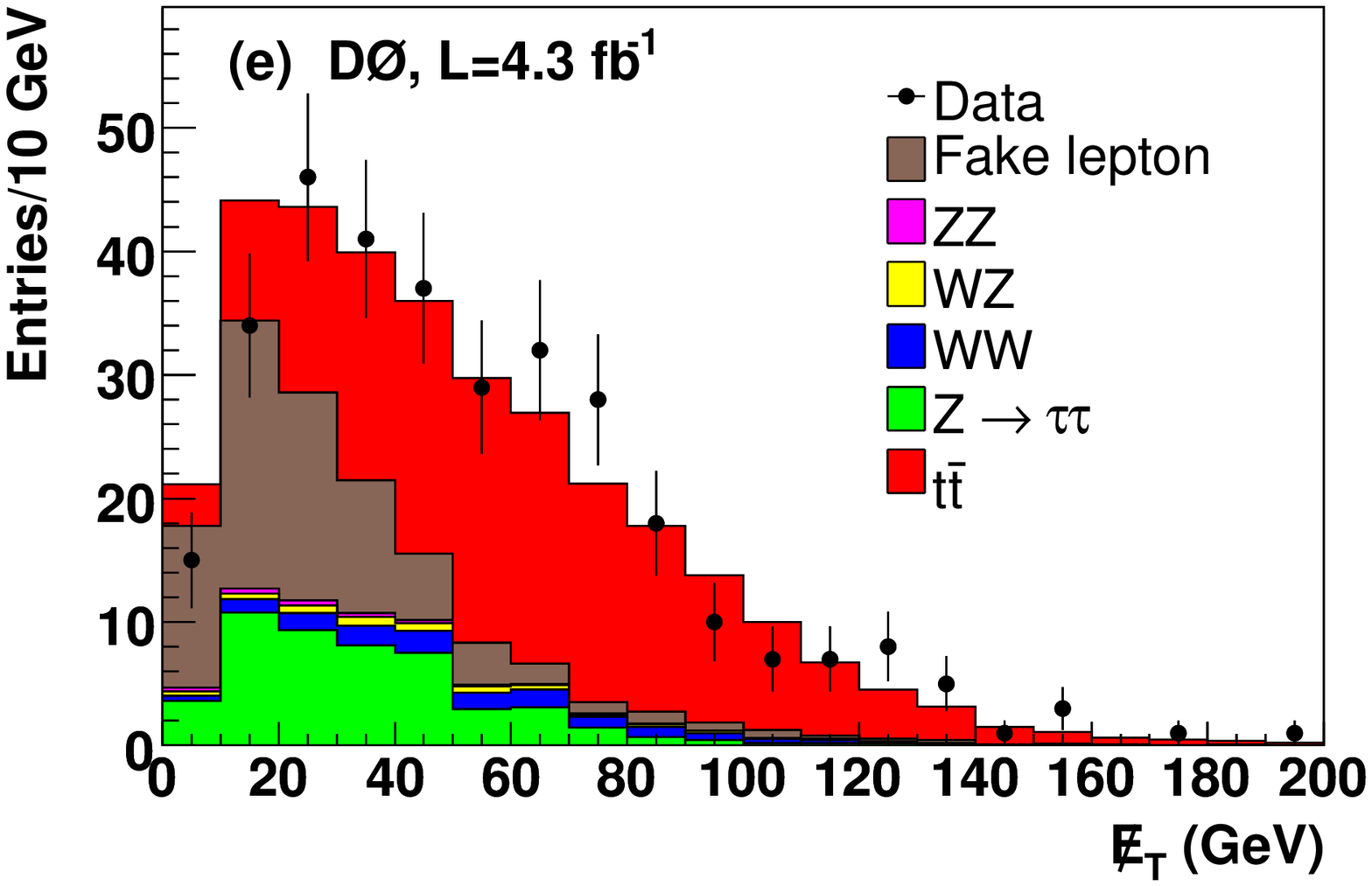}
\includegraphics[scale=0.4]{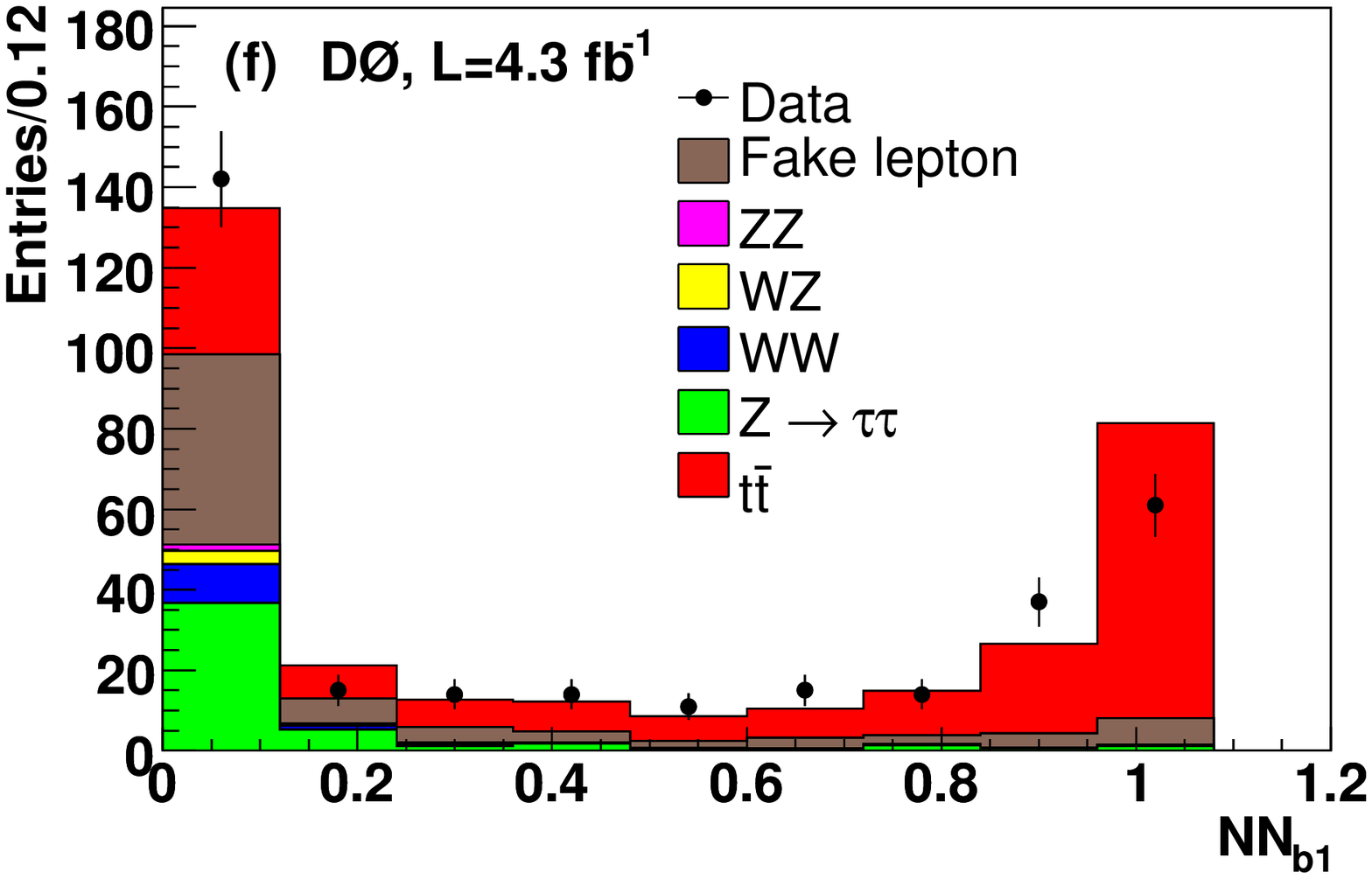}
\includegraphics[scale=0.4]{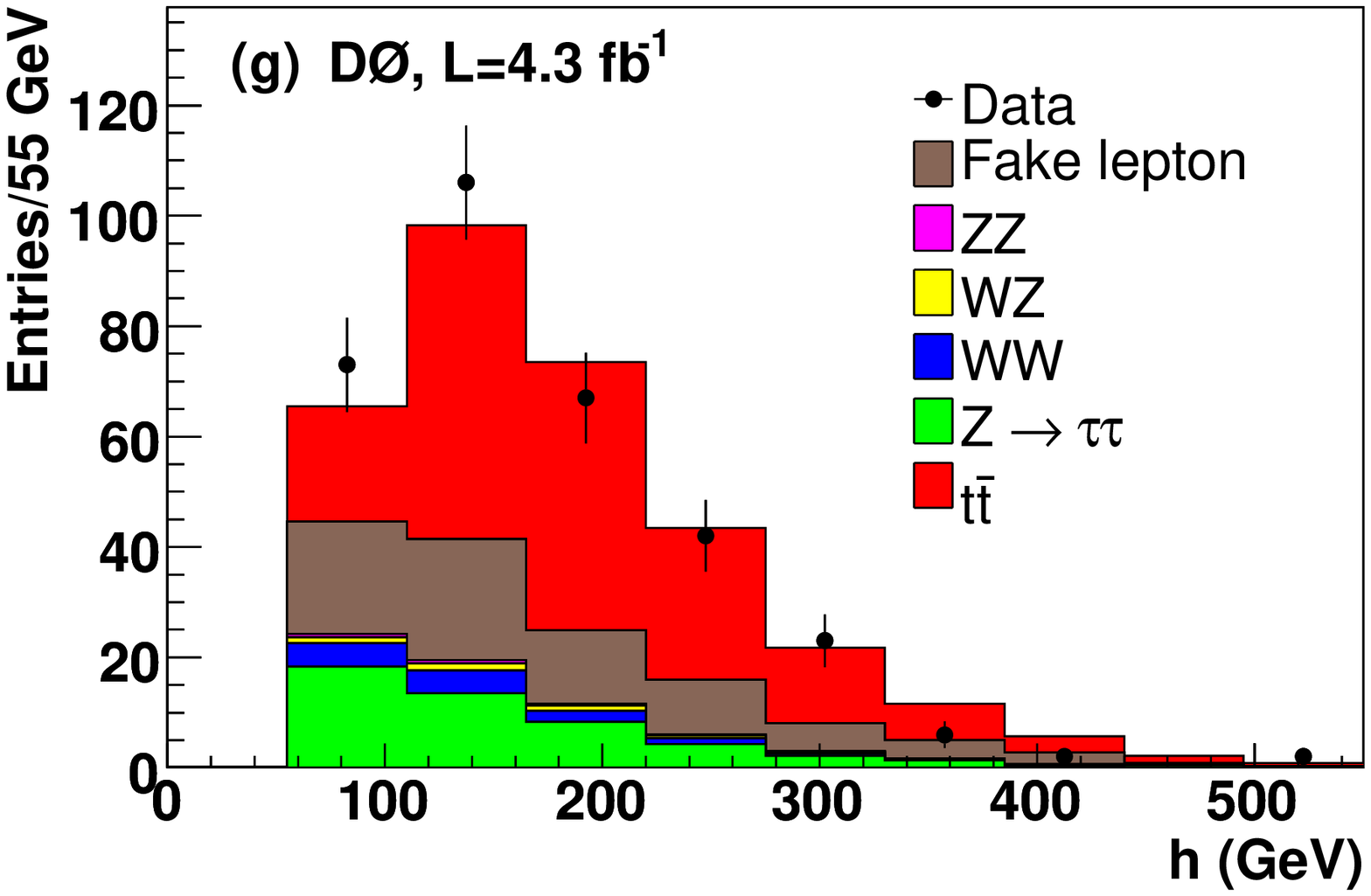}
\includegraphics[scale=0.4]{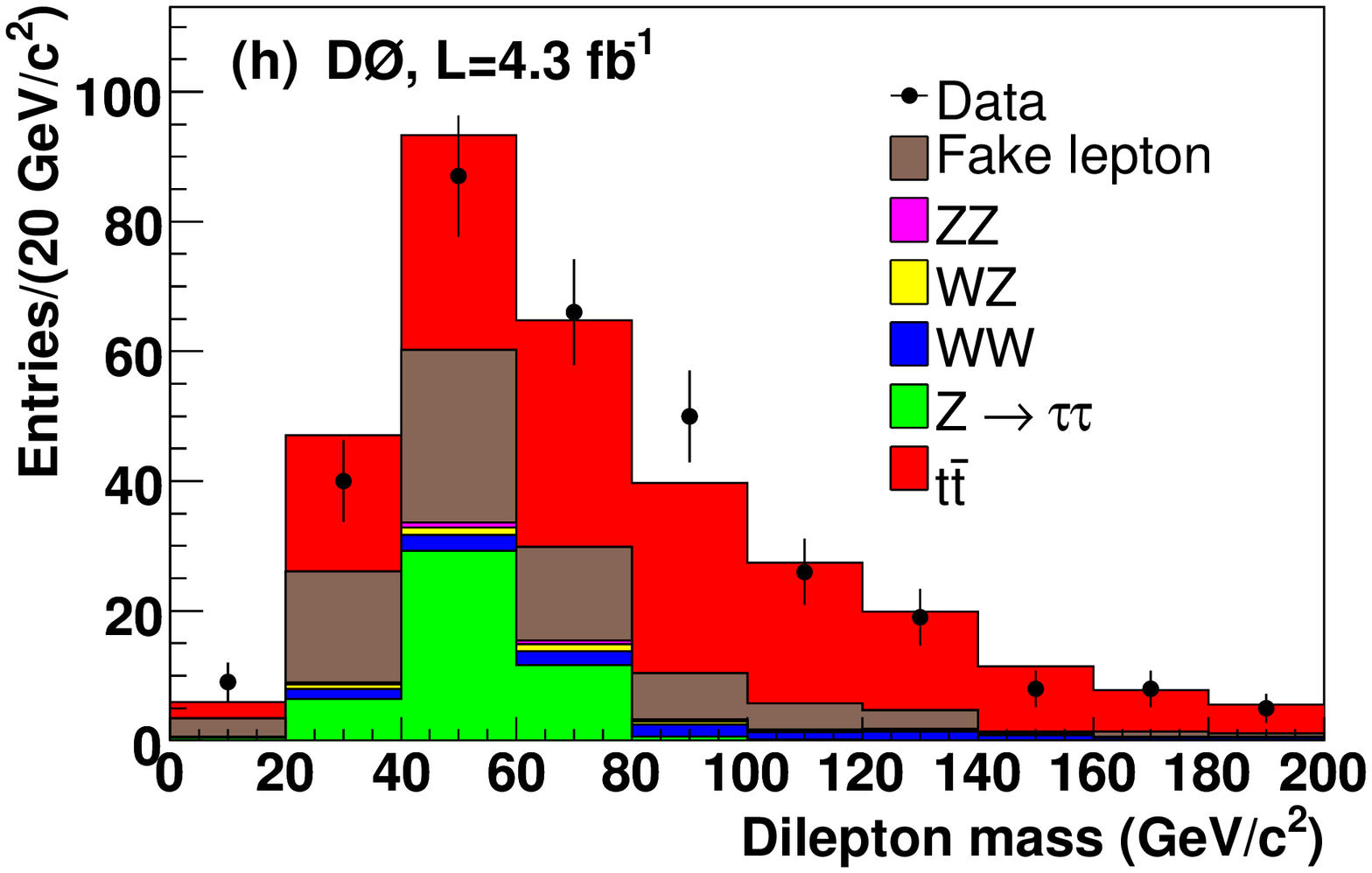}
\caption{(Color online) Comparison of data and MC of the variables for preselected events, chosen for the best likelihood discriminant $L_{t}$ in the $e\mu$ channel: (a) ${\cal A}$, (b) ${\cal S}$, (c)  $m_{jj\text{min}}$, (d) ${K_{T\text{min}}^\prime}$, (e) \met, (f) NN$_{b1}$, (g) $h$, and (h) $m_{\ell\ell}$.  The uncertainties on the data points are statistical only.}
\label{fig:emu_cent_spher}
\end{center}
\end{figure*}

\begin{figure*}
\begin{center}
\includegraphics[scale=0.4]{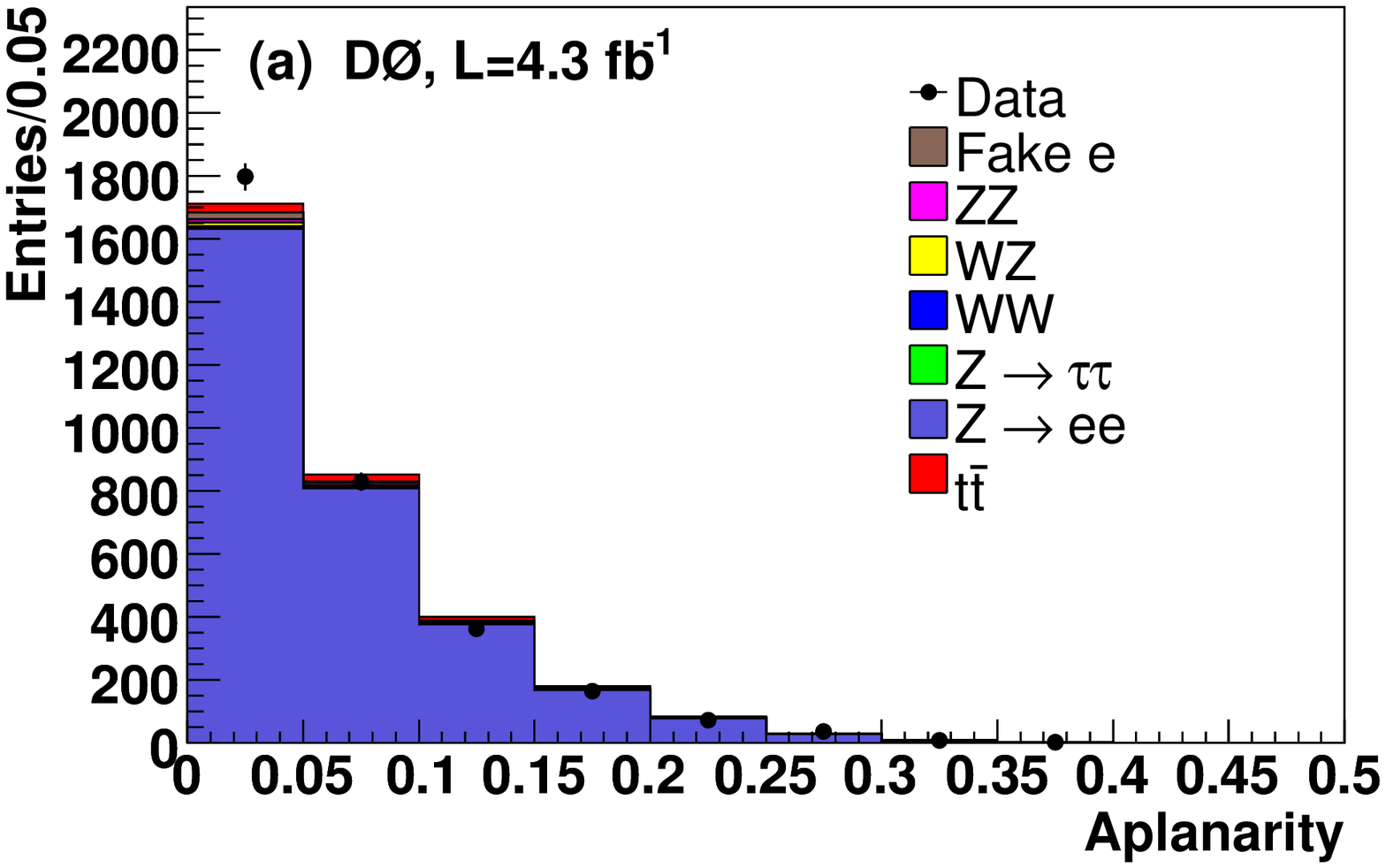}
\includegraphics[scale=0.4]{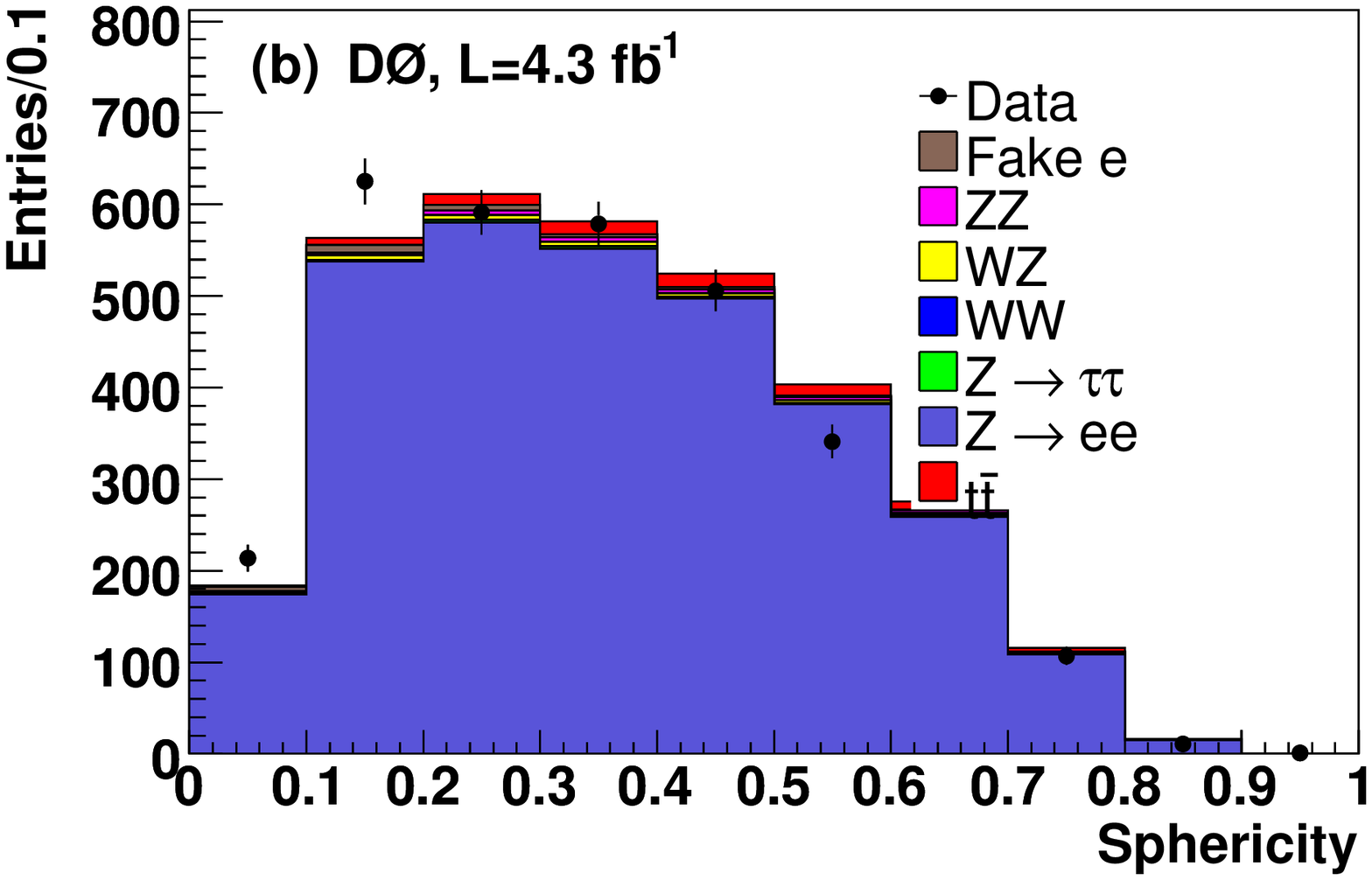}
\includegraphics[scale=0.4]{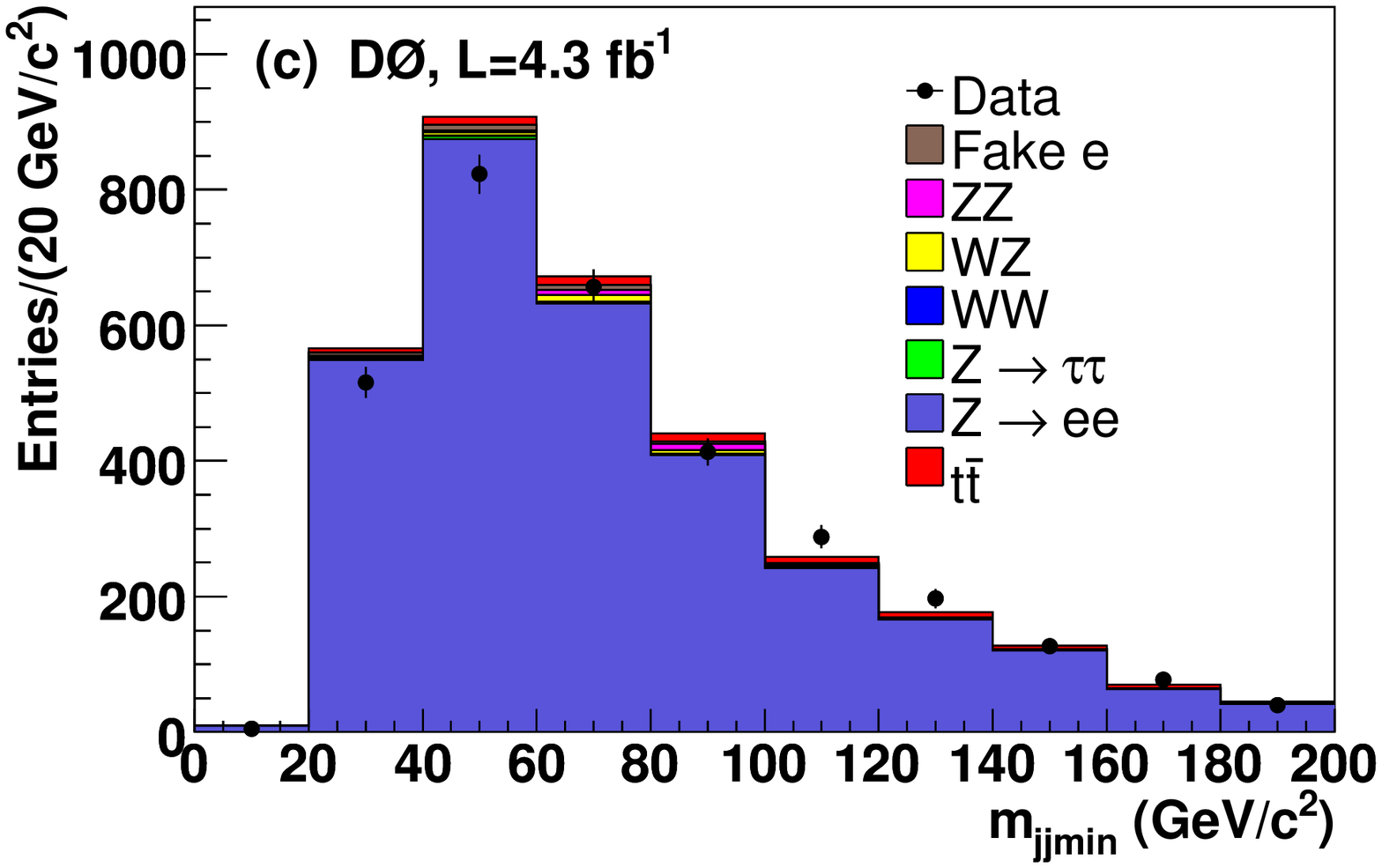}
\includegraphics[scale=0.4]{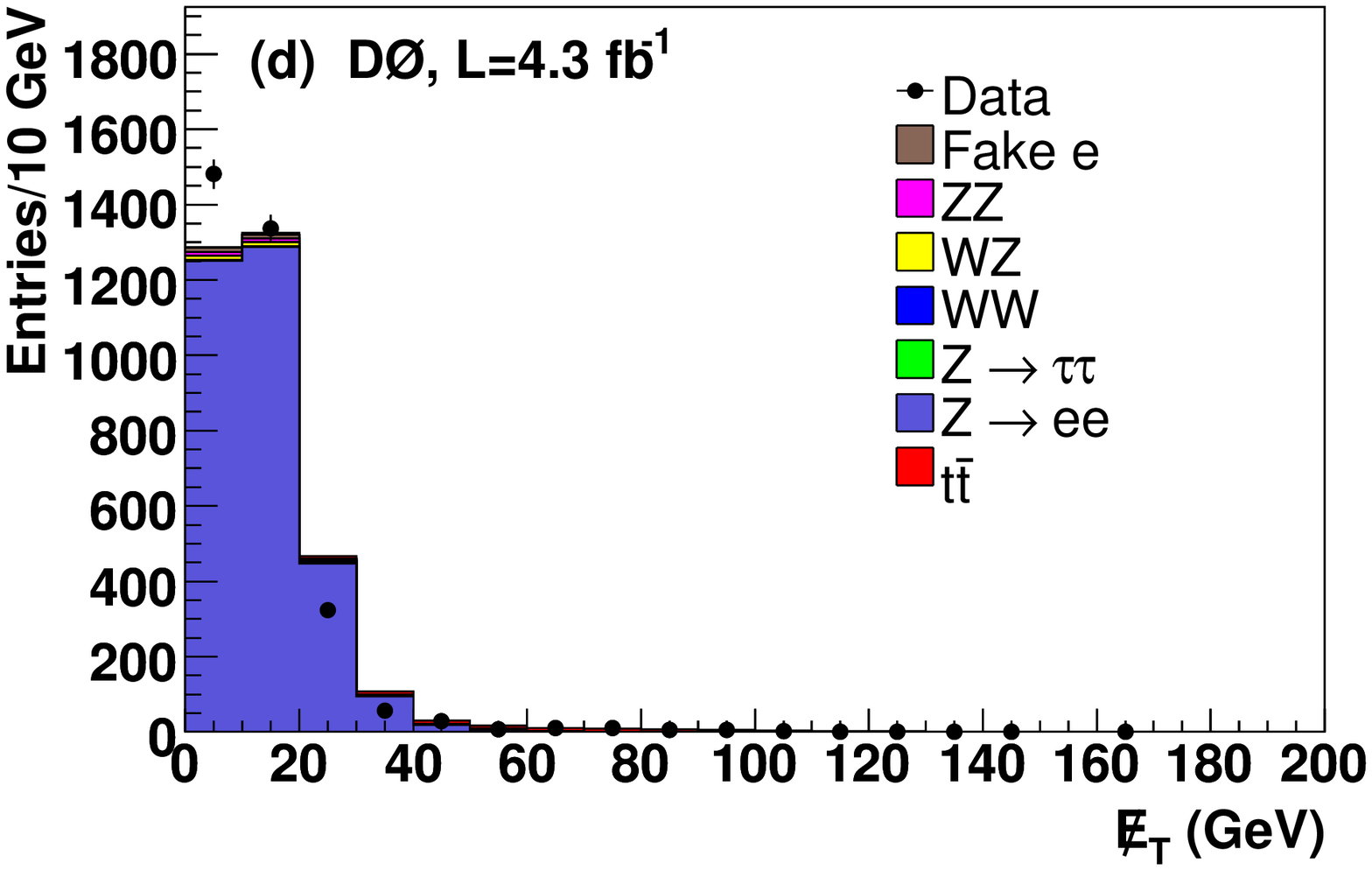}
\includegraphics[scale=0.4]{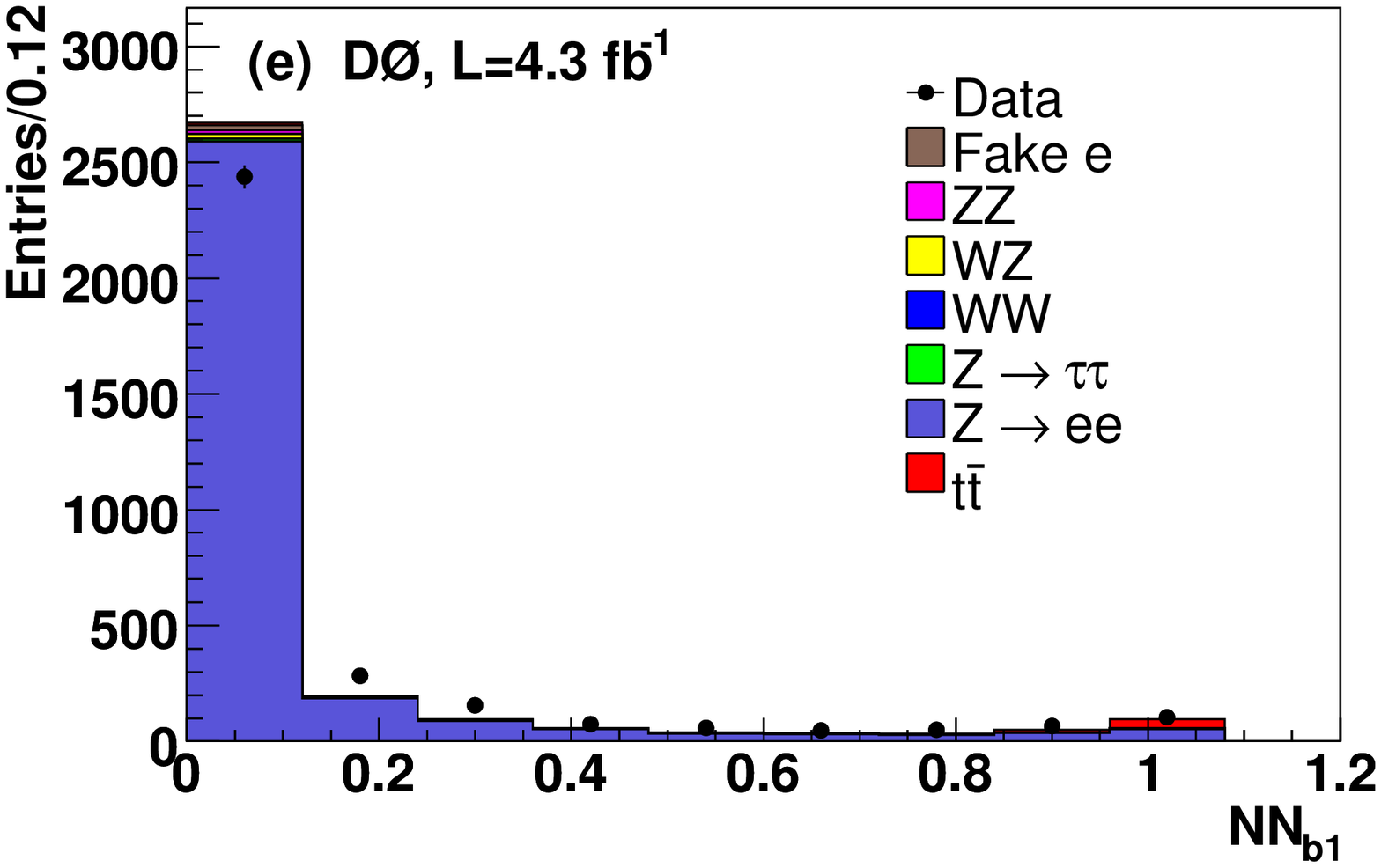}
\includegraphics[scale=0.4]{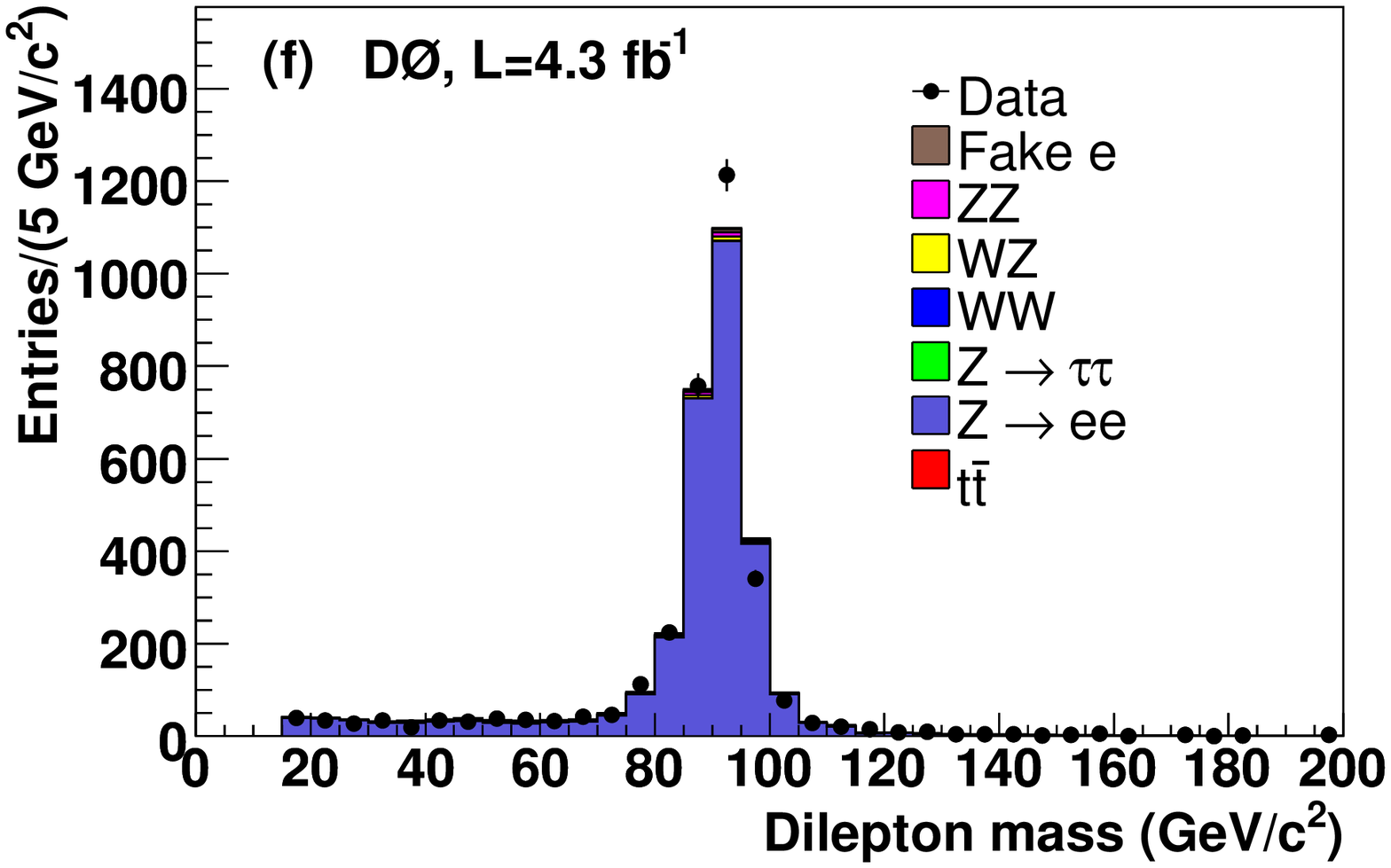}
\caption{(Color online) Comparison of data and MC of the variables for preselected events, chosen for the best likelihood discriminant $L_{t}$ in the $ee$ channel: (a) ${\cal A}$, (b) ${\cal S}$, (c)  $m_{jj\text{min}}$, (d) \met, (e) NN$_{b1}$, and (f) $m_{\ell\ell}$.  The uncertainties on the data points are statistical only.}
\label{fig:ee_apla_spher}
\end{center}
\end{figure*}

\begin{figure*}
\begin{center}
\includegraphics[scale=0.4]{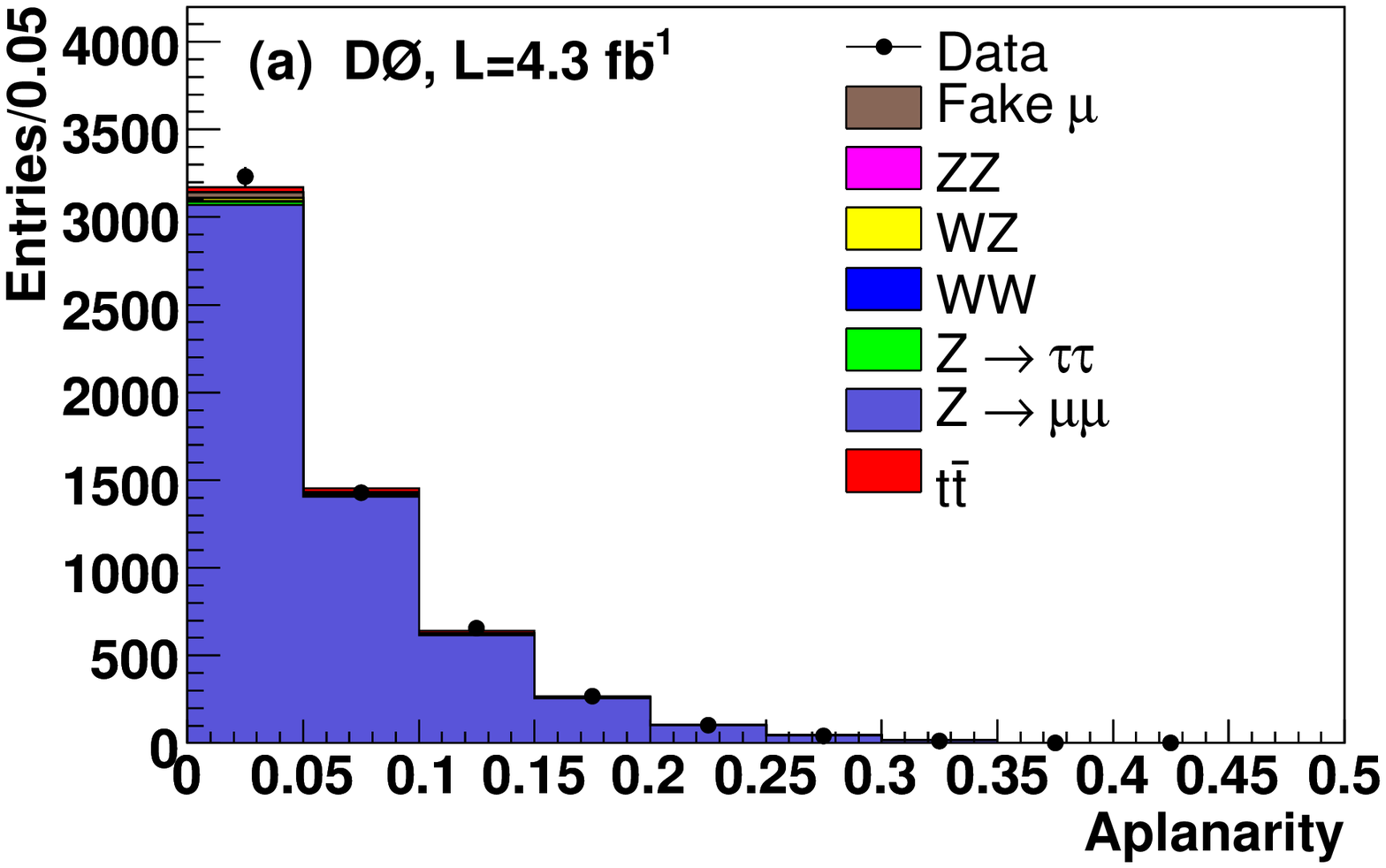}
\includegraphics[scale=0.4]{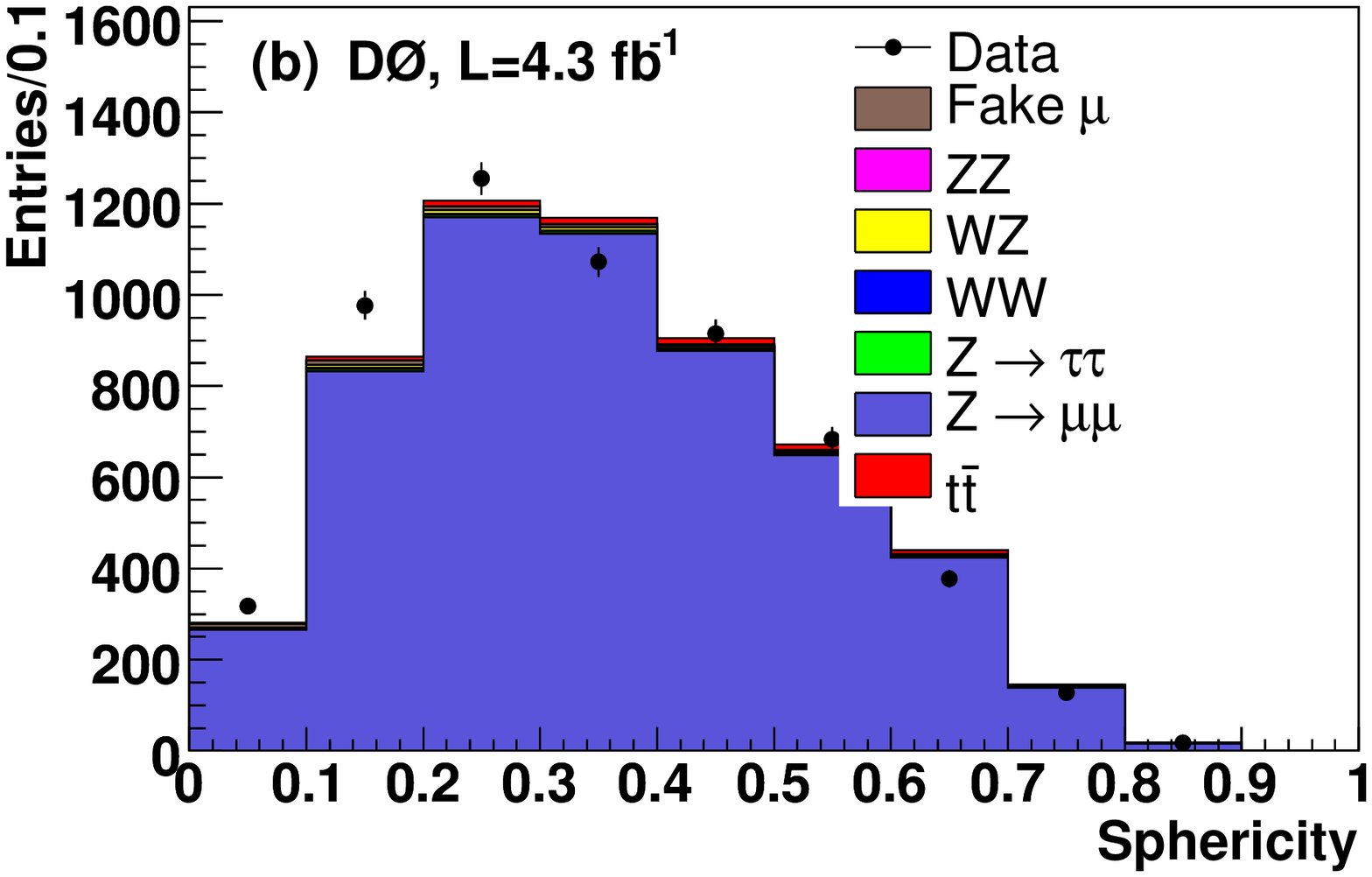}
\includegraphics[scale=0.4]{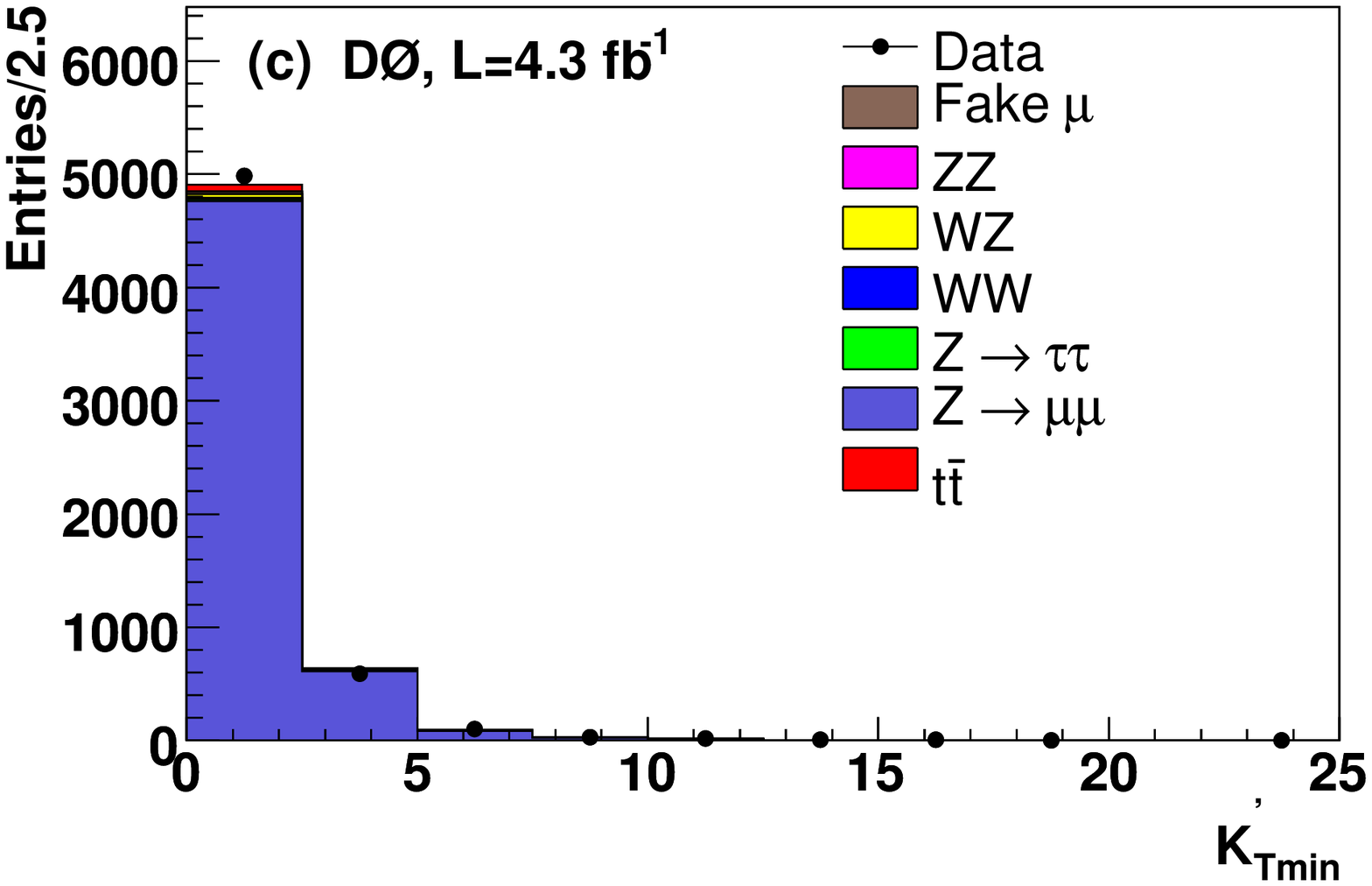}
\includegraphics[scale=0.4]{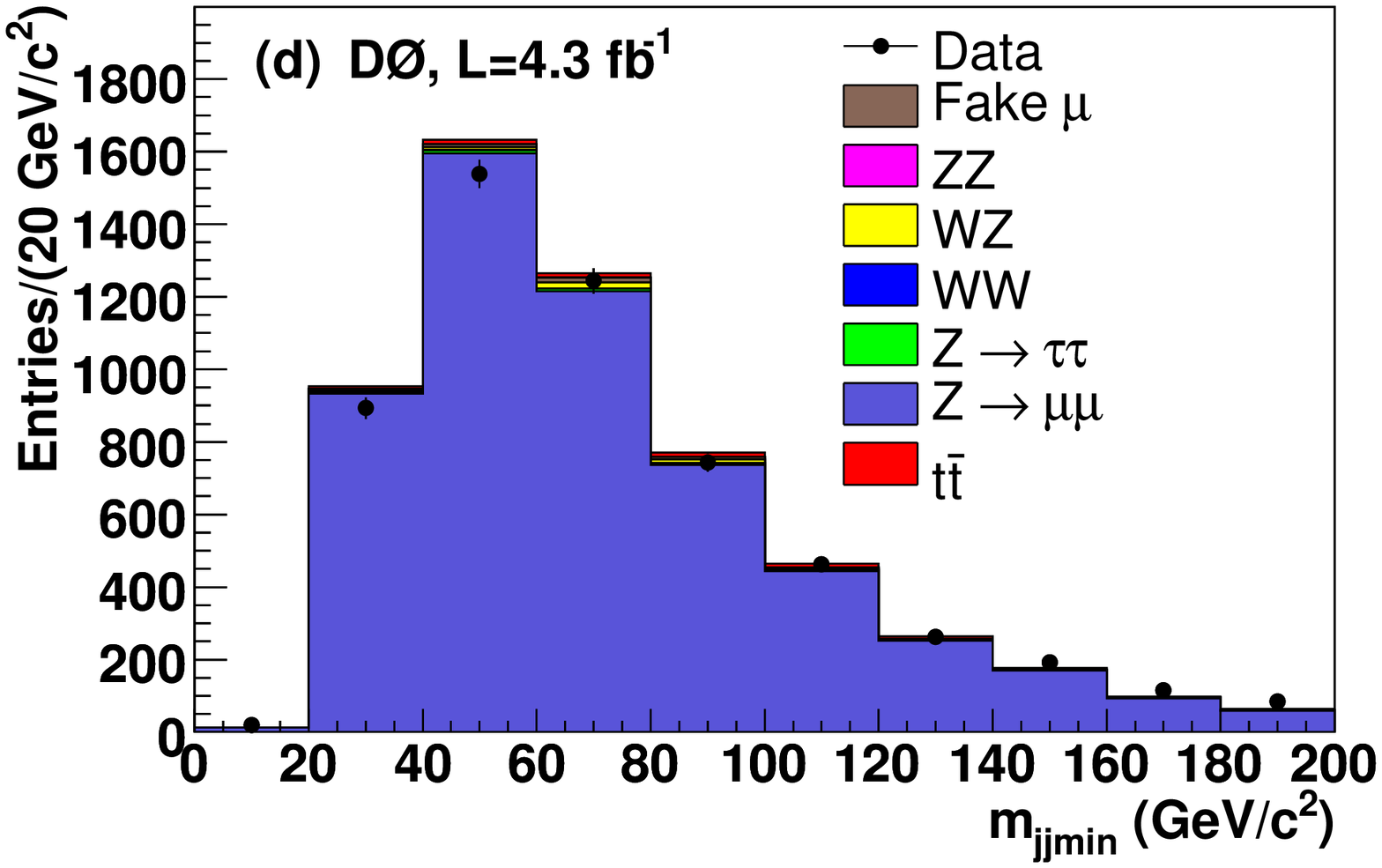}
\includegraphics[scale=0.4]{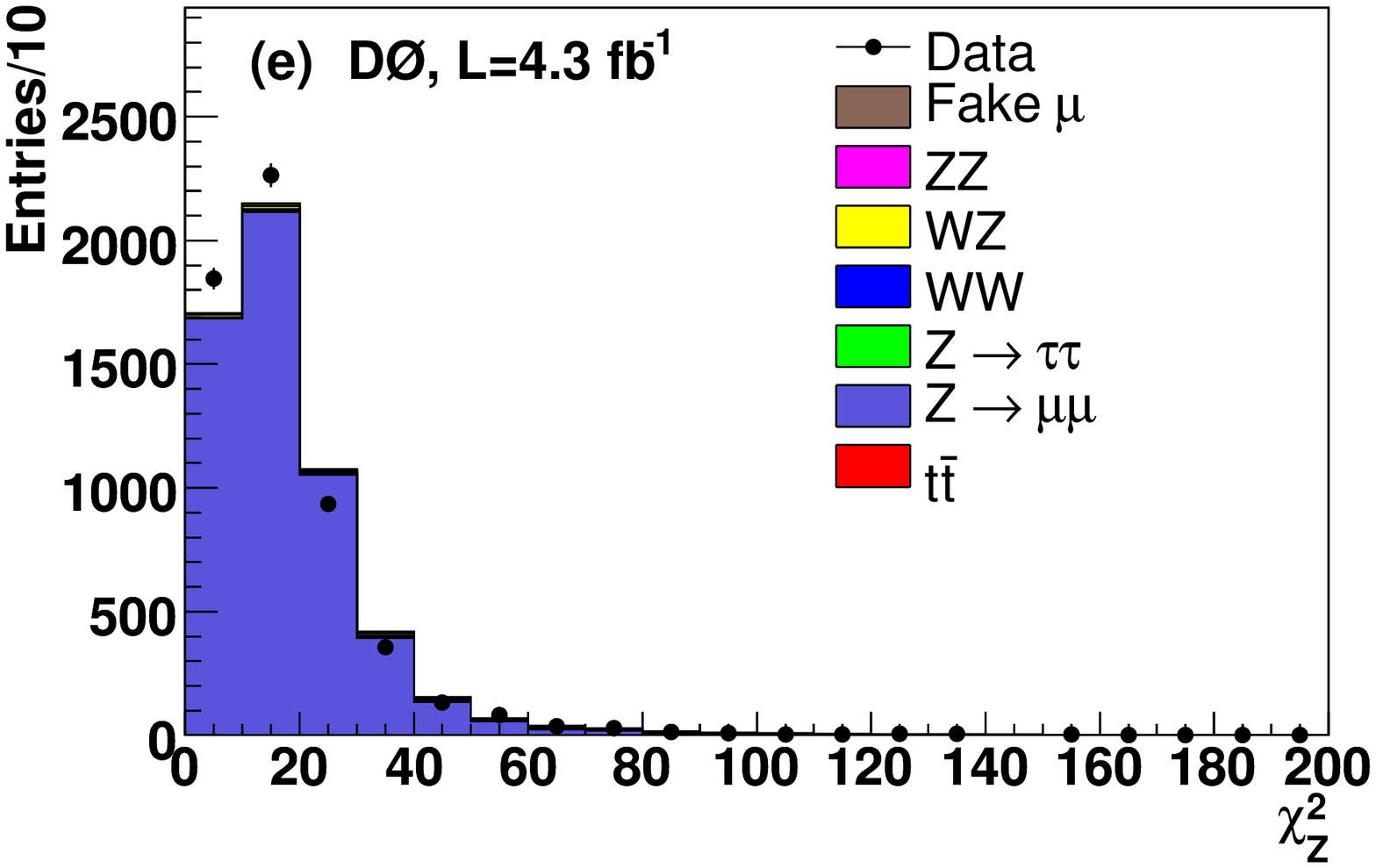}
\includegraphics[scale=0.4]{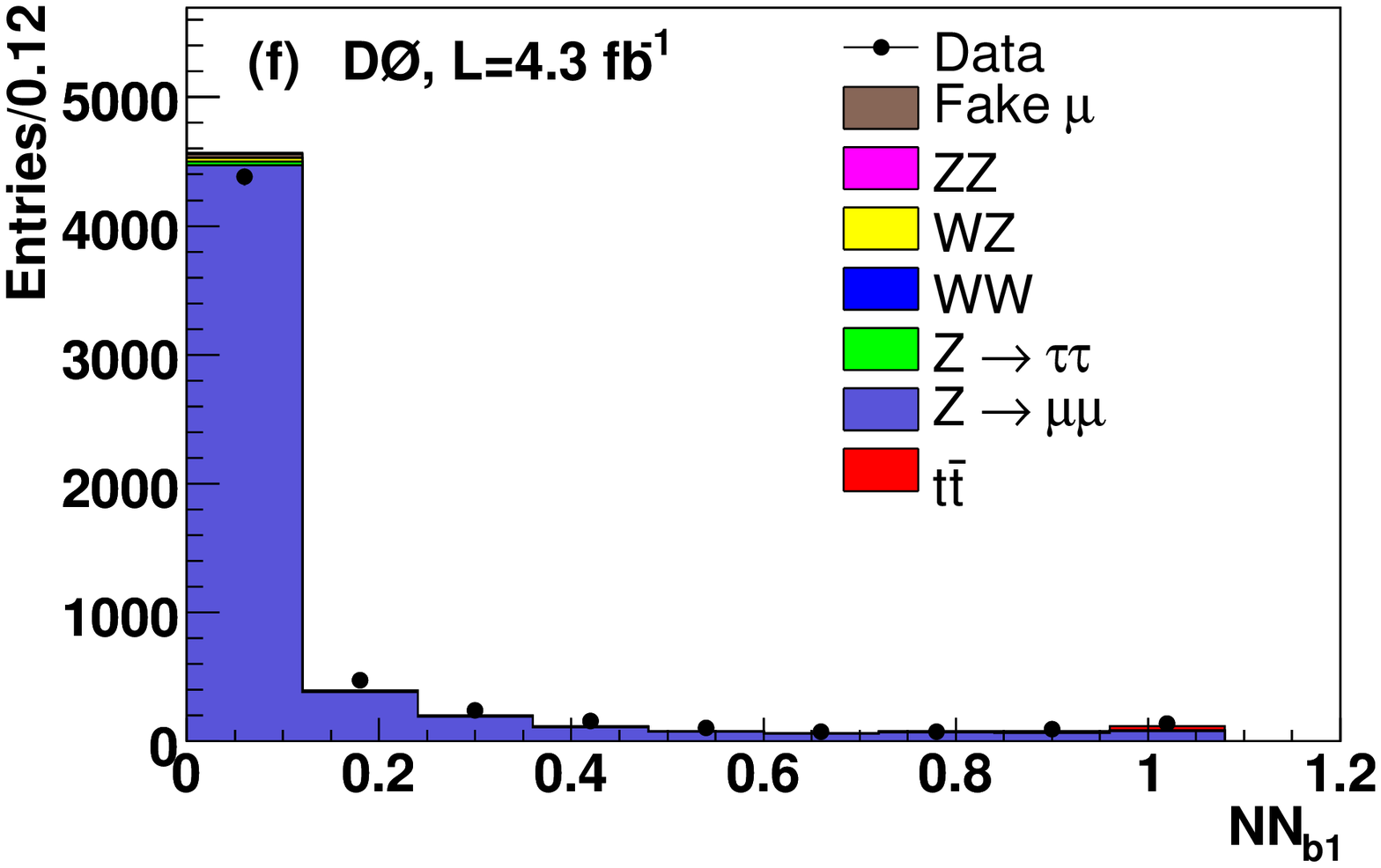}
\caption{(Color online) Comparison of data and MC of the variables for preselected events, chosen for the best likelihood discriminant $L_{t}$ in the $\mu\mu$ channel: (a) ${\cal A}$, (b) ${\cal S}$, (c)  ${K_{T\text{min}}^\prime}$, (d) $m_{jj\text{min}}$, (e) $\chi^2_Z$,  and (f) NN$_{b1}$.  The uncertainties on the data points are statistical only.}
\label{fig:mumu_apla_spher}
\end{center}
\end{figure*}

\begin{figure}
\includegraphics[scale=0.40]{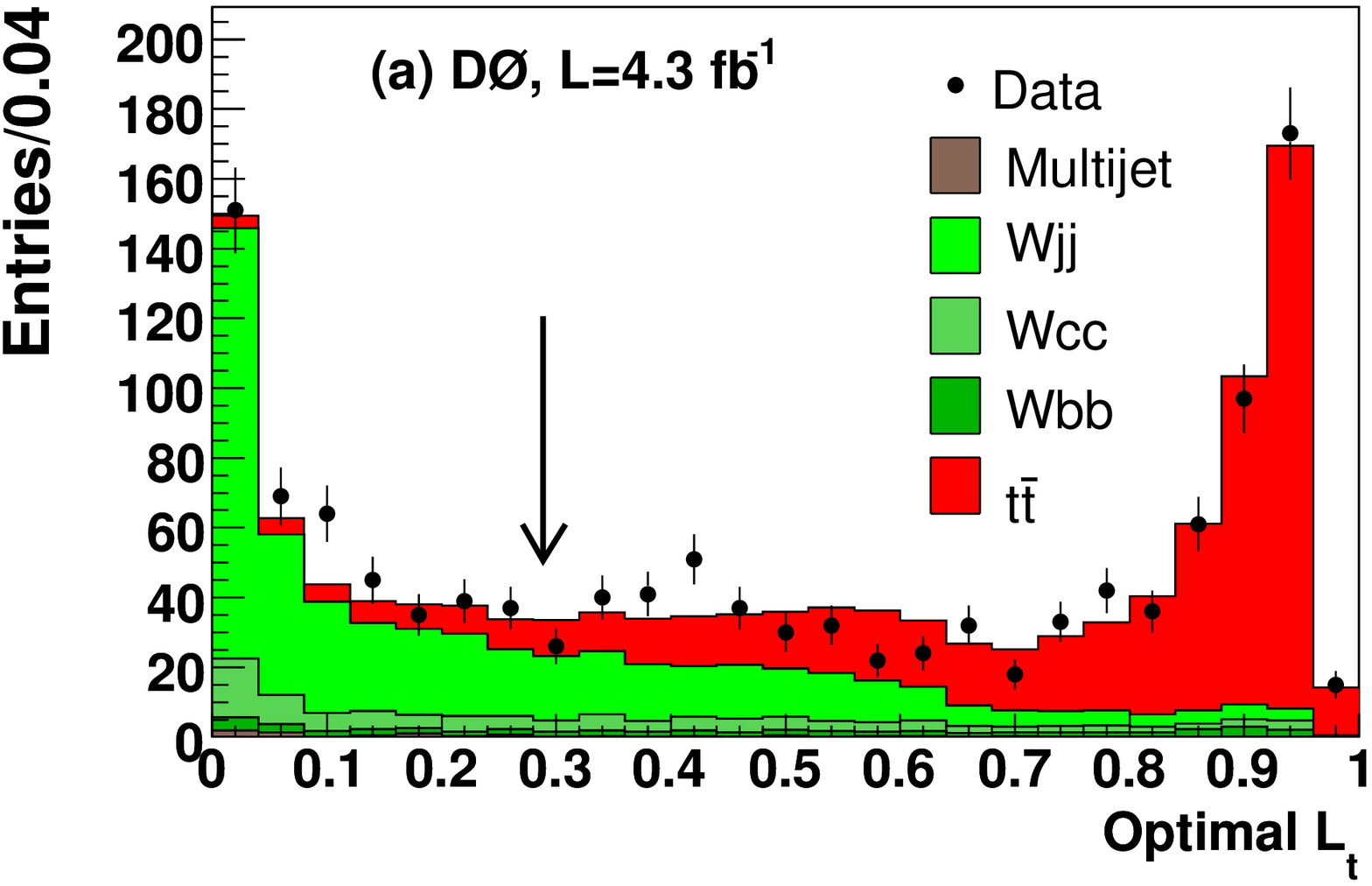}
\includegraphics[scale=0.40]{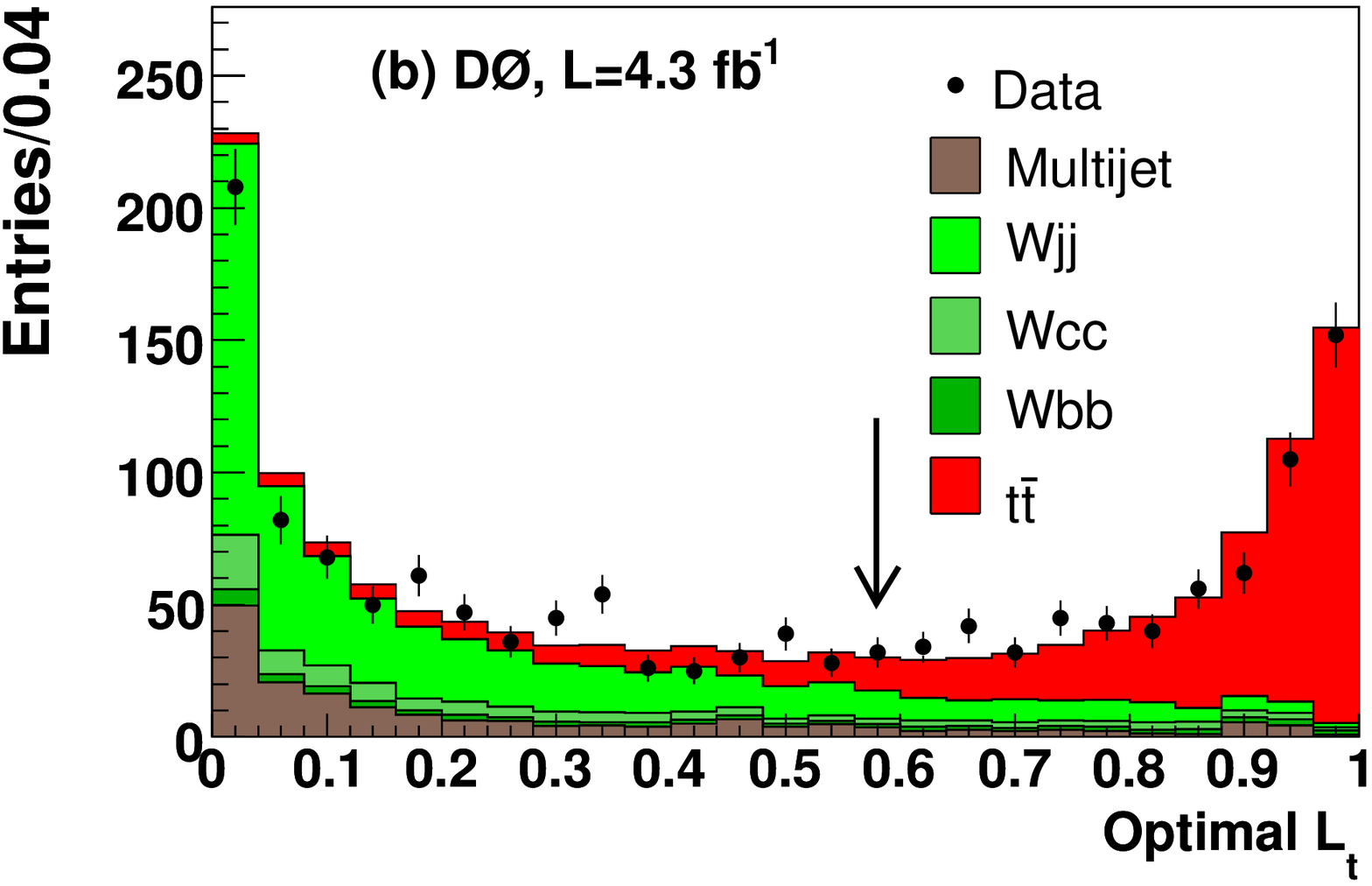}
\caption{\label{fig:BestLt} (Color online) Best $L_t$ variable for the (a) $\mu+$jets and (b) $e+$jets channels. The normalization of the signal and background models is determined by the Poisson maximum likelihood fit to the $L_t$ distribution. The arrows mark the required $L_t$ values for events in each channel.}
\end{figure}

\begin{figure}
\includegraphics[scale=0.40]{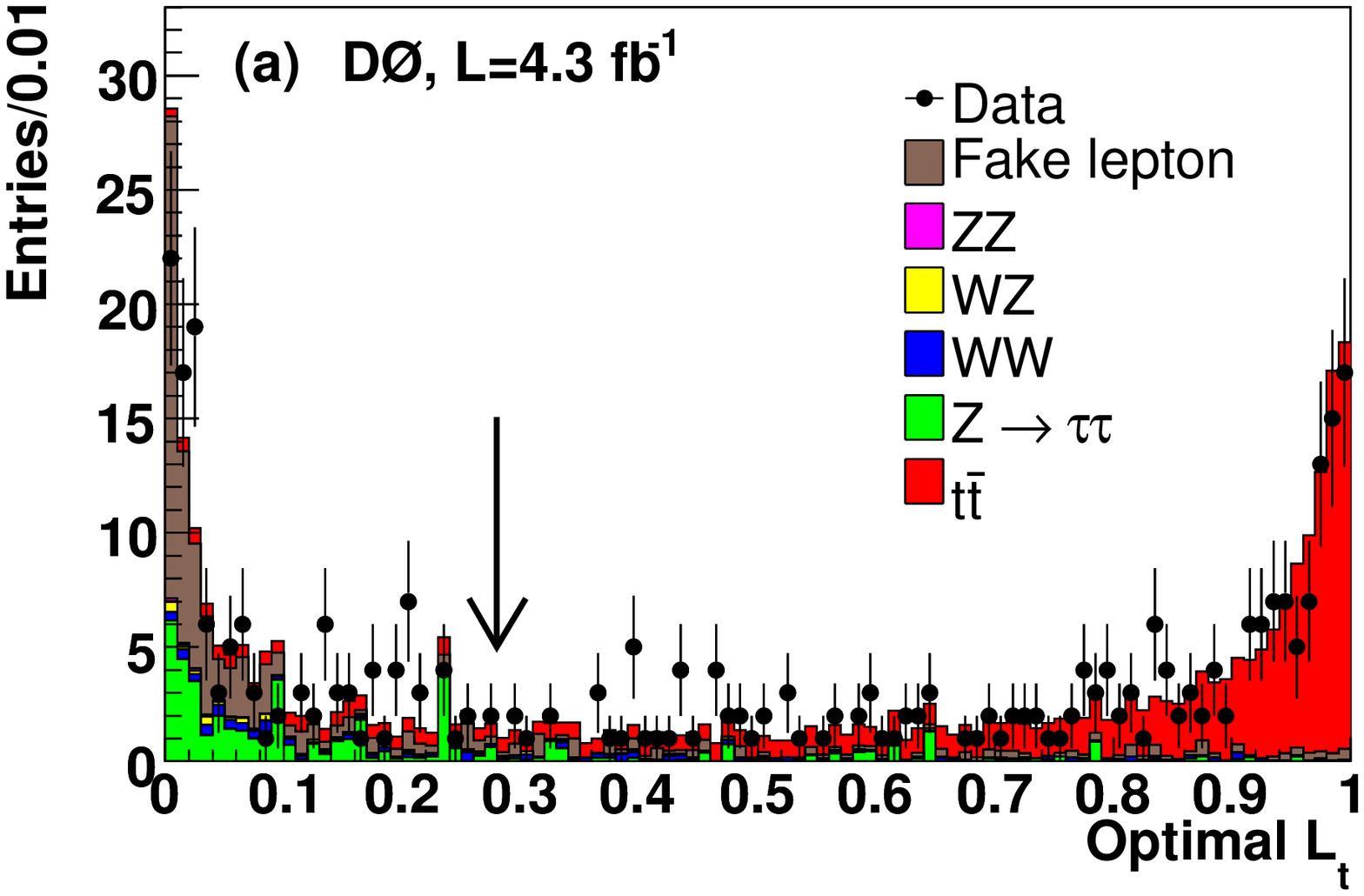}
\includegraphics[scale=0.40]{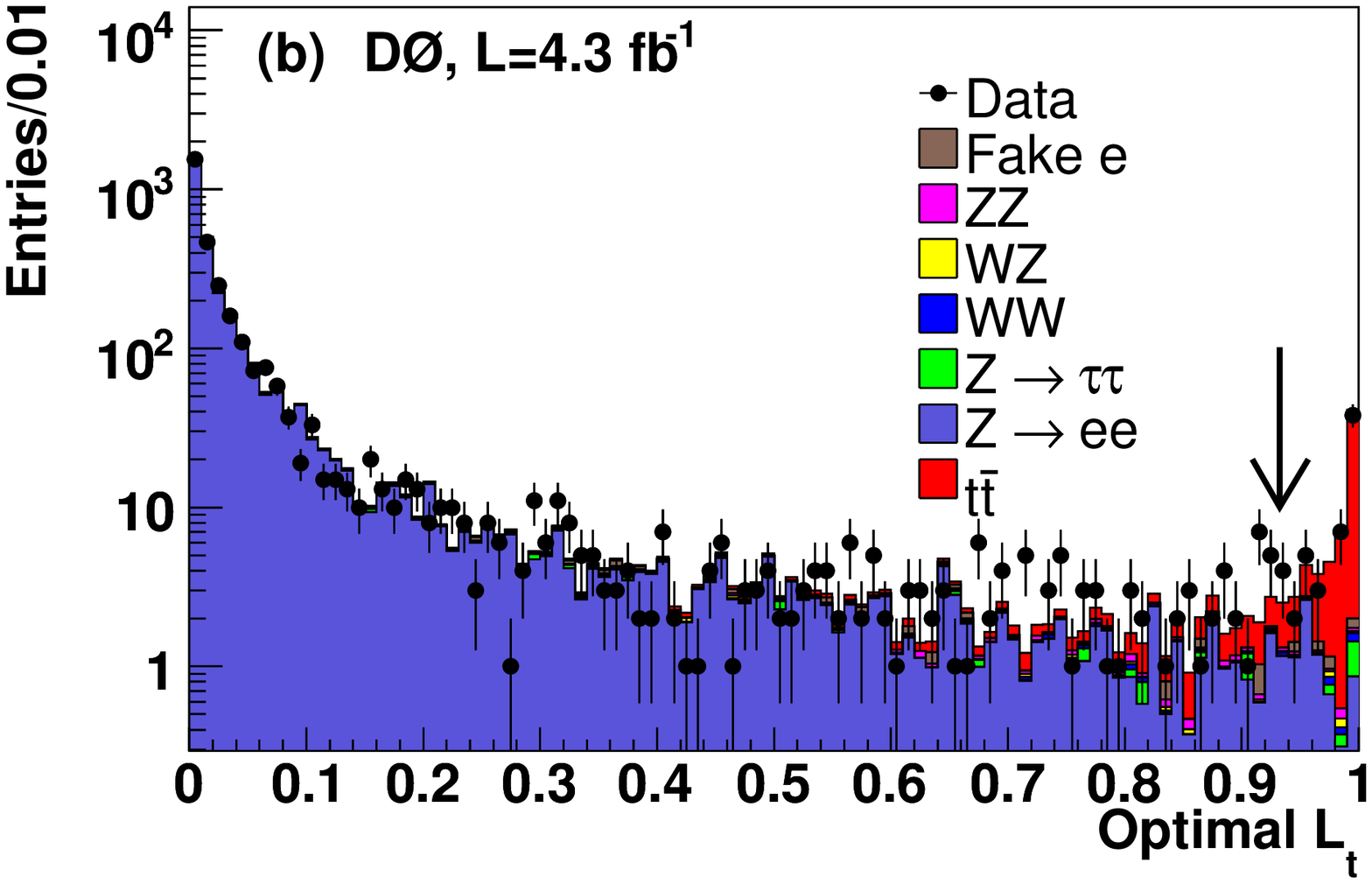}
\includegraphics[scale=0.40]{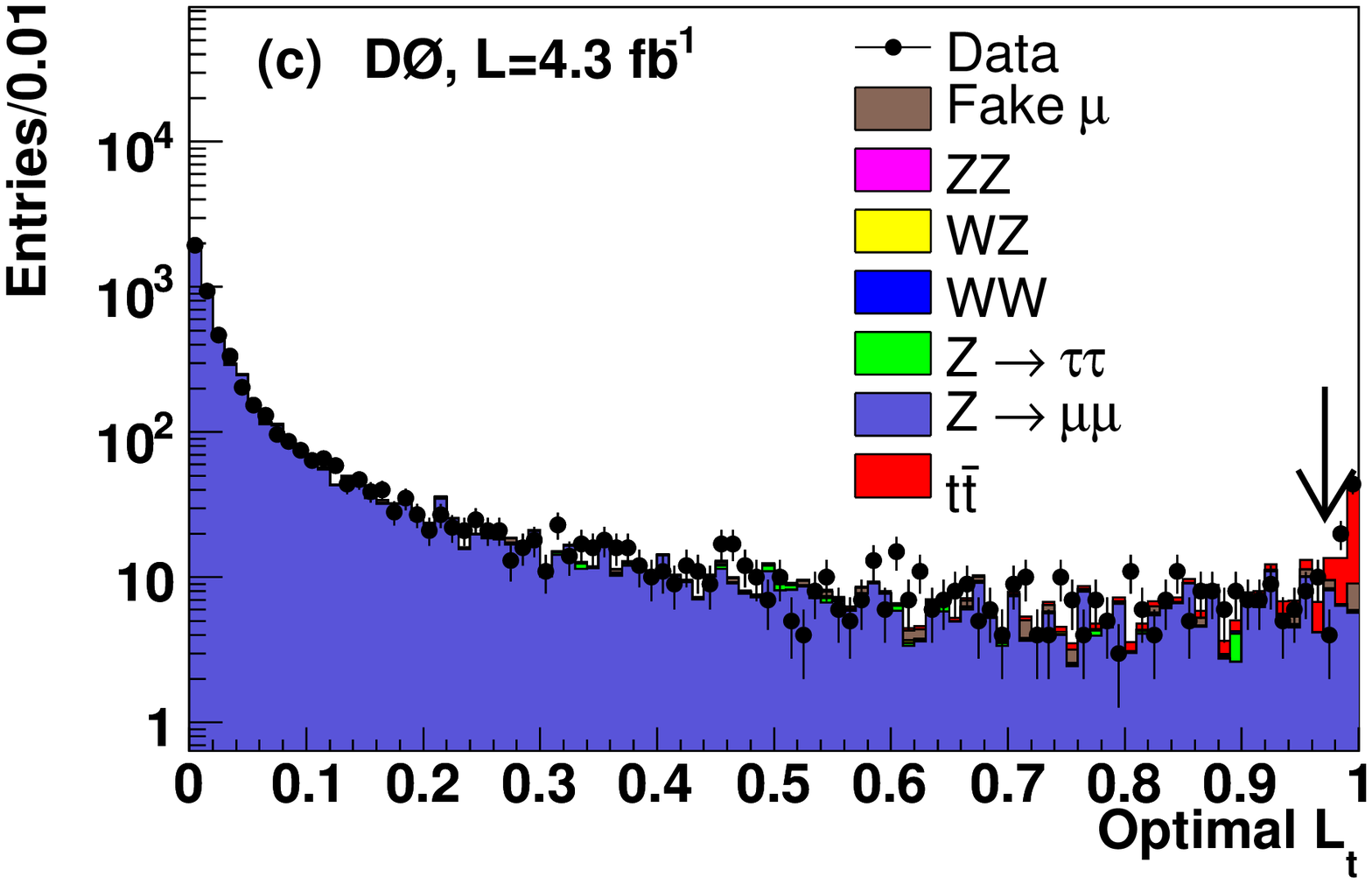}
\caption{\label{fig:BestLtll} (Color online) Best $L_t$ variable for the (a) $e\mu$, (b) $ee$  and (c) $\mu\mu$ decay channels.  The normalization of the signal and background models is determined by the Poisson maximum likelihood fit to the $L_t$ distribution.  The arrows mark the required $L_t$ values for events in each channel.}
\end{figure}

  The sets of variables and $L_t$ selection criteria that maximize the FOM defined in Eq.~\ref{eq:FOM} for each \ttbar\ final state are shown in Tables~\ref{tab:optimization} and~\ref{tab:ll_ltfits}. Figures~\ref{fig:input_ejets}-\ref{fig:mumu_apla_spher} show the distributions of the variables in the best likelihood discriminant $L_t$ for the events passing the preselection cuts, where the signal and background contributions are normalized as described below. In addition, we use $L_t$ to determine the signal and background content of the initially-selected sample by performing a binned Poisson maximum likelihood fit to the $L_t$ distribution where the signal and total background normalizations are free parameters. The $W+$jets contribution is determined by the fit to the $L_t$ distribution, while the multijet component is constrained to be consistent with the value determined from Eqs.~\ref{eq:matrix1} and~\ref{eq:matrix2}.  In the dilepton channels the relative contributions of the different background sources are fixed according to their expected yield, but the total background is allowed to float.  The signal and background yields in the initially-selected sample for the $\mathbf{\ell+}$jets channels are listed in Table~\ref{tab:optimization}, and for the dilepton channels in Table~\ref{tab:ll_ltfits}. Figures~\ref{fig:BestLt} and~\ref{fig:BestLtll} show the distribution of the best likelihood discriminant for each channel, where the signal and background contributions are normalized according to the values returned by the fit. Tables~\ref{tab:data_selection} and~\ref{tab:llfinal} show the optimal $L_t$ cut value for each channel and the final number of events in data and the expected numbers of signal and background events after applying the $L_t$ requirement.

\begin{table}[hhh]
\caption{\label{tab:data_selection} Expected background and $t\bar{t}$ yields, and the number of events observed, after the selection on $L_t$ in the $\ell+$jets decay channels.}
\begin{tabular}{lr@{$\,\pm \,$}lr@{$\,\pm \,$}l}
\hline
\hline
                                       & \multicolumn{2}{c}{$e+$jets}  & \multicolumn{2}{c}{$\mu+$jets}      \\ \hline
Optimized $L_t$ requirement               & \multicolumn{2}{c}{$>$ 0.58} & \multicolumn{2}{c}{$>$ 0.29} \\
\hline
\ttbar                   & 484.4 & 41.4 & 567.2 & 47.3\\
$W+$jets                 & 111.7 & 12.6 & 227.7 & 19.2\\
Multijet                      &  58.1 & 3.9 & 4.0 & 3.1\\ \hline
Total                 &  656.2 & 43.4 & 798.9 & 51.2\\ \hline
Observed                 &  \multicolumn{2}{c}{628} & \multicolumn{2}{c}{803}           \\
\hline
\hline
\end{tabular}
\end{table}

\begin{center}
\begin{table}
\caption{\label{tab:llfinal}  Expected background and $t\bar{t}$ yields, and the number of events observed, after the selection on $L_t$ in the dilepton decay channels.}
\begin{tabular}{lr@{$\,\pm \,$}lr@{$\,\pm \,$}lr@{$\,\pm \,$}l}
\hline
\hline
Source & \multicolumn{2}{c}{$e\mu$} & \multicolumn{2}{c}{$ee$} & \multicolumn{2}{c}{$\mu\mu$} \\ \hline
Optimized $L_t$ requirement   & \multicolumn{2}{c}{$>$ 0.28}            & \multicolumn{2}{c}{$>$ 0.934} & \multicolumn{2}{c}{$> 0.972$} \\
\hline 
$t\bar{t}$ & 186.6 & 0.4 & 44.5 & 0.3  &  43.6 & 0.3  \\ 
$Z/\gamma^* \rightarrow \ell^+\ell^-$  & \multicolumn{2}{c}{N/A} & 7.4 & 1.0 &  19.1 & 1.3\\
$Z/\gamma^* \rightarrow \tau\tau$      & 11.2  & 3.7  & 0.8 & 0.3 & 0.35 & 0.05 \\
$WW$                          & 5.6 & 1.4   & 0.3 &  0.1& 0.13 & 0.05\\
$WZ$                          &  1.5 & 0.5 & 0.28 & 0.04 & 0.16 & 0.01\\
$ZZ$                          &  1.0 & 0.5 & 0.34 & 0.04 & 0.57 & 0.04\\
Misidentified jets                   & 15.9 & 3.1       & 0.54 & 0.48 & 3.7 & 2.5\\ \hline
Total & 221.7 & 5.1 & 54.2 & 1.2 & 67.7 & 3.9 \\ \hline
Observed &  \multicolumn{2}{c}{193}   & \multicolumn{2}{c}{58}  &  \multicolumn{2}{c}{68} \\ \hline
\hline
\end{tabular}
\end{table}
\end{center}

\section{\label{sec:template}Templates}

After the final event selection,  \coss\ is calculated for each event by using the reconstructed top quark and $W$ boson four-momenta. In the $\ell+$jets decay channel, the four-momenta are reconstructed using a kinemetic fit with the constraints: (i) two jets should give the invariant mass of the $W$ boson (80.4 GeV/$c^2$), (ii) the invariant mass of the  lepton and neutrino should be the  $W$ boson mass, (iii) the mass of the reconstructed top and anti-top quark should be 172.5 GeV/$c^2$, and (iv) the ${\vec p_T}$ of the $t\bar{t}$ system should be opposite that of the unclustered energy in the event. The four highest-$p_T$ jets in each event are used in the fit, and among the twelve possible permutations in the assignment of the jets to initial partons, the solution with the highest probability is chosen, considering both the NN$_b$ values of the four jets and $\chi^2_k$.  This procedure selects the correct jet assignment in 59\% of MC $t\bar{t}$ events.  With the jet assigned, the complete kinematics of the \ttbar\ decay products (i.e., including the neutrino) are determined, allowing us to boost to the rest frames of each $W$ boson in the event.  We compute \coss\ for the $W$ boson that decays leptonically.  The hadronic $W$ boson decay from the other top quark in the event also contains information about the helicity of that $W$ boson, but since we do not distinguish between jets formed from up-type and down-type quarks, we can not identify the down-type fermion to calculate \coss.  We therefore calculate only $|\coss|$, which is identical for both jets in the rest frame of the hadronically decaying $W$ boson.   Left-handed and right-handed $W$ bosons have identical $|\coss|$ distributions, but we can distinguish either of those states from longitudinal $W$ bosons, thereby improving the precision of the measurement. 

In the dilepton decay channel, the presence of two neutrinos prevents a constrained kinematic fit, but with the assumption that the top quark mass is 
172.5~GeV/$c^2$, an algebraic solution for the neutrino momenta can be obtained (up to a two-fold ambiguity in pairing the jets and leptons, and a four-fold solution ambiguity).  To account for the lepton and jet energy resolutions, the procedure described above is repeated 500 times with the energies fluctuated according to their uncertainties,  and the average of all the solutions is used as the value of the \coss ~for each top quark.  

\begin{figure}
\includegraphics[scale=0.45]{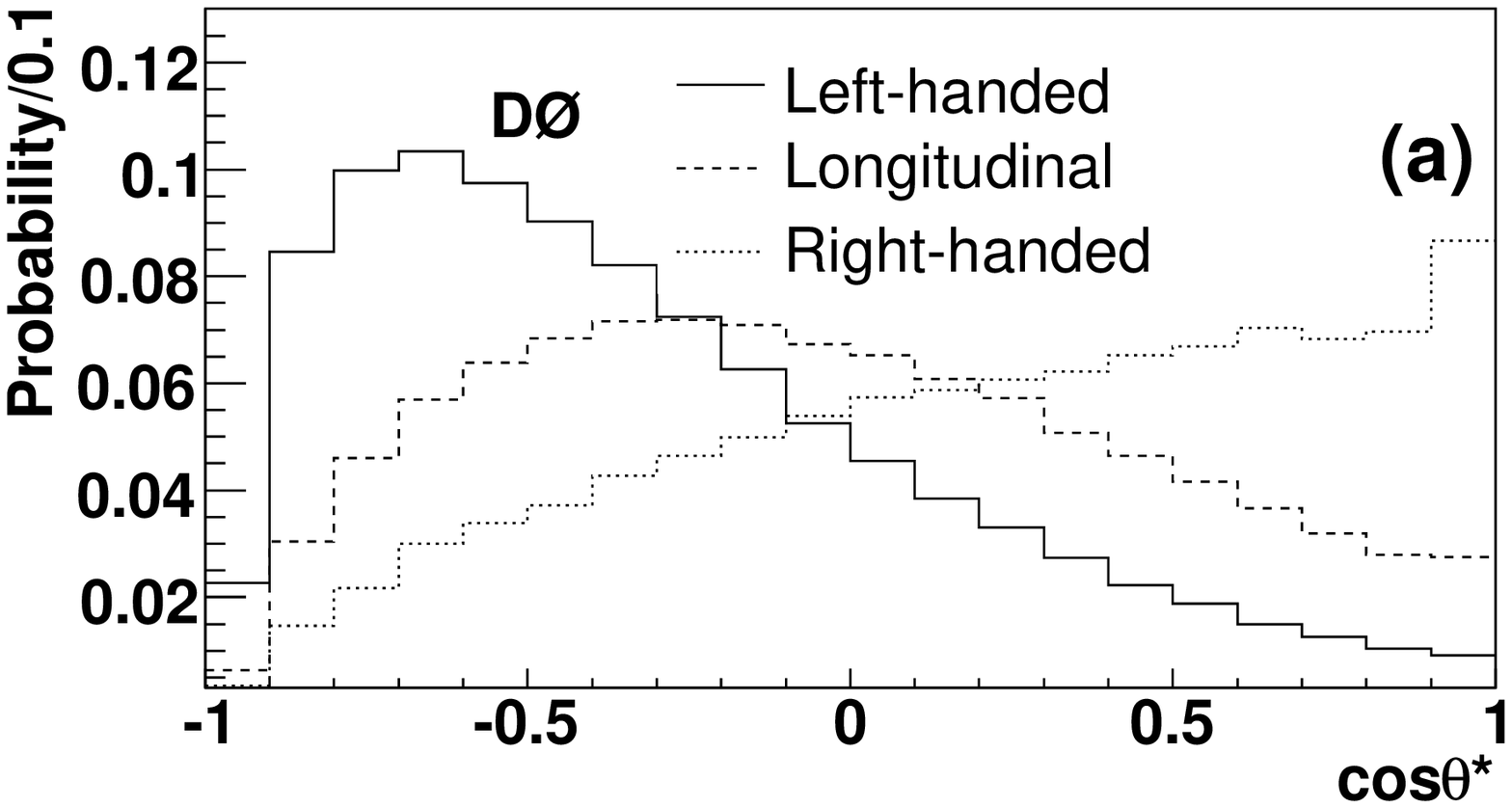}
\includegraphics[scale=0.45]{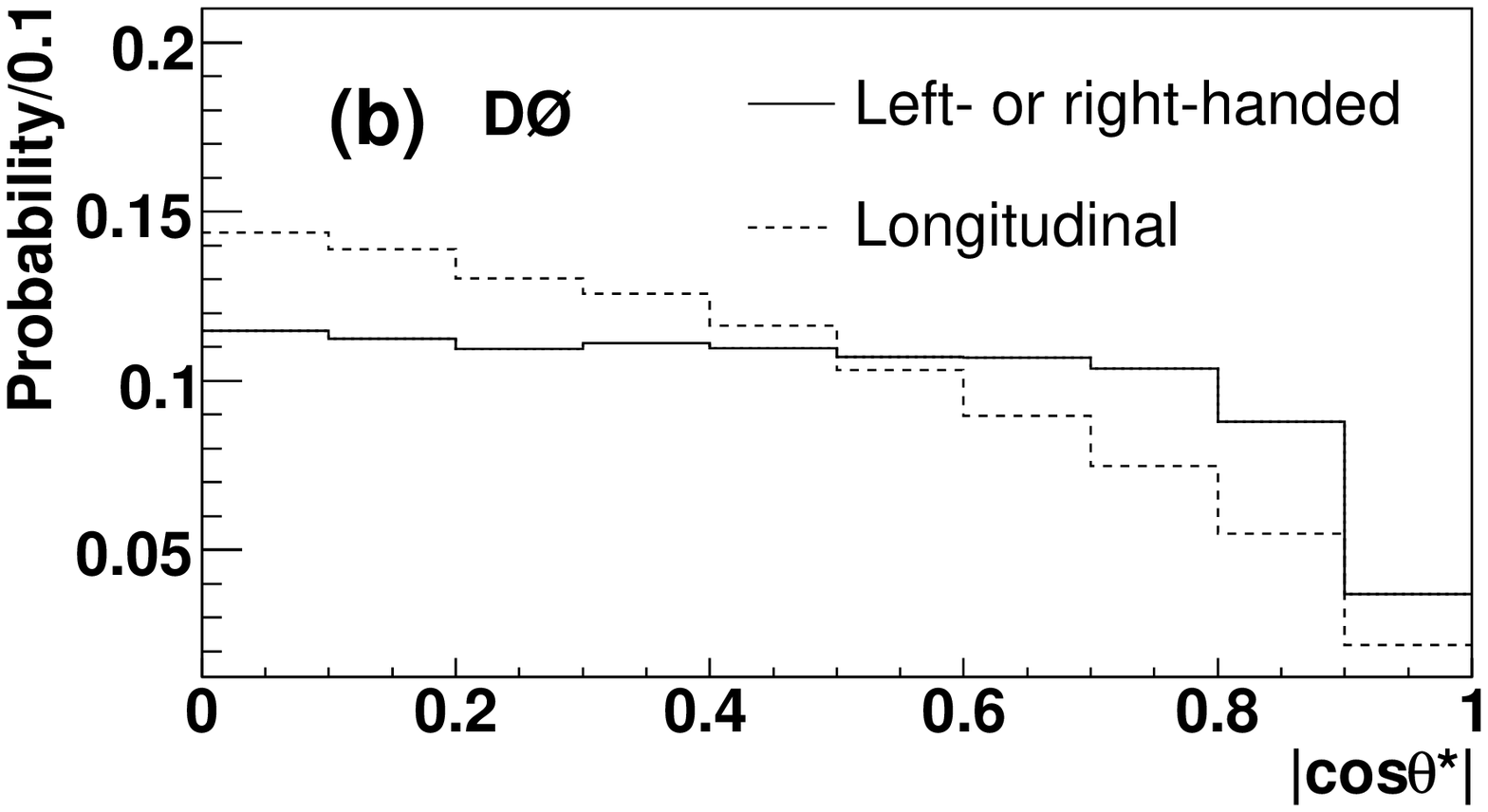}
\includegraphics[scale=0.45]{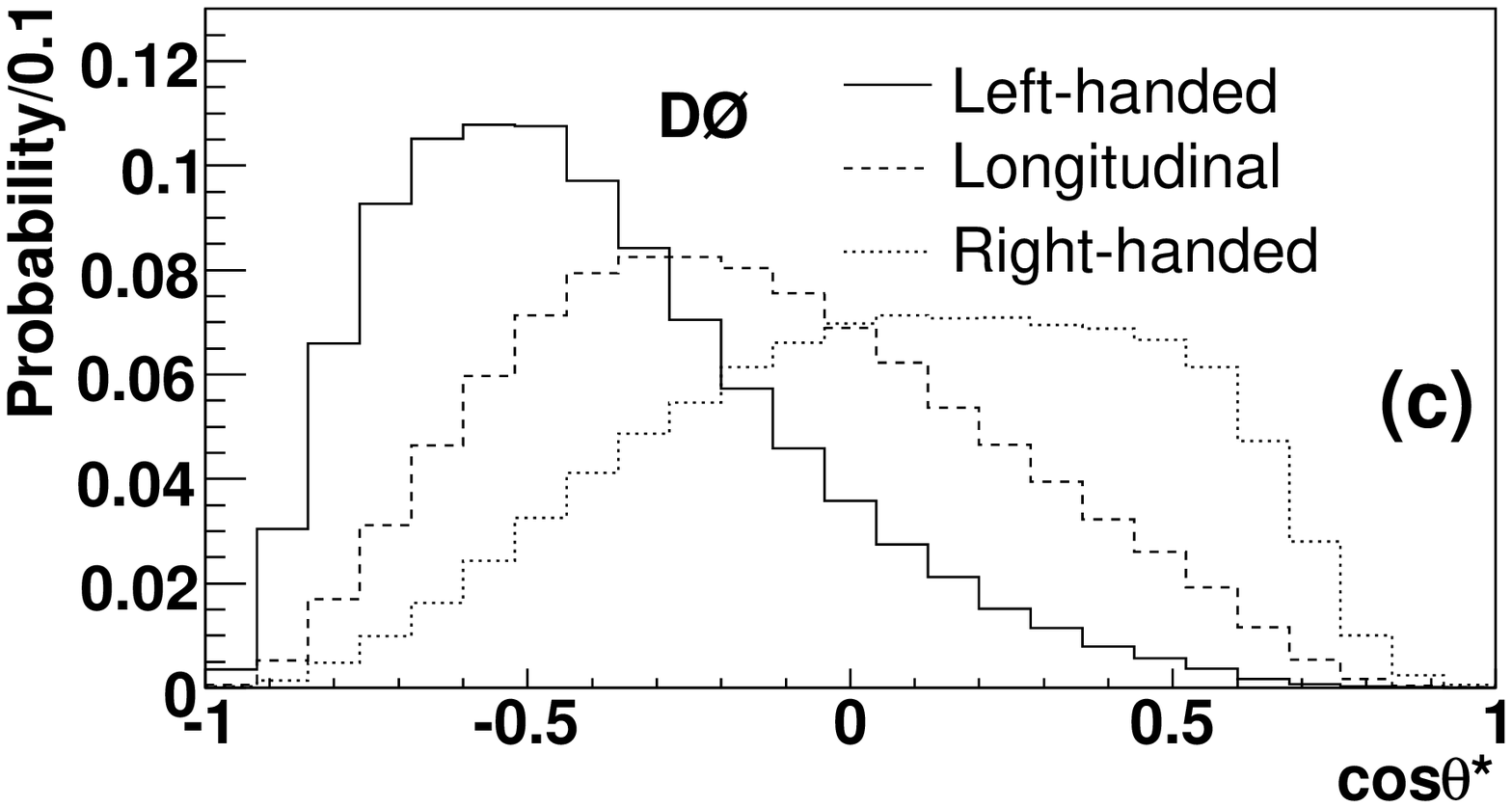}
\caption{\label{fig:pureTemplates} Distribution of \coss\ in \ttbar\ MC 
samples that were reweighted to derive the distributions for purely left-handed, longitudinal, or right-handed $W$ bosons.  The distribution for leptonically- and hadronically-decaying $W$ bosons in $\ell+$jets events are shown in (a) and (b), respectively, and the distribution for dilepton events is shown in (c). For hadronically decaying $W$ bosons the \coss\ distribution for left- and right-handed $W$ bosons are identical. All of the distributions are normalized to unity.
}
\end{figure}

\begin{figure}
\includegraphics[scale=0.40]{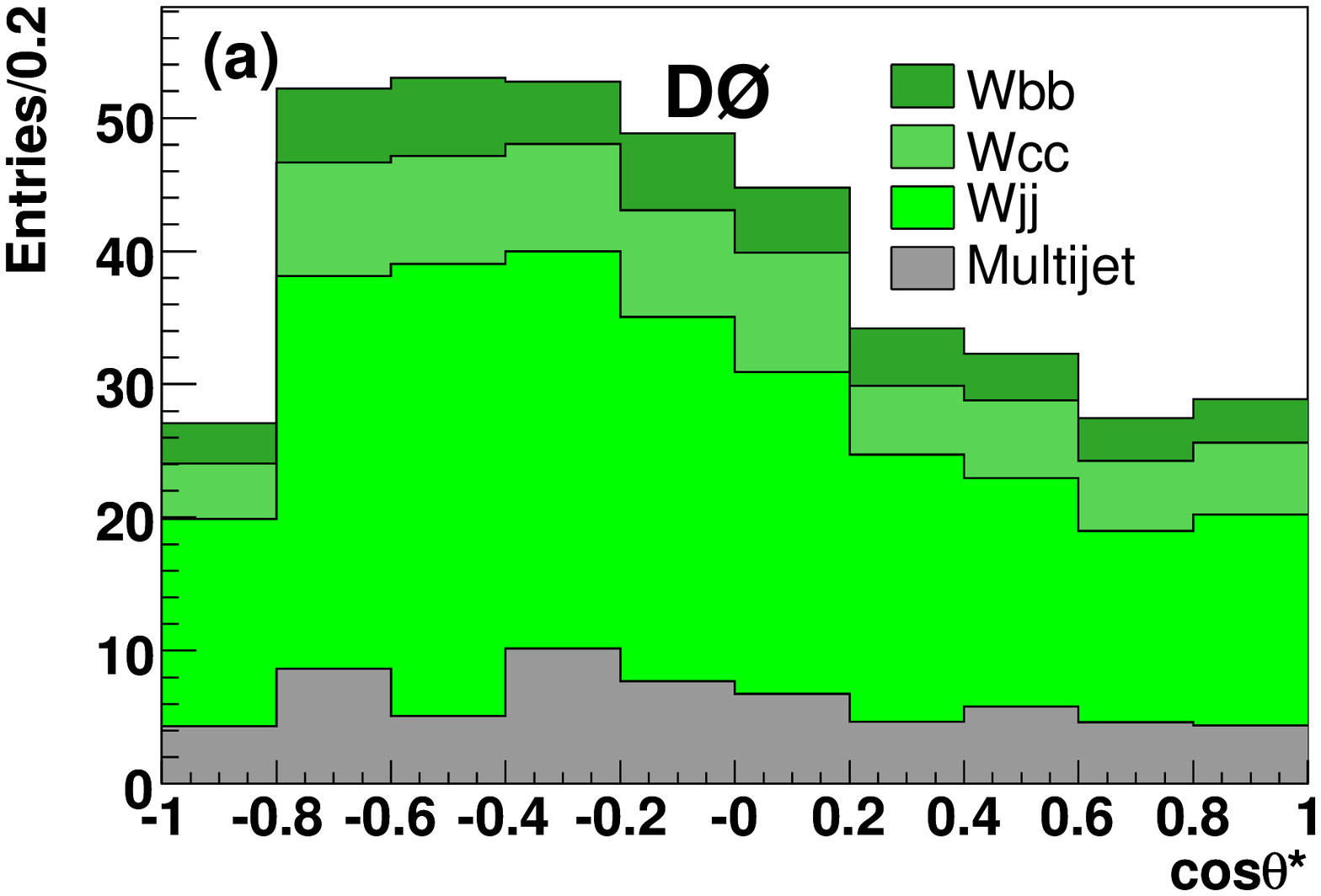}
\includegraphics[scale=0.40]{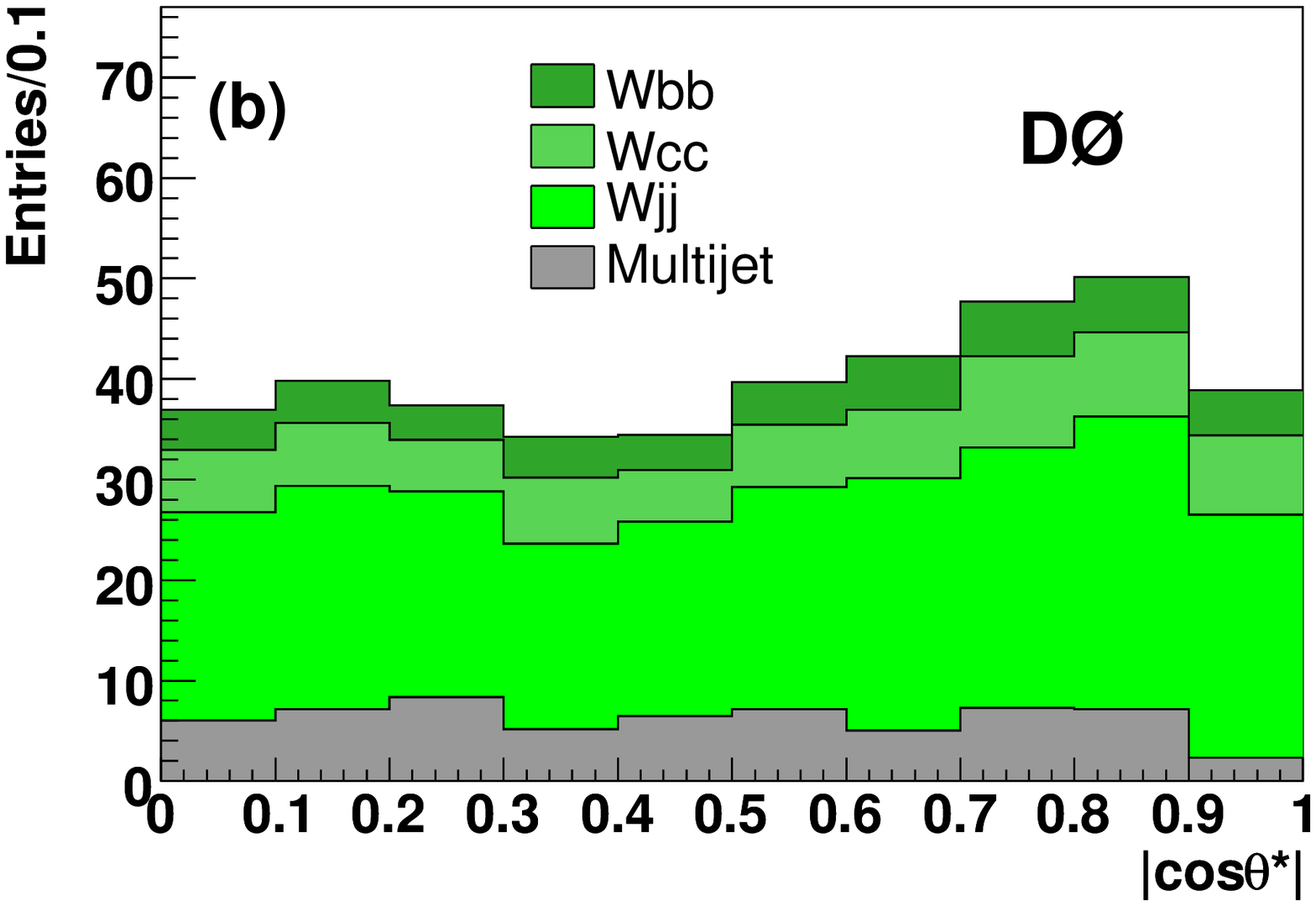}
\includegraphics[scale=0.40]{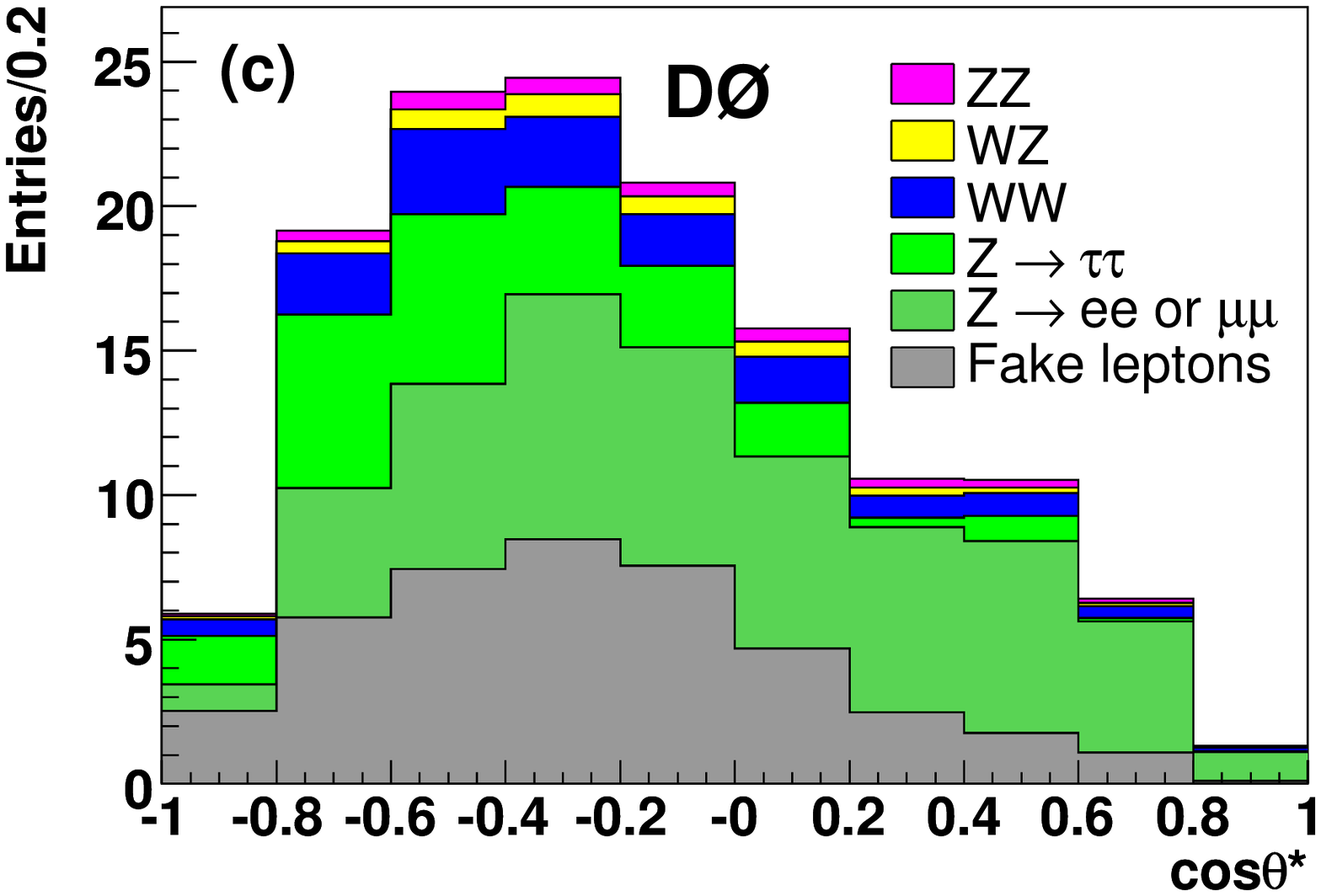}
\caption{\label{fig:bkgTemplates} (Color online) Distribution of \coss\ in background samples.  The distribution for leptonically- and hadronically-decaying $W$ bosons in $\ell+$jets events are shown in (a) and (b), respectively, and the distribution for dilepton events is shown in (c).  All of the distributions are normalized to the expected yield for each source of background.
}
\end{figure}

  As mentioned above, the extraction of both $f_0$ and $f_+$ requires comparing the data with the MC models in which both of these values are varied. Since {\sc alpgen} can only produce linear combinations of  $V-A$ and $V+A$ $tWb$ couplings, it is unable to produce non-SM $f_0$ values, and can produce $f_+$ values only in the range $[0, 0.30]$. We therefore start with {\sc alpgen} $V-A$ and $V+A$ samples, and divide the samples in bins of parton-level \coss.  For each bin, we note the efficiency for the event to satisfy the event selection and the distribution of reconstructed \coss\ values.  With this information we  determine the expected distribution of reconstructed \coss\ values for any assumed $W$ helicity fractions, and in particular we choose to derive the 
  distributions expected for purely left-handed, longitudinal, or right-handed $W$ boson, as shown
in 
Fig.~\ref{fig:pureTemplates}.   The deficit of entries near $\coss\ = -1$ relative to the expectation from
Eq.~\ref{eq:expcost} is due to the $p_T$ requirement imposed when selecting leptons.   We verify the reweighting procedure by comparing the generated $V\pm A$ {\sc alpgen} samples with the combination of reweighted distributions expected for $V\pm A$ couplings, and find that these distributions agree within the MC statistics.
The templates for background samples are obtained directly from the relevant MC or data background samples, and are shown in
Fig.~\ref{fig:bkgTemplates}.

\section{\label{sec:2Dfit}Model-independent $W$ Helicity Fit}

The $W$ boson helicity fractions are extracted by computing a binned Poisson likelihood $L(f_0,f_+)$ with the distribution of \coss\ in the data to be consistent with the sum of signal and background templates. The likelihood is a function of the $W$ boson helicity fractions $f_0$ and $f_+$, defined as
\begin{eqnarray}
L(f_0,f_+) & = & \prod_{i=1}^{N_{\rm chan}}\prod_{j=1}^{N_{{\rm bkg},i}}e^{-(n_{b,ij}-\overline{n}_{b,ij})^2/2\sigma_{b,ij}^2} \times \nonumber \\                   
& & \prod_{k=1}^{N_{{\rm bins},i}} P(d_{ik};n_{ik}) 
\label{eq:lhood}
\end{eqnarray}
where $P(d_{ik};n_{ik})$ is the Poisson probability for observing $d_{ik}$ events given a mean expectation value $n_{ik}$, $N_{\rm chan}$ is the number of channels in the fit (a maximum of five in this analysis: $e+$jets, $\mu+$jets, $e\mu$, $ee$, and $\mu\mu$), $N_{{\rm bkg},i}$ is the number of background sources in the $i^{\rm th}$ channel, ${N_{{\rm bins},i}}$ is the number of bins in the \coss\ distribution for any given channel (plus the number of bins in the $|\coss|$ distribution for hadronic $W$ boson decays in the $\ell+$jets channels), $\overline{n}_{b,ij}$ is the nominal number of \coss\ measurements from the $j^{\rm th}$ background contributing to the $i^{\rm th}$ channel, $\sigma_{b,ij}$ is the uncertainty on $\overline{n}_{b,ij}$, ${n_{b,ij}}$ is the fitted number of events for this background, $d_{ik}$ is the number of data events in the $k^{\rm th}$ bin of  \coss ~for the $i^{\rm th}$ channel, and $n_{ik}$ is the predicted sum of signal and background events in that bin. The $n_{ik}$ can be expressed as
\begin{eqnarray}
n_{ik} & = & n_{s,i}{\varepsilon_0 f_0p_{0,ik} + \varepsilon_+ f_+p_{+,ik} + \varepsilon_-(1-f_0-f_+)p_{-,ik} \over  f_- \varepsilon_- + f_0 \varepsilon_0 + f_+ \varepsilon_+ } \nonumber \\ & & + \sum_{j=1}^{N_{\rm bkg}}n_{b,ij}p_{b,ijk}
\end{eqnarray}
where $n_{s,i}$ represents the number of \coss\ measurements from signal events in a given channel, the $p$ represent the probabilities for an event from some source to appear in bin $k$ for channel $i$ (as determined from the templates), and the subscripts 0, $+$, $-$ refer to the templates for \ttbar\ events in which the $W$ bosons have zero, negative, or positive helicity, and the subscript $b,i$ refers to the templates for the $i^{\rm th}$ background source. The efficiency for a \ttbar\ event to satisfy the selection criteria depends upon the helicity states of the two $W$ bosons in the event; the $\varepsilon$ are therefore necessary to translate the fractions of events with different helicity states in the selected sample to the fractions that were produced.  The quantity $\varepsilon_\lambda$ is defined as
\begin{eqnarray}
 \varepsilon_{\lambda} =   \sum_{\lambda^\prime} f_{\lambda\prime} \varepsilon_{\lambda \lambda^\prime}
\end{eqnarray}
where $ \varepsilon_{\lambda \lambda^\prime}$ is the relative efficiency for events with $W$ bosons 
in the $\lambda$ and $\lambda^\prime$ helicity states to satisfy the selection criteria.
The values of $ \varepsilon_{\lambda \lambda^\prime}$ for each \ttbar\ decay channel are given in 
Table~\ref{tab:relEff}. While performing the fit, both $f_0$ and $f_+$ are allowed to float freely, and the measured $W$ helicity fractions correspond to those leading to the highest likelihood value. 

\begin{table}
\caption{\label{tab:relEff} Efficiencies of different $W$ boson helicity configurations in \ttbar\ events
to pass the selection criteria, relative to the efficiencies for a mixture of $V-A$ and $V+A$ events.   The indices $-,0$ and $+$ correspond to the helicity states of the two $W$ bosons, and their order is  leptonic $W$, hadronic $W$ for the \ljets\ channel, and arbitrary for dilepton channels (where there is no distinction between the two $W$ bosons in the event).  Small differences in values in the dilepton channels under interchange of the indices are from variations in MC statistics.}
\begin{tabular}{cccccc}
\hline
\hline
         & $e$+jets & $\mu$+jets & $e\mu$ & $ee$ & $\mu\mu$  \\ \hline
$\varepsilon_{--}$  & 0.76 & 0.73 & 0.67 &  0.68  & 0.68 \\
$\varepsilon_{-0}$ &  0.87 & 0.83  & 0.84  & 0.86 & 0.85 \\
$\varepsilon_{-+}$ &  0.76 & 0.73 &  0.88  & 0.89 & 0.89 \\
$\varepsilon_{0-}$  & 0.94 & 0.95 & 0.85 & 0.86 & 0.87 \\ 
$\varepsilon_{00}$  & 1.08 & 1.09 & 1.06  & 1.05 & 1.05 \\ 
$\varepsilon_{0+}$ & 0.94 &  0.95 & 1.10 & 1.05 & 1.05 \\ 
$\varepsilon_{+-}$ &0.92 & 0.96  & 0.89  & 0.88 & 0.91 \\ 
$\varepsilon_{+0}$ & 1.06 & 1.11 & 1.12 & 1.03 & 1.07 \\ 
$\varepsilon_{++}$ & 0.92 & 0.96 & 1.15 & 0.99 & 1.03 \\  \hline
\hline
\end{tabular}
\end{table}

\begin{figure}
\includegraphics[scale=0.4]{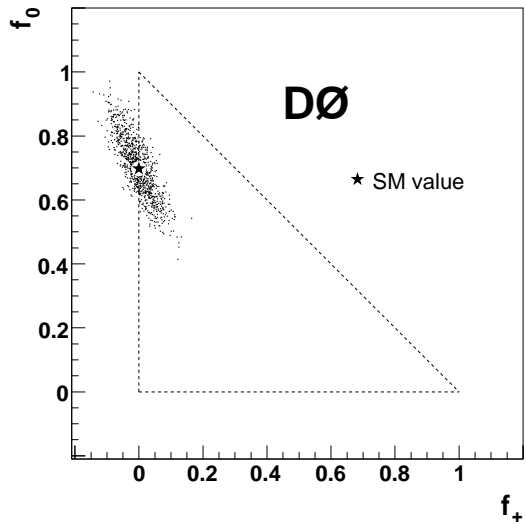}
\caption{\label{fig:ensemExample}  Fit values for  $f_0$ and $f_+$ obtained with 1000 MC simulations of the $W$ boson helicity measurement.   The SM helicity fractions, marked by the
star, were taken as input to the simulations.  The triangle corresponds to the physically allowed region where $f_0 + f_+ \le 1$.}
\end{figure}

We check the performance of the fit using simulated ensembles of events, with all values of $f_0$ and $f_+$ from 0 through 1 as inputs in increments of 0.1, with the sum of $f_0$ and $f_+$ not exceeding unity. We simulate input data distributions for the various values by combining the pure left-handed, longitudinal, and right-handed templates in the assumed proportions.  In these ensembles, we draw a random subset of the simulated events, with the number of events chosen in each channel fixed to the number observed in data.  Within the constant total number of events, the numbers of signal and background events are fluctuated binomially around the expected values.   Each of these sets of simulated events is passed through the maximum likelihood fit using the standard \coss\ templates.  We find that the average fit output value is close to the input value across the entire range of possible values for the helicity fractions, with the small differences between the input and output values being consistent with statistical fluctuations in the ensembles.
As an example, the set of $f_0$ and $f_+$ values obtained when $t\bar{t}$ events are drawn
in the proportions expected in the SM is shown in Fig.~\ref{fig:ensemExample}.

\section{\label{sec:syst}Systematic Uncertainties}

Systematic uncertainties are evaluated using simulated event ensembles in which both changes in the background yield and changes in the shape of the \coss\ templates in signal and background are considered.  The simulated samples from which the events are drawn can be either the nominal samples or samples in which the systematic effect under study has been shifted away from the nominal value.  In general, the systematic uncertainties assigned to $f_0$ and $f_+$ are determined by taking an average of the absolute values of the differences in the average fit output values between the nominal and shifted $V-A$ and $V+A$ samples.

The jet energy scale, jet energy resolution, and jet identification efficiency each have relatively small uncertainties that are difficult to observe above fluctuations in the MC samples.  To make the effects more visible, we vary these quantities by $\pm$5 standard deviations, and then divide the resulting differences in the average fit output by 5.  The top quark mass uncertainty corresponds to shifting $m_t$ by 1.4 GeV/$c^2$, which is the sum in quadrature of the uncertainty on the world average $m_t$ (1.1 GeV/$c^2$) and the difference between the world average value (173.3  GeV/$c^2$) and the value assumed in the analysis (172.5 GeV/$c^2$).  We evaluate the contribution of template statistics to the uncertainty by repeating the fit to the data 1000 times, fluctuating the signal and background distributions according to their statistics in each fit.  The uncertainties due to the modeling of \ttbar ~events are separated into several categories and evaluated
using special-purpose MC samples.  The uncertainty in the model of gluon radiation is assessed using {\sc pythia} MC samples in which the amount of gluon radiation is shifted upwards and downwards;  the impact of NLO effects is assessed by comparing the default leading-order {\sc alpgen} generator with the NLO generator {\sc mc@nlo}~\cite{mcatnlo};  the uncertainty in the hadronic showering model is assessed by comparing {\sc alpgen} events showered with {\sc pythia} and with {\sc herwig}~\cite{herwig}; and lastly, the impact of color reconnection effects is assessed by comparing {\sc pythia} samples where the underlying event model does and does not include color reconnection.  The uncertainty due to data and MC differences in the background \coss ~distribution is derived by taking the ratio of the data and the MC distribution for a background-enriched sample (defined by requiring that events have low values of $L_t$) and then using that ratio to re-weight the distribution of background MC events that satisfy the standard selection.   The uncertainty in the heavy flavor content of the background is estimated by varying the fraction of background events with heavy flavor jets by $\pm 20$\%.  Uncertainties due to the fragmentation of $b$ jets are evaluated by comparing the default fragmentation model, the Bowler scheme~\cite{bowler} tuned to data collected at the CERN LEP collider, with an alternate model tuned to data collected by the SLD collaboration~\cite{bfragtuning}.  Uncertainties in the parton distribution functions (PDFs) are estimated using the set of $2\times20$ errors provided for the CTEQ6M~\cite{cteq6m} PDF.  The analysis consistency uncertainty reflects the typical difference between the input helicity fractions and the average output values observed in fits to simulated event ensembles. Finally,  we include an uncertainty corresponding to muon triggers and identification, as control samples indicate some substantial data/MC discrepancies for the loose selection we use.   All the systematic uncertainties are summarized in Table~\ref{tab:2dsyst}.

\begin{table}[hhh]
\caption{\label{tab:2dsyst} Summary of the absolute systematic uncertainties on $f_+$ and $f_{0}$.}
\begin{tabular}{lcc}
\hline
\hline
Source & Uncertainty ($f_+$) & Uncertainty ($f_0$)  \\ \hline
Jet energy scale       & 0.007  & 0.009 \\
Jet energy resolution  & 0.004 & 0.009 \\
Jet ID        & 0.004 & 0.004 \\
Top quark mass          & 0.011 & 0.009 \\
Template statistics   & 0.012 & 0.023 \\
\ttbar ~model          & 0.022 & 0.033 \\
Background model       & 0.006   & 0.017  \\
Heavy flavor fraction   & 0.011 & 0.026 \\
$b$ fragmentation      & 0.000 &  0.001 \\
PDF                 & 0.000 & 0.000 \\
Analysis consistency   & 0.004 &  0.006\\
Muon ID            &  0.003 & 0.021 \\
Muon trigger     &  0.004 & 0.020 \\  \hline
Total                  & 0.032 &   0.060 \\ \hline
\hline
\end{tabular}
\end{table}

\section{\label{sec:p20}Result}
Applying the model independent fit to the Run IIb data, we find
\begin{eqnarray}
f_0 &=& 0.739 \pm 0.091 \hbox{ (stat.)} \pm  0.060  \hbox{ (syst.)}   \\ \nonumber
f_+ &=& -0.002 \pm 0.045 \hbox{ (stat.)} \pm  0.032 \hbox{ (syst.)}.
\end{eqnarray}

\begin{figure}
\includegraphics[scale=0.45]{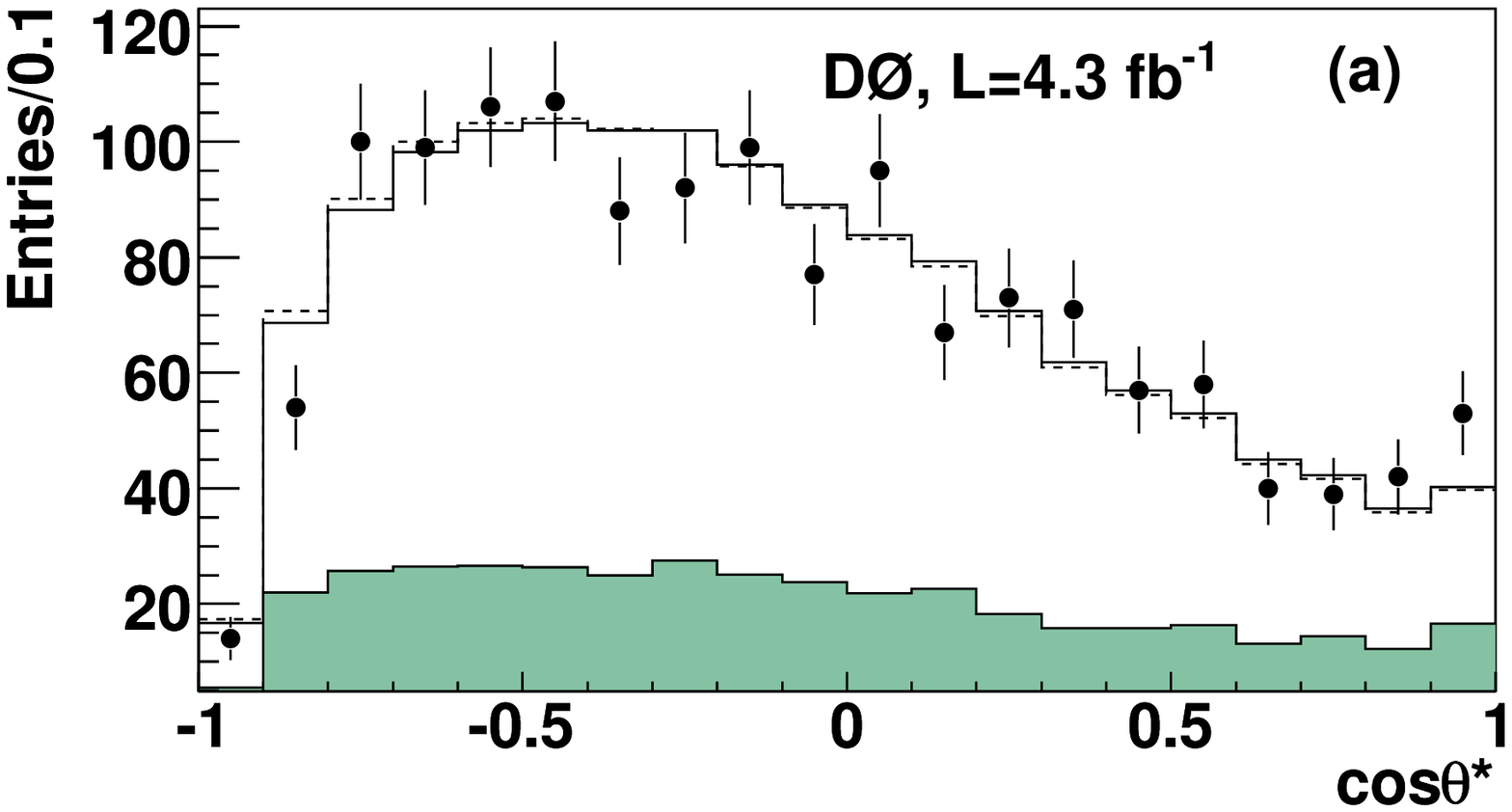} \\
\includegraphics[scale=0.45]{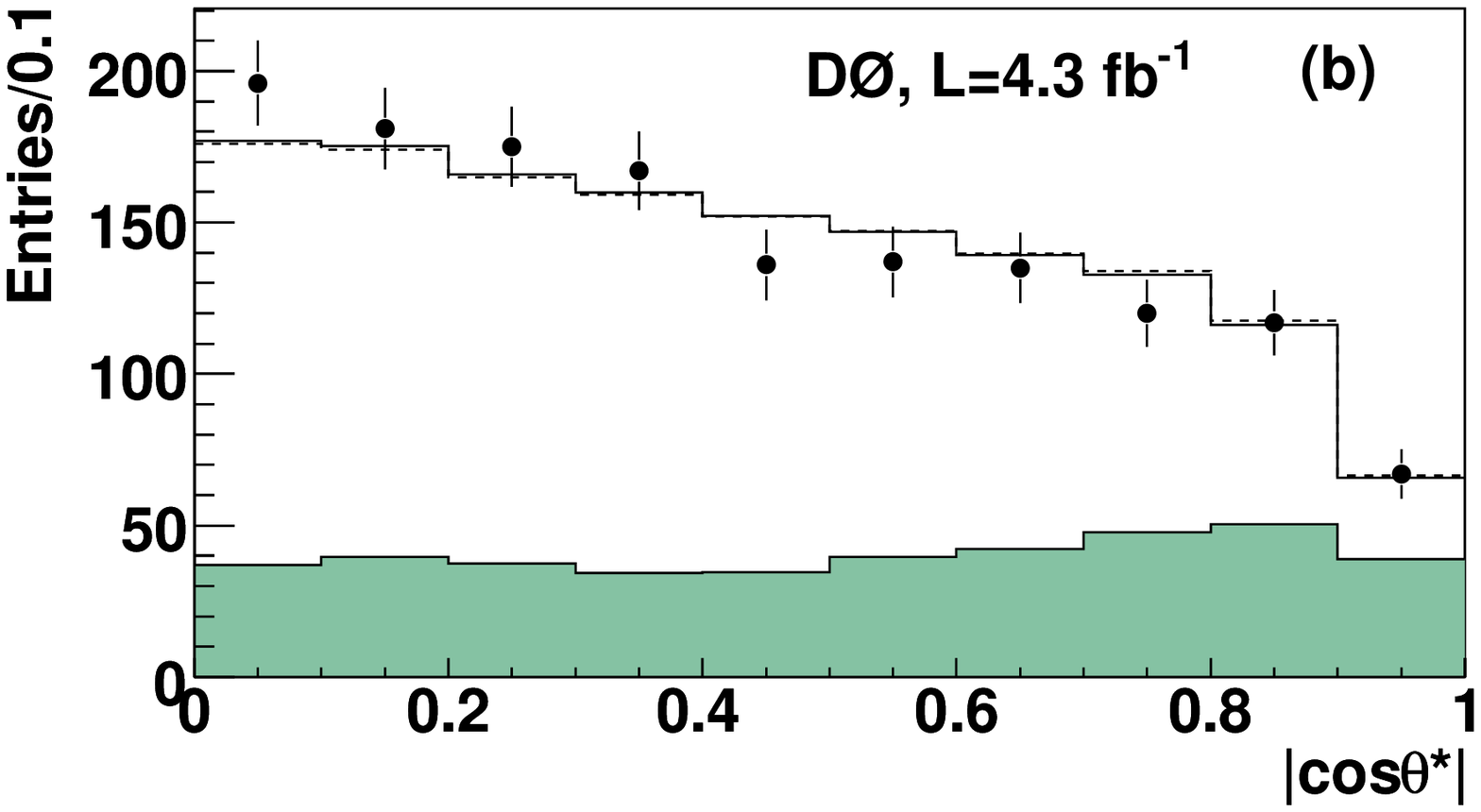}\\
\includegraphics[scale=0.45]{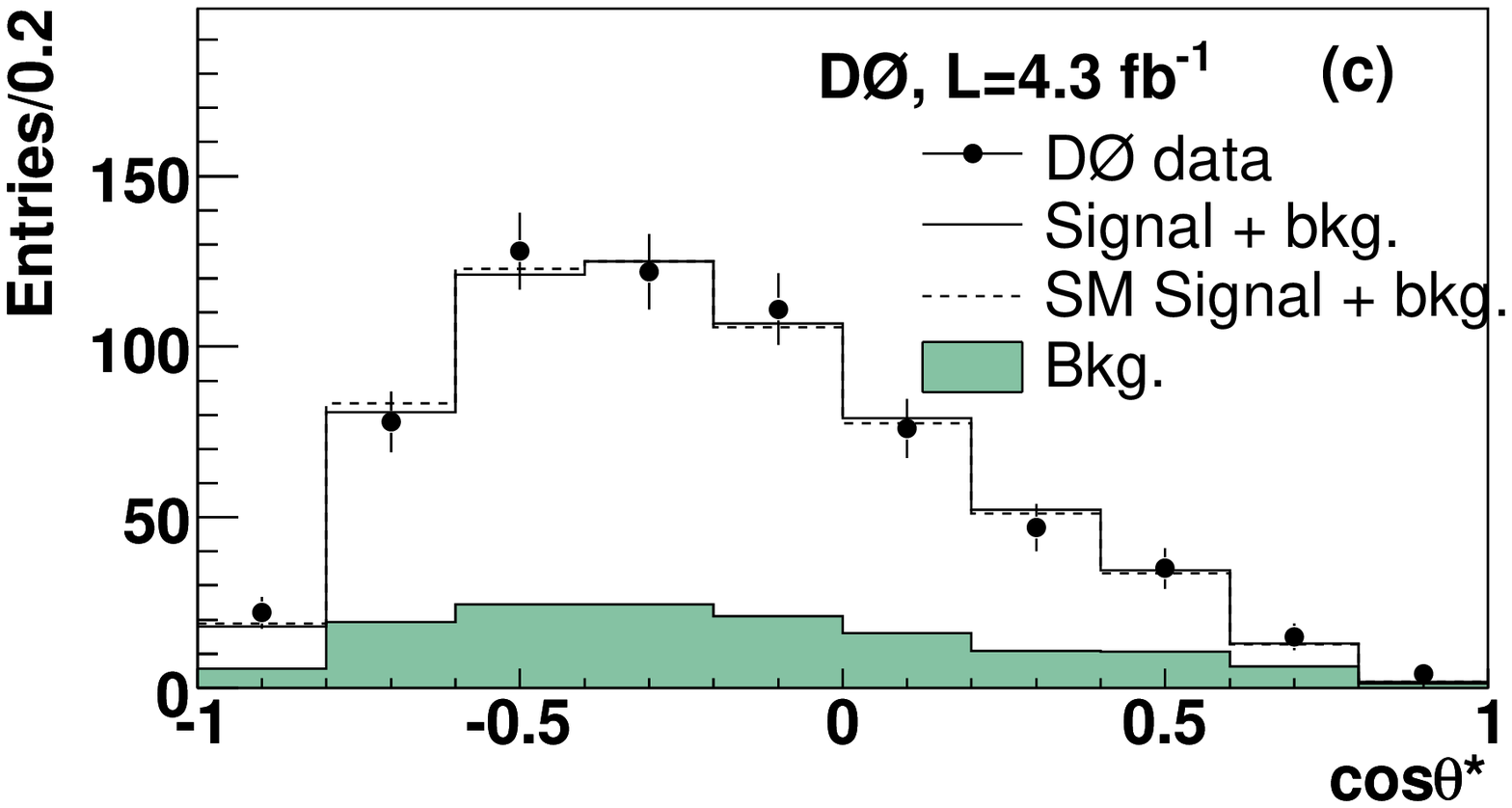}
\caption{\label{fig:data2dmodel} 
(Color online) Comparison of the \coss\ distribution in Run IIb data and the global best-fit model (solid line) and the SM (dashed line) for (a) leptonic $W$ boson decays in \ljets\ events, (b) hadronic $W$ boson decays in \ljets\ events, and (c) dilepton events.}
\end{figure}

\begin{figure*}
\includegraphics[scale=0.40]{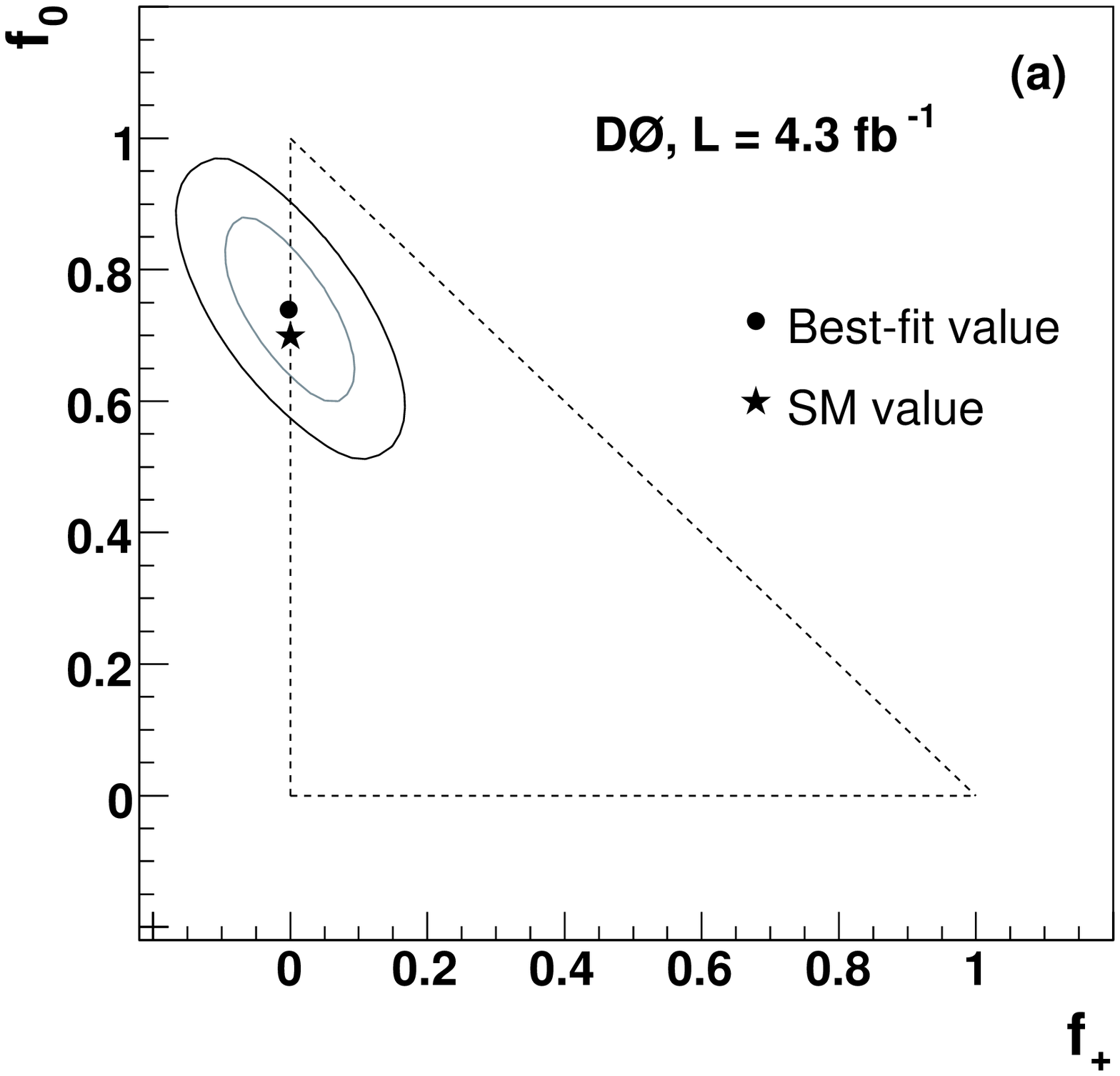}
\includegraphics[scale=0.40]{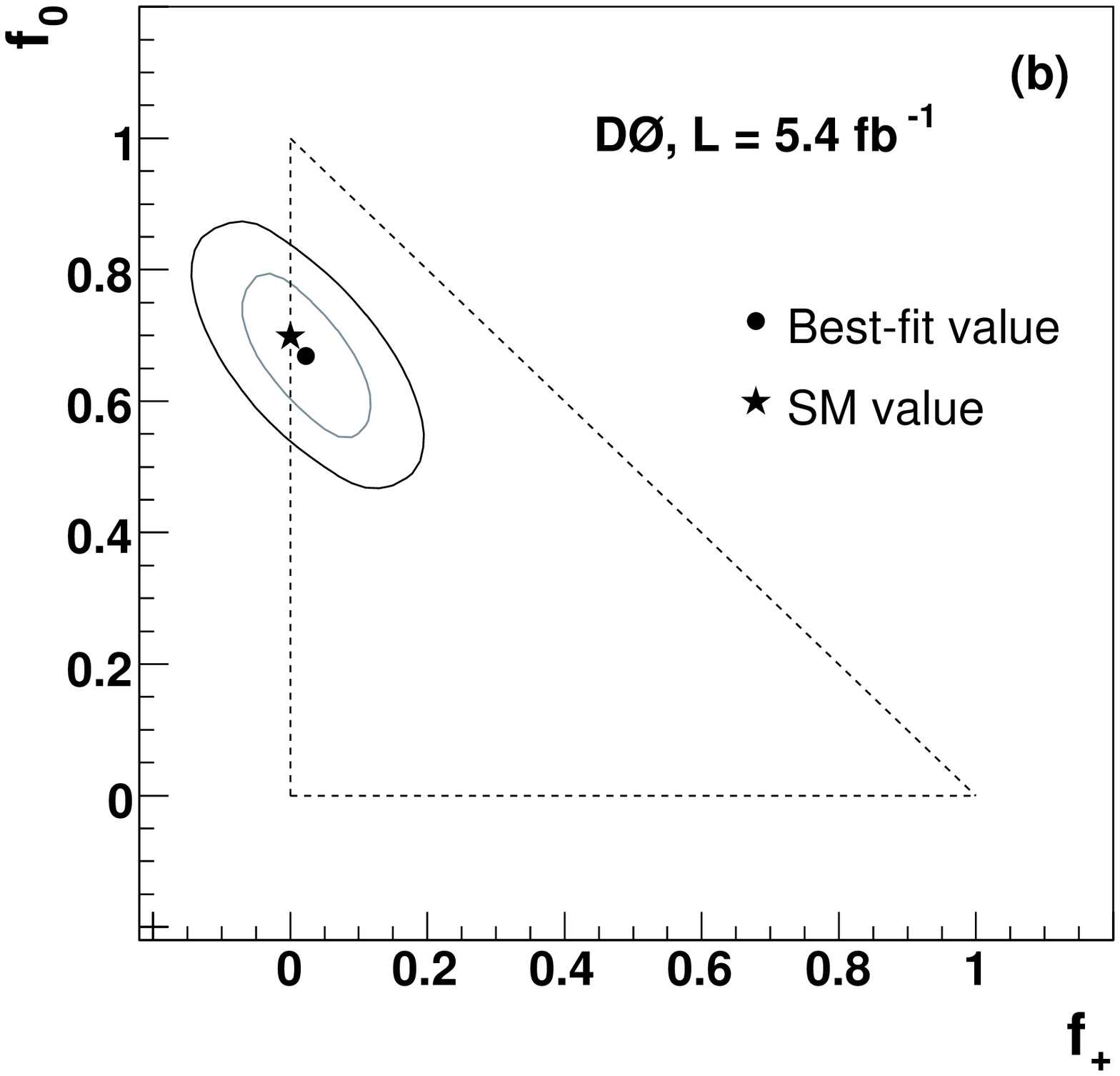}
\caption{\label{fig:data2dfit}  Result of the model-independent $W$ boson helicity fit for (a) the Run IIb data sample and (b) the combined Run IIa and Run IIb data sample.  In both plots, the ellipses indicate the 68\% and  95\% C.L.  contours, the dot shows the best-fit value, the triangle corresponds to the physically allowed region where $f_0 + f_+ \le 1$, and the star marks the expectation from the SM.}
\end{figure*}

The comparison between the best-fit model and the data is shown in Fig.~\ref{fig:data2dmodel}, and
the 68\% and 95\% C.L. contours in the $(f_+,f_0)$ plane are shown in Fig.~\ref{fig:data2dfit}(a). To account for systematic uncertainties, we perform a MC smearing of the $L$ distribution, where the width of the smearing in $f_0$ and $f_+$ is given by the systematic uncertainty on each helicity fraction, and the correlation coefficient of $-0.83$ between them is taken into account.

To assess the consistency of the result with the SM, we note that the change in $-\ln L(f_0,f_+)$ (Eq.~\ref{eq:lhood}) between the best fit and the SM points is 0.24 considering only statistical uncertainties and 0.16 when systematic uncertainties are included.  The probability of observing a greater deviation from the SM due to 
fluctuations in the data is 78\% when only the statistical uncertainty is considered and 85\% when both
statistical and systematic uncertainties are considered.

We have also split the data sample in various ways to check the internal consistency of the measurement.  Using $\ell+$jets events only, we find
\begin{eqnarray}
f_0 &=& 0.767 \pm 0.117 \hbox{ (stat.)}, \\  \nonumber
f_+ &=& 0.018 \pm 0.061 \hbox{ (stat.)}; 
\end{eqnarray}

and when using only dilepton events we find
\begin{eqnarray}
f_0 &=& 0.677 \pm 0.144 \hbox{ (stat.)}, \\ \nonumber
f_+ &=& -0.013 \pm 0.065 \hbox{ (stat.)}.
\end{eqnarray}

We also divide the sample into events with only electrons ($e+$jets and $ee$) and events with only muons ($\mu+$jets and $\mu\mu$).  The results for electrons only are\\*
\begin{eqnarray}
f_0 &=& 0.816 \pm 0.142 \hbox{ (stat.)}, \\ \nonumber
f_+ &=& -0.063 \pm 0.066 \hbox{ (stat.)},
\end{eqnarray}
\\*
and for muons only are
\begin{eqnarray}
f_0 &=& 0.618 \pm 0.150 \hbox{ (stat.)}, \\ \nonumber
f_+ &=& 0.130 \pm 0.081 \hbox{ (stat.)}.
\end{eqnarray}

Finally, we perform fits in which one of the two helicity fractions is fixed to its SM value.
Constraining $f_0$, we find
\begin{eqnarray}
f_+ = 0.014 \pm 0.025 \pm \hbox{ (stat.)} \pm 0.028  \hbox{(syst.)} ,
\end{eqnarray}

We also constrain $f_+$ and measure $f_0$, finding
\begin{eqnarray}
f_0 = 0.735 \pm 0.051 \hbox{ (stat.)} \pm 0.051   \hbox{(syst.)}.
\end{eqnarray}

\section{\label{sec:p17p20}Combination with Our Previous Measurement}

To combine this result with the previous measurement from Ref.~\cite{prevd0result}, we repeat the maximum likelihood fit with the earlier and current data samples and their respective MC models, treating them as separate channels in the fit.  This is equivalent to multiplying the two-dimensional likelihood distributions in $f_0$ and $f_+$ corresponding to the two data sets.  We determine the systematic uncertainty on the combined result by treating most uncertainties as correlated (the exception is template statistics) and propagating the uncertainties to the combined result.  The results are presented in Table~\ref{tab:p17p20combsyst}.

\begin{table}[hhh]
\caption{\label{tab:p17p20combsyst} Summary of the combined systematic uncertainties on $f_+$ and $f_{0}$ for Run IIa and Run IIb.}
\begin{tabular}{lcc}
\hline
\hline
Source & Uncertainty ($f_+$) & Uncertainty ($f_0$)  \\ \hline
Jet energy scale       & 0.009  & 0.010 \\
Jet energy resolution  & 0.004 & 0.008 \\
Jet ID        & 0.005 & 0.007 \\
Top mass          & 0.012 & 0.009\\
Template statistics   & 0.011 & 0.021 \\
\ttbar ~model          & 0.024 & 0.039 \\
Background model       & 0.008   & 0.023  \\
Heavy flavor fraction   & 0.010 & 0.022 \\
$b$ fragmentation  & 0.002 &  0.004 \\
PDF                 & 0.000 & 0.001 \\
Analysis consistency   & 0.004 & 0.006 \\
Muon ID         &   0.002 & 0.017 \\
Muon trigger   &  0.003 & 0.024 \\ \hline
Total                  & 0.034  &   0.065 \\ \hline
\hline
\end{tabular}
\end{table}

The combined result for the entire 5.4 fb$^{-1}$ sample is
\begin{eqnarray}
f_0 &=& 0.669 \pm 0.078 \hbox{ (stat.)} \pm  0.065 \hbox{ (syst.)}, \\ \nonumber
f_+ &=& 0.023 \pm 0.041 \hbox{ (stat.)} \pm  0.034 \hbox{ (syst.)}.
\end{eqnarray}
The combined likelihood distribution is presented in Figs.~\ref{fig:data2dfit}(b).  The probability of observing a greater deviation from the SM due to fluctuations in the data is 83\% when only statistical uncertainties are considered and 98\% when systematic uncertainties are included.

Constraining $f_0$ to the SM value,  we find
\begin{eqnarray}
f_+ = 0.010 \pm 0.022 \hbox{ (stat.)} \pm 0.030  \hbox{ (syst.)}
\end{eqnarray}
and constraining $f_+$ to the SM value gives
\begin{eqnarray}
f_0 = 0.708 \pm 0.044 \hbox{ (stat.)} \pm 0.048 \hbox{ (syst.)}.
\end{eqnarray}

\section{\label{sec:summ}Conclusion}
We have measured the helicity of $W$ bosons arising from top quark decay in \ttbar\ events using both the $\ell+$jets and dilepton decay channels and find
\begin{align}
f_0 = 0.669 & \pm  0.102 \\ \nonumber
                    &[ \pm  0.078 \hbox{ (stat.)} \pm  0.065 \hbox{ (syst.)}], \\ \nonumber  
f_+ = 0.023 & \pm   0.053 \\ \nonumber
                     &[ \pm  0.041 \hbox{ (stat.)} \pm  0.034  \hbox{ (syst.)}].
\end{align}
in a model-independent fit.  The consistency of this measurement with the SM values $f_0 = 0.698$, $f_+=3.6\times10^{-4}$ is 98\%.  Therefore, we report no evidence for new physics at the $tWb$ decay vertex.

\section{Acknowledgement}
%
We thank the staffs at Fermilab and collaborating institutions,
and acknowledge support from the
DOE and NSF (USA);
CEA and CNRS/IN2P3 (France);
FASI, Rosatom and RFBR (Russia);
CNPq, FAPERJ, FAPESP and FUNDUNESP (Brazil);
DAE and DST (India);
Colciencias (Colombia);
CONACyT (Mexico);
KRF and KOSEF (Korea);
CONICET and UBACyT (Argentina);
FOM (The Netherlands);
STFC and the Royal Society (United Kingdom);
MSMT and GACR (Czech Republic);
CRC Program and NSERC (Canada);
BMBF and DFG (Germany);
SFI (Ireland);
The Swedish Research Council (Sweden);
and
CAS and CNSF (China).
%

\end{document}